\documentclass[
aps,
prd,
showkeys,
showpacs,
superscriptaddress,
groupedaddress,
amsfonts
]{revtex4-1}
\usepackage{subcaption}
\usepackage{graphicx}  
\usepackage{blindtext}
\usepackage{amssymb}
\usepackage{url}
\usepackage{amsmath}
\usepackage{amsthm}
\usepackage{sidecap}

\theoremstyle{remark}

\theoremstyle{definition}

\usepackage{hyperref}
\usepackage{amsmath,amsfonts,amssymb}
\hypersetup{dvips,dvipdfm,colorlinks=true,urlcolor=green,filecolor=magenta,linktoc=page,citecolor=red,linkcolor=blue,bookmarks=true}
\usepackage{graphicx,epstopdf}
\usepackage{dcolumn}   
\usepackage{subcaption}
\usepackage{multirow,booktabs}
\usepackage{caption}

\DeclareCaptionJustification{justified}{\leftskip=0pt \rightskip=0pt \parfillskip=0pt plus 1fil}

\begin{document}
	
	\title{A Dynamical System Analysis of cosmic evolution with coupled phantom dark energy with dark matter}
	\author{Soumya Chakraborty}
	\email{soumyachakraborty150@gmail.com}
	\author{Sudip Mishra\thanks{corresponding author}}
	\email[corresponding author: ]{sudipcmiiitmath@gmail.com}
	\author{Subenoy Chakraborty}
	\email{schakraborty.math@gmail.com}
	\affiliation{ Department of Mathematics, Jadavpur University, Kolkata- 700032, WB, India.}
	
	\begin{abstract}
	The present work is an example of the application of the dynamical system analysis in the context of cosmology.  Here cosmic evolution is considered in the background of homogeneous and isotropic flat Friedmann-Lema\^{i}tre-Robertson-Walker space-time with interacting dark energy and varying mass dark matter as the matter content.  The DE is chosen as phantom scalar field with self-interacting potential while the DM is in the form of dust.  The potential of the scalar field and the mass function of dark matter are chosen as exponential or power-law form or in their product form.  Using suitable dimensionless variables the Einstein field equations and the conservation equations constitute an autonomous system.  The stability of the non-hyperbolic critical points are analyzed by using center manifold theory.  Finally, cosmological phase transitions have been detected through bifurcation analysis which has been done by Poincar\'{e} index theory.
	\end{abstract}

	
	
	\maketitle

	\section{Introduction}
The unexpected accelerated expansion of the universe as predicted by recent series of observations is speculated by cosmologist as a smooth transition from decelerated era in recent past \cite{Riess:1998cb,Perlmutter:1998np,Spergel:2003cb,Allen:2004cd,Riess:2004nr}.  The cosmologists are divided in opinion about the cause of this transition.  One group has the opinion of modification of the gravity theory while others are in favour introducing exotic matter component.  Due to two severe drawbacks \cite{RevModPhys.61.1} of the cosmological constant as a DE candidate dynamical DE models namely quintessence field (canonical scalar field), phantom field \cite{Caldwell:2003vq,Vikman:2004dc,Nojiri:2005sr,Saridakis:2009pj,Setare:2008mb} (ghost scalar field) or a unifield model named quintom \cite{Feng:2004ad,Guo:2004fq,Feng:2004ff,Feng:2004ff} are popular in the literature.\par    
	However, a new cosmological problem arises due to the dynamical nature of the DE although vacuum energy and DM scale independently during cosmic evolution but why their energy densities are nearly equal today.  To resolve this coincidence problem cosmologists introduce interaction between the DE and DM.  As the choice of this interaction is purely phenomenological so various models appear to match the observational prediction.  Although these models may resolve the above coincidence problem but a non-trivial, almost tuned sequence of cosmological eras \cite{Amendola:2006qi} appear as a result.  Further, the interacting phantom DE models \cite{Chen:2008ft,Nunes:2004wn,Clifton:2007tn,Xu:2012jf,Zhang:2005jj,Fadragas:2014mra,Gonzalez:2007ht} deal with some special coupling forms, alleviating the coincidence problem.\par
	Alternatively cosmologists put forward with a special type of interaction between DE and DM where the DM particles has variable mass, depending on the scalar field representing the DE \cite{Anderson:1997un}.  Such type of interacting model is physically more sound as scalar field dependent varying mass model appears in string theory or scalar-tensor theory \cite{PhysRevLett.64.123}.  This type of interacting model in cosmology considers mass variation as linear \cite{Farrar:2003uw,Anderson:1997un,Hoffman:2003ru}, power law \cite{Zhang:2005rg} or exponential \cite{Berger:2006db,PhysRevD.66.043528,PhysRevD.67.103523,PhysRevD.75.083506,Amendola:1999er,Comelli:2003cv,PhysRevD.69.063517} on the scalar field.  Among these the exponential dependence is most suitable as it not only solves the coincidence problem but also gives stable scaling behaviour.\par
	In the present work, varying mass interacting DE/DM model is considered in the background of homogeneous and isotropic space-time model.  Due to highly coupled nonlinear nature of the Einstein field equations it is not possible to have any analytic solution.  So by using suitable dimensionless variables the field equations are converted to an autonomous system.  The phase space analysis of non-hyperbolic equilibrium points has been done by center manifold theory (CMT)  for various choices of the mass functions and the scalar field potentials.  The paper is organized as follows:  Section \ref{BES} deals with basic equations for the varying mass interacting dark energy and dark matter cosmological model.  Autonomous system is formed and critical points are determined in Section \ref{FASC}.  Also stability analysis of all critical points for various choices of the involving parameters are shown in this section.  Possible bifurcation scenarios \cite{10.1140/epjc/s10052-019-6839-8, 1950261, 1812.01975} by Poincar\'{e} index theory and global cosmological evolution have been examined in Section \ref{BAPGCE}.  Finally, brief discussion and important concluding remarks of the present work is proposed in Section \ref{conclusion}.

	\section{Varying mass interacting dark energy and dark matter cosmological model : Basic Equations\label{BES}}
Throughout this paper, we assume a homogeneous and isotropic universe with the flat Friedmann-Lema\^{i}tre-Robertson-Walker (FLRW) metric	as follows:
\begin{equation}
ds^2=-dt^2+a^2(t)~d{\Sigma}^2,
\end{equation}
where `$t$' is the comoving time;  $a(t)$ is the scale factor; $d{\Sigma}^2$ is the 3D flat space line element.\\
The	Friedmann equations in the background of flat FLRW metric can be expressed as
\begin{eqnarray}
3H^2&=&k^2(\rho_\phi +\rho_{_{DM}}),\label{equn2}\\
2\dot{H}&=&-k^2(\rho_\phi +p_\phi +\rho_{_{DM}}),\label{equn3}
\end{eqnarray}
where `$\cdot $' denotes the derivative with respect to $t$; $\kappa(=\sqrt{8\pi G}$) is the gravitational coupling; $\{\rho_\phi,p_\phi\}$ are the energy density and thermodynamic pressure of the phantom scalar field $\phi$ (considered as DE) having expressions                       	
	\begin{align}
	\begin{split}
	\rho_{\phi}&=-\frac{1}{2}\dot{\phi}^2+V(\phi),\\
	p_\phi&=-\frac{1}{2}\dot{\phi}^2-V(\phi),\label{equn4}
	\end{split}
	\end{align}
	and $\rho_{_{DM}}$ is the energy density for the dark matter in the form of dust having expression 
		\begin{align}
	\rho_{_{DM}}=M_{_{DM}}(\phi)n_{_{DM}},\label{equn5}
	\end{align}
where $n_{_{DM}}$, the number density \cite{Leon:2009dt} for DM satisfies the number conservation equation	
	\begin{align}
	\dot{n}_{_{DM}}+3H n_{_{DM}}=0.\label{equn6}
	\end{align}
Now differentiating (\ref{equn5}) and using (\ref{equn6}) one has the DM conservation equation as
\begin{align}
\dot{\rho}_{_{DM}}+3H\rho_{_{DM}}=\frac{d}{d\phi}\left\{\ln M_{_{DM}}(\phi)\right\}\dot{\phi}\rho_{_{DM}},\label{equn7}
\end{align}
which shows that mass varying DM (in the form of dust) can be interpreted as a barotropic fluid with variable equation of state : $\omega_{_{DM}}=\frac{d}{d\phi}\left\{\ln M_{_{DM}}(\phi)\right\}\dot{\phi}$.  Now due to Bianchi identity, using the Einstein field equations (\ref{equn2}) and (\ref{equn3}) the conservation equation for DE takes the form
	\begin{align}
\dot{\rho}_{\phi}+ 3H(\rho_{\phi}+p_{\phi})=-\frac{d}{d\phi}\left\{\ln M_{_{DM}}(\phi)~\right\}\dot{\phi}\rho_{_{DM}}.\label{equn8}
\end{align}	
or using (\ref{equn4}) one has
\begin{align}
\ddot{\phi}+3H\dot{\phi}-\frac{\partial V}{\partial \phi}=\frac{d}{d\phi}\left\{\ln M_{_{DM}}(\phi)\right\}\rho_{_{DM}}.\label{equn9}
\end{align}		
The combination of the conservation equations (\ref{equn7})	and (\ref{equn8}) for DM (dust) and phantom DE (scalar) shows that the interaction between these two matter components depends purely on the mass variation, i.e., $Q=\frac{d}{d\phi}\left\{\ln M_{_{DM}}(\phi)\right\}\rho_{_{DM}}$.  So, if $M_{_{DM}}$ is an increasing function of $\phi$, i.e., $Q>0$ then energy is exchanged from DE to DM while in the opposite way if $M_{_{DM}}$ is a decreasing function of $\phi$.  Further, combining equations (\ref{equn7})	and (\ref{equn8}) the total matter $\rho_{tot}=\rho_{DM}+\rho_{DE}$ satisfies
\begin{align}
\dot{\rho}_{tot}+3H(\rho_{tot}+p_{tot})=0
\end{align}
with
\begin{align}
\omega_{tot}=\frac{p_{\phi}}{\rho_{\phi}+\rho_{_{DM}}}=\omega_{\phi}\Omega_{\phi}.
\end{align}	
Here $\omega_{\phi}=\frac{p_{\phi}}{\rho_{\phi}}$ is the equation of state parameter for phantom field and  $\Omega_{\phi}=\frac{\rho_{\phi}}{\frac{3H^2}{\kappa^2}}$ is the density parameter for DE.

\section{Formation of Autonomous System : Critical point and stability analysis\label{FASC}}
    	In the present work the dimensionless variables can be taken as \cite{Leon:2009dt}
    		\begin{eqnarray}
    		x:&=&\frac{\kappa\dot{\phi}}{\sqrt{6}H}, \\
    		y:&=&\frac{\kappa\sqrt{V(\phi)}}{\sqrt{3}H}, \\
    		z:&=&\frac{\sqrt{6}}{\kappa \phi}
    		\end{eqnarray}
    		together with $N=\ln a$  and the expression of the cosmological parameters can be written as
    	\begin{align}
    		\Omega_{\phi}\equiv \frac{{\kappa}^2\rho_{\phi}}{3H^2}&=-x^2+y^2,\label{eq4}
    		 \end{align}
    		 \begin{equation}
    		\omega_{\phi}= \frac{-x^2-y^2}{-x^2+y^2}
    		\end{equation}
    		and
    		\begin{equation}
    		\omega_{tot}=-x^2-y^2.
    		\end{equation}
   For the scalar field potential we consider two well
   studied cases in the literature, namely the power-law
    	\begin{equation}
    V(\phi)=V_0 \phi^{-\lambda}
    \end{equation}
    and the exponential dependence as
    \begin{equation}
     V(\phi)=V_1 e^{-\kappa\lambda \phi}.
    \end{equation}
   For the dark matter particle mass we also consider power-law 
    	    \begin{eqnarray}
    	    M_{_{DM}}(\phi)&=& M_0 {\phi}^{-\mu}
    	    \end{eqnarray} 
    	    and the exponential dependence as
    	    \begin{eqnarray}
    	    M_{_{DM}}(\phi)&=& M_1 e^{-\kappa\mu \phi},
    	    \end{eqnarray}
    	    where $V_0,V_1,M_0,M_1 (>0)$ and $\lambda,\mu$ are constant parameters.  Here we study the dynamical analysis of this cosmological system for five possible models.  In Model $1$ (\ref{M1}) we consider $V(\phi)=V_0\phi^{-\lambda}, M_{_{DM}}(\phi)=M_0\phi^{-\mu}$, in Model $2$ (\ref{M2}) we consider  $V(\phi)=V_0\phi^{-\lambda}, M_{_{DM}}(\phi)=M_1e^{-\kappa\mu\phi}$, in Model $3$ (\ref{M3}) we consider $V(\phi)=V_1e^{-\kappa\lambda\phi}, M_{_{DM}}(\phi)=M_0\phi^{-\mu}$, in Model $4$ (\ref{M4}) we consider $V(\phi)=V_1 e ^{-\kappa\lambda\phi}, M_{_{DM}}(\phi)=M_1e^{-\kappa\mu\phi}$ and lastly in Model $5$ (\ref{M5}) we consider $V(\phi)=V_2\phi^{-\lambda} e ^{-\kappa\lambda\phi}, M_{_{DM}}(\phi)=M_2\phi^{-\mu}e^{-\kappa\mu\phi}$, where $V_2=V_0V_1$ and $M_2=M_0M_1$.
    	  
    	    \subsection{Model 1: Power-law potential and power-law-dependent dark-matter particle mass \label{M1}}
    		 In this consideration evolution equations in Section \ref{BES} can be converted to an autonomous system as follows 
    			\begin{eqnarray}
    			x'&=&-3x+\frac{3}{2}x(1-x^2-y^2)-\frac{\lambda y^2 z}{2}-\frac{\mu}{2}z(1+x^2-y^2),\label{eq9} \\
    			y'&=&\frac{3}{2}y(1-x^2-y^2)-\frac{\lambda xyz}{2},\label{eq10}  \\
    			z'&=&-xz^2,\label{eq11} 
    			\end{eqnarray}
    				where $\lq$dash' over a variable denotes differentiation with respect to $ N=\ln a $.\bigbreak
 To obtain the stability analysis of the critical points corresponding to the autonomous system $(\ref{eq9}-\ref{eq11})$, we consider four possible choices of $\mu$ and $\lambda$ as
$(i)$ $\mu\neq0$ and $\lambda\neq0$, $~~~(ii)$ $\mu\neq0$ and $\lambda=0$, $(iii)$ $\mu=0$ and $\lambda\neq0$, $(iv)$ $\mu=0$ and $\lambda=0$.
\subsubsection*{Case-(i)$~$\underline{$\mu\neq0$ and $\lambda\neq0$}}	
    In this case we have three real and physically meaningful critical points $A_1(0, 0, 0)$, $A_2(0, 1, 0)$ and $A_3(0, -1, 0)$.  First we determine the Jacobian matrix at these critical points corresponding to the autonomous system $(\ref{eq9}-\ref{eq11})$.  Then we shall find the eigenvalues and corresponding eigenvectors of the Jacobian matrix.  After that we shall obtain the nature of the vector field near the origin for every critical points.  If the critical point is hyperbolic in nature we use Hartman-Grobman theorem and if the critical point is non-hyperbolic in nature we use Center Manifold Theory \cite{Chakraborty:2020vkp}.  At every critical points the eigenvalues of the Jacobian matrix corresponding to the autonomous system $(\ref{eq9}-\ref{eq11})$, value of cosmological parameters and the nature of the critical points are shown in Table \ref{TI}.
    	\begin{table}[h]
    			\caption{\label{TI}Table shows the eigenvalues, cosmological parameters and nature of the critical points corresponding to each critical points $(A_1-A_3)$.}
    			
    			\begin{tabular}{|c|c c c|c|c|c| c|c|}
    				\hline
    				\hline	
    				\begin{tabular}{@{}c@{}}$~~$\\$~Critical~ Points$\\$~~$\end{tabular}  ~~  &  $ \lambda_1 $  ~~ & $\lambda_2$ ~~ & $\lambda_3$& $~\Omega_\phi~$&$~\omega_\phi~$ &$~\omega_{tot}~$& $~q~$ & $Nature~of~critical~points$ \\ \hline\hline
    				~ & ~ & ~& ~& ~ & ~ & ~ & ~ & ~\\
    				$A_1(0,0,0)$ & $-\frac{3}{2}$ & $\frac{3}{2}$ & 0 & 0 & Undetermined & 0 &$\frac{1}{2}$& Non-hyperbolic\\ 
    				~ & ~ & ~& ~& ~ & ~ & ~ & ~ & ~\\\hline
    					~ & ~ & ~& ~& ~ & ~ & ~ & ~ & ~\\
    				$A_2(0,1,0)$ & $-3$ & $-3$ & 0 & 1 & $-1$ & $-1$&$-1$& Non-hyperbolic\\
    				~ & ~ & ~& ~& ~ & ~ & ~ & ~& ~\\ \hline
    					~ & ~ & ~& ~& ~ & ~ & ~ & ~ & ~\\
    				$A_3(0,-1,0)$ & $-3$ & $-3$ & $0$ & $1$ & $-1$ & $-1$&$-1$& Non-hyperbolic\\
    				~ & ~ & ~& ~& ~ & ~ & ~ & ~ & ~\\ \hline
    			\end{tabular}
    		    \end{table}

\begin{figure}[h]
	\centering
	\includegraphics[width=1\textwidth]{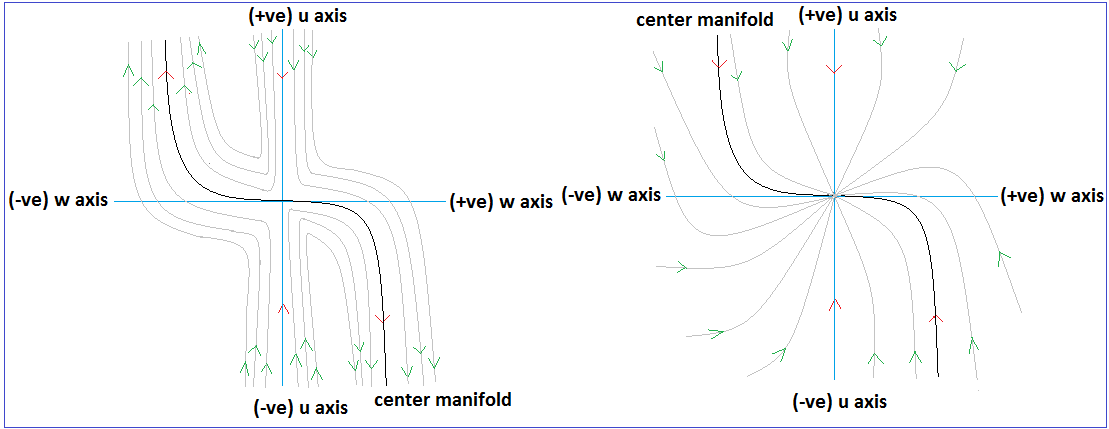}
	\caption{Vector field near the origin for the Critical point $A_3$ in $(uw)$-plane.  L.H.S. figure is for $\lambda>0$ and R.H.S. figure is for $\lambda<0$.}
	\label{24}
\end{figure}

\begin{figure}[h]
	\includegraphics[width=1\textwidth]{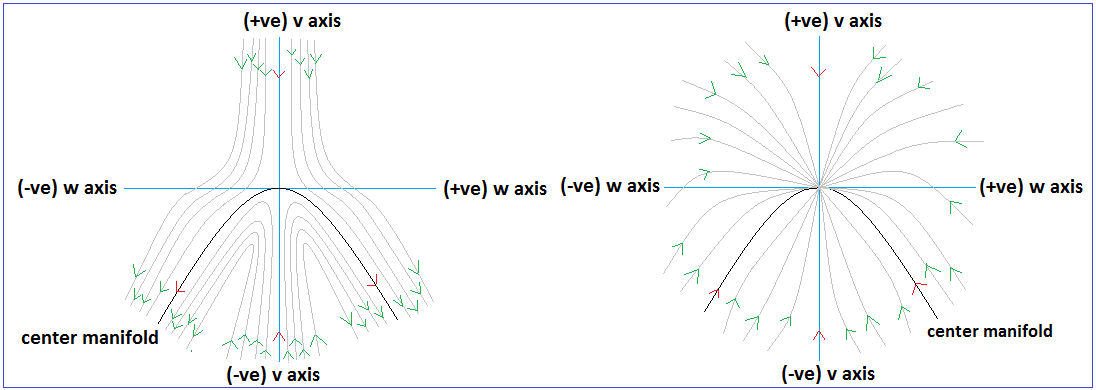}
	\caption{Vector field near the origin for the Critical point $A_3$ in $(vw)$-plane.  L.H.S. figure is for $\lambda>0$ and R.H.S. figure is for $\lambda<0$.}
	\label{25}
\end{figure}

    	       \begin{center}
    	     $1.~Critical~Point~A_1$
    	       \end{center}
    			The Jacobian matrix at the critical point $A_1$ can be put as
    			\begin{equation}\renewcommand{\arraystretch}{1.5}	
    			J(A_1)=\begin{bmatrix}
    			-\frac{3}{2} & 0 & -\frac{\mu}{2}\\	
    			~~0  &  \frac{3}{2} & ~~0\\
    			~~0 & 0 & ~~0 
    			\end{bmatrix}.\label{eq12}	
    			\end{equation}	
    			The eigenvalues of $J(A_1)$ are $-\frac{3}{2}$, $\frac{3}{2}$ and $0$.  $[1, 0, 0]^T$ ,  $[0, 1, 0]^T$ and $[-\frac{\mu}{3}, 0, 1]^T$ are the eigenvectors corresponding to the eigenvalues $-\frac{3}{2}$, $\frac{3}{2}$ and 0 respectively.  Since the critical point $A_1$ is non-hyperbolic in nature, so we use Center Manifold Theory for analyzing the stability of this critical point.  From the entries of the Jacobian matrix we can see that there is a linear term of $z$ corresponding to the eqn.(\ref{eq9}) of the autonomous system $(\ref{eq9}-\ref{eq11})$.  But the eigen value $0$ of the Jacobian matrix (\ref{eq12}) is corresponding to (\ref{eq11}).  So we have to introduce another coordinate system $(X,~Y,~Z)$ in terms of $(x,~y,~z)$.  By using the eigenvectors of the Jacobian matrix (\ref{eq12}), we introduce the following coordinate system
    			\begin{equation}\renewcommand{\arraystretch}{1.5}	
    			\begin{bmatrix}
    			X\\
    			Y\\
    			Z
    			\end{bmatrix}\renewcommand{\arraystretch}{1.5}
    			=\begin{bmatrix}
    			1 & 0 & \frac{\mu}{3} \\	
    			0 &  1 & 0 \\
    			0 & 0 & 1
    			\end{bmatrix}\renewcommand{\arraystretch}{1.5}
    			\begin{bmatrix}
    			x\\
    			y\\
    			z
    			\end{bmatrix}\label{eq15}	
    			\end{equation}		
    			and in these new coordinate system the equations $(\ref{eq9}-\ref{eq11})$ are transformed into	
    			\begin{equation}\renewcommand{\arraystretch}{1.5}	
    			\begin{bmatrix}
    			X'\\
    			Y'\\
    			Z'
    			\end{bmatrix}\renewcommand{\arraystretch}{1.5}
    			=\begin{bmatrix}
    			-\frac{3}{2} & 0 & 0 \\	
    			~~0 &  \frac{3}{2} & 0 \\
    			~~0 & 0 & 0
    			\end{bmatrix}
    			\begin{bmatrix}
    			X\\
    			Y\\
    			Z
    			\end{bmatrix}		
    			+\renewcommand{\arraystretch}{1.5}	
    			\begin{bmatrix}
    			non\\
    			linear\\
    			terms
    			\end{bmatrix}.	
    			\end{equation}	
    			By Center Manifold Theory there exists a continuously differentiable function 	$h$:$\mathbb{R}$$\rightarrow$$\mathbb{R}^2$ such that 
    			\begin{align}\renewcommand{\arraystretch}{1.5}
    			h(Z)=\begin{bmatrix}
    			X \\
    			Y \\
    			\end{bmatrix}
    			=\begin{bmatrix}
    			a_1Z^2+a_2Z^3+a_3Z^4 +\mathcal{O}(Z^5)\\
    			b_1Z^2+b_2Z^3+a_3Z^4 +\mathcal{O}(Z^5) 
    			\end{bmatrix}.
    			\end{align}
    			Differentiating both side with respect to $N$, we get 
    			\begin{eqnarray}
    			X'&=&(2a_1Z+3a_2Z^2+4a_3Z^3)Z',\\
    			Y'&=&(2b_1Z+3b_2Z^2+4b_3Z^3)Z',
    			\end{eqnarray}	
    			where $a_i$, $b_i$ $\in\mathbb{R}$.  We only concern about the non-zero coefficients of the lowest power terms in CMT as we analyze arbitrary small neighbourhood of the origin.  Comparing coefficients corresponding to power of Z we get, 
    			$a_1$=0, $a_2=\frac{2\mu^2}{27}$, $a_3=0$ and $b_i$=0 for all $i$.	
    			So, the center manifold is given by 
    			\begin{eqnarray}
    			X&=&\frac{2\mu^2}{27}Z^3,\label{eq18}\\
    			Y&=&0\label{eq19}
    			\end{eqnarray} 
    			and the flow on the Center manifold is determined by
    			\begin{eqnarray}
    			Z'&=&\frac{\mu}{3}Z^3+\mathcal{O}(Z^5).\label{eq20}
    			\end{eqnarray}
    			\begin{figure}[h]
    				\includegraphics[width=1\textwidth]{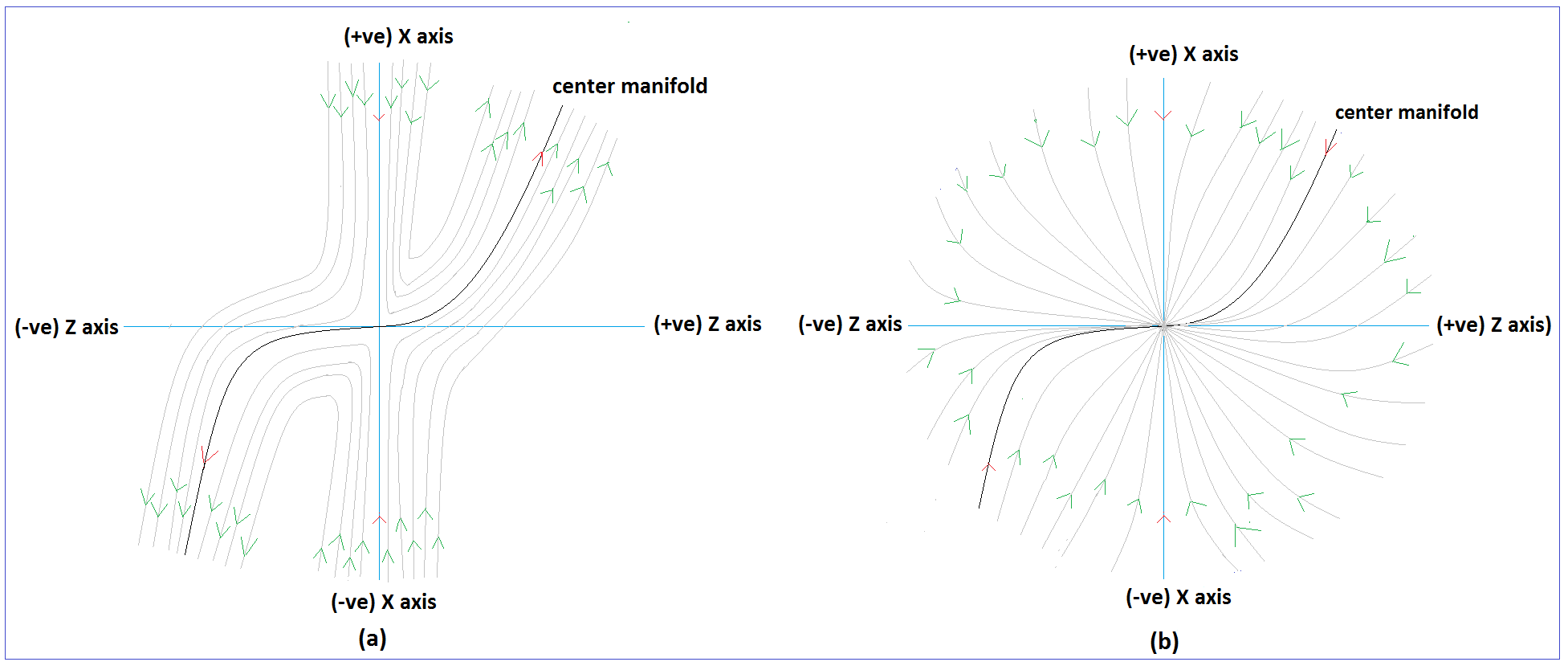}
    				\caption{Vector field near the origin for the critical point $A_1$ in $XZ$-plane.  L.H.S. figure is for $\mu>0$ and R.H.S. figure is for $\mu<0$. }
    				\label{A_1}
    			\end{figure}
    The flow on the center manifold depends on the sign of $\mu$.  If $\mu>0$ then $Z'>0$ for $Z>0$ and $Z'<0$ for $Z<0$.  Hence, we conclude that for $\mu>0$ the origin is a saddle node and unstable in nature (FIG.\ref{A_1}(a)).  Again if $\mu<0$ then $Z'<0$ for $Z>0$ and $Z'>0$ for $Z<0$.  So, we conclude that for $\mu<0$ the origin is a stable node, i.e., stable in nature (FIG.\ref{A_1}(b)).	\bigbreak	
	\begin{center}
    			$2.~Critical~Point~A_2$
    		\end{center}
    			The Jacobian matrix  at $A_2$ can be put as
    			\begin{equation}\renewcommand{\arraystretch}{1.5}	
    			J(A_2)=\begin{bmatrix}
    		-3 & ~~0 & -\frac{\lambda}{2}\\	
    			~~0  & -3 & ~~0\\
    			~~0 & ~~0 & ~~0 
    			\end{bmatrix}\label{eq21}.	
    			\end{equation}
    			The eigenvalues of the above matrix are $-3$, $-3$ and $0$.  $[1, 0, 0]^T$ and $[0, 1, 0]^T$ are the eigenvectors corresponding to the eigenvalue $-3$ and $\left[-\frac{\lambda}{6}, 0, 1\right]^T$ be the eigenvector corresponding to the eigenvalue $0$.  Since the critical point $A_2$ is non-hyperbolic in nature, so we use Center Manifold Theory for analyzing the stability of this critical point.  We first transform the coordinates into a new system $x=X,~ y=Y+1,~ z=Z$, such that the critical point $A_2$ moves to the origin. By using the eigenvectors of the Jacobian matrix $J(A_2)$, we introduce another set of new coordinates $(u,~v,~w)$ in terms of $(X,~Y,~Z)$ as
    			\begin{equation}\renewcommand{\arraystretch}{1.5}	
    			\begin{bmatrix}
    			u\\
    			v\\
    			w
    			\end{bmatrix}\renewcommand{\arraystretch}{1.5}
    			=\begin{bmatrix}
    			1 & 0 & \frac{\lambda}{6} \\	
    			0 &  1 & 0 \\
    			0 & 0 & 1
    			\end{bmatrix}\renewcommand{\arraystretch}{1.5}
    			\begin{bmatrix}
    			X\\
    			Y\\
    			Z
    			\end{bmatrix}\label{eq24}
    			\end{equation}		
    			and in these new coordinates the equations $(\ref{eq9}-\ref{eq11})$ are transformed into	
    			\begin{equation}	\renewcommand{\arraystretch}{1.5}
    			\begin{bmatrix}
    			u'\\
    			v'\\
    			w'
    			\end{bmatrix}
    			=\begin{bmatrix}
    			-3 & ~~0 & 0 \\	
    			~~0 & -3 & 0 \\
    			~~0 & ~~0 & 0
    			\end{bmatrix}
    			\begin{bmatrix}
    			u\\
    			v\\
    			w
    			\end{bmatrix}		
    			+	
    			\begin{bmatrix}
    			non\\
    			linear\\
    			terms
    			\end{bmatrix}.	
    			\end{equation}	
    			By center manifold theory there exists a continuously differentiable function 	$h$:$\mathbb{R}$$\rightarrow$$\mathbb{R}^2$ such that 
    			\begin{align}\renewcommand{\arraystretch}{1.5}
    			h(w)=\begin{bmatrix}
    			u \\
    			v \\
    			\end{bmatrix}
    			=\begin{bmatrix}
    			a_1w^2+a_2w^3 +\mathcal{O}(w^4)\\
    			b_1w^2+b_2w^3 +\mathcal{O}(w^4) 
    			\end{bmatrix}.
    			\end{align}
    			Differentiating both side with respect to $N$, we get 
    			\begin{eqnarray}
    			u'&=&(2a_1w+3a_2w^2)w'+\mathcal{O}(w^3)\label{eq25}\\
    			v'&=&(2b_1w+3b_2w^2)w'+\mathcal{O}(w^3)\label{eq26}
    			\end{eqnarray}
    			where $a_i$, $b_i$ $\in\mathbb{R}$.  We only concern about the non-zero coefficients of the lowest power terms in CMT as we analyze arbitrary small neighbourhood of the origin.  Comparing coefficients corresponding to power of $w$ both sides of (\ref{eq25}) and (\ref{eq26}), we get	
    			$a_1$=0, $a_2=\frac{\lambda^2}{108}$ and $b_1=\frac{\lambda^2}{72}$, $b_2=0$.  So, the center manifold can be written as 
    			\begin{eqnarray}
    			u&=&\frac{\lambda^2}{108}w^3,\label{eqn27}\\
    			v&=&\frac{\lambda^2}{72}w^2\label{eqn28}
    			\end{eqnarray} 
    				\begin{figure}
    				\includegraphics[width=1\textwidth]{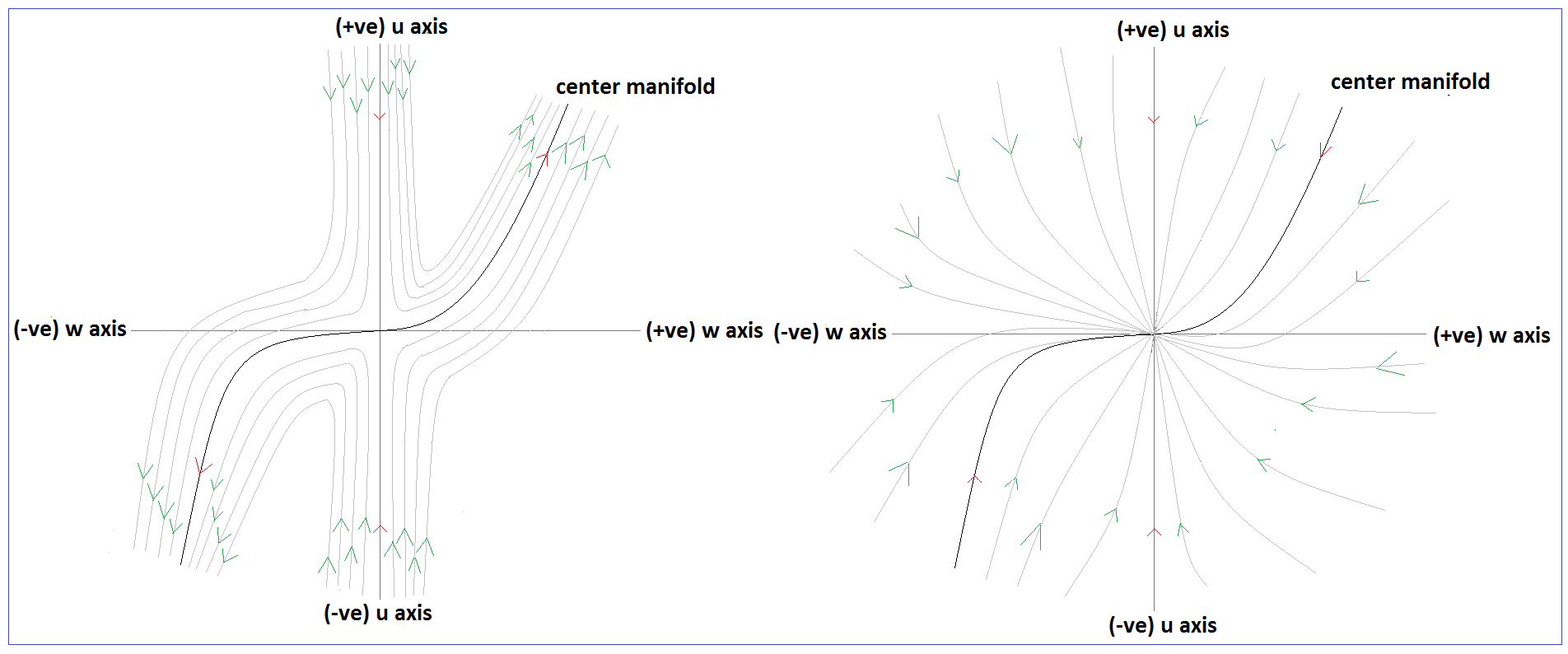}
    				\caption{Vector field near the origin for the critical point $A_2$ in (uw)-plane.  L.H.S. figure is for $\lambda>0$ and R.H.S. figure is for $\lambda<0$. }
    				\label{19}
    			\end{figure}
    			\begin{figure}
    				\includegraphics[width=1\textwidth]{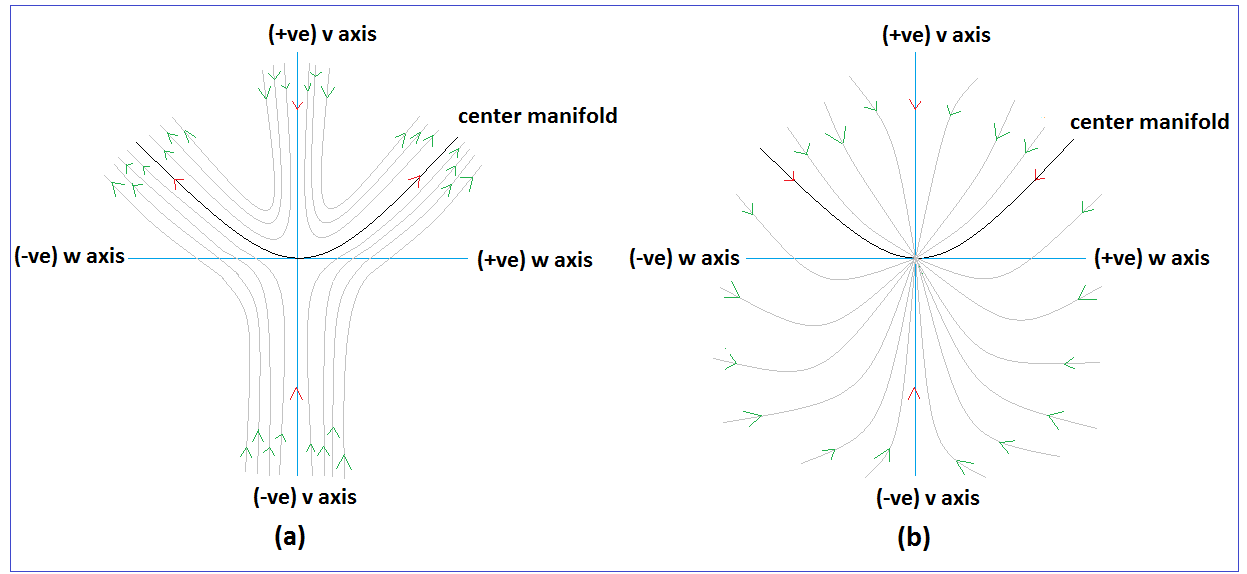}
    				\caption{Vector field near the origin for the critical point $A_2$ in $(vw)$-plane.  L.H.S. figure is for $\lambda>0$ and R.H.S. figure is for $\lambda<0$.}
    				\label{20}
    			\end{figure}
    			and the flow on the center manifold is determined by
    			\begin{eqnarray}
    			w'&=&\frac{\lambda}{6}w^3+\mathcal{O}(w^4) .\label{eq29}
    			\end{eqnarray}
    			Here we see the center manifold and the flow on the center manifold is completely same as the center manifold and the flow which was determined in \cite{1111.6247} and the stability of the vector field near the origin depends on the sign of $\lambda$.  If $\lambda<0$ then $w'<0$ for $w>0$ and $w'>0$ for $w<0$.  So, for $\lambda<0$ the origin is a stable node, i.e., stable in nature.  Again if $\lambda>0$ then $w'>0$ for $w>0$ and $w'<0$ for $w<0$.  So, for $\lambda>0$ the origin is a saddle node, i.e., unstable in nature.
    			  The vector field near the origin are shown as in FIG.\ref{19} and FIG.\ref{20} separately for $(wu)$-plane and $(wv)$-plane respectively.  As the new coordinate system $(u,~v,~w)$ is topologically equivalent to the old one, hence the origin in the new coordinate system, i.e., the critical point $A_2$ in the old coordinate system $(x,~y,~z)$ is a stable node for $\lambda<0$ and a saddle node for $\lambda>0$. 
\begin{center}
	$3.~Critical~Point~A_3$
\end{center}
    			The Jacobian matrix at the critical point $A_3$ is same as (\ref{eq21}).  So, the eigenvalues and corresponding eigenvectors are also same as above.  Now we transform the coordinates into a new system $x=X,~ y=Y-1,~ z=Z$, such that the critical point is at the origin.  Then by using the matrix transformation (\ref{eq24}) and after putting similar arguments as above, the expressions of the center manifold can be written as
    			\begin{eqnarray}
    			u&=&-\frac{\lambda^2}{108}w^3\label{eqn30},\\
    			v&=&-\frac{\lambda^2}{72}w^2\label{eqn31}
    			\end{eqnarray} 
    			and the flow on the center manifold is determined by
    			\begin{eqnarray}
    			w'&=&\frac{\lambda}{6}w^3+\mathcal{O}(w^4) .\label{eqn32}
    			\end{eqnarray}
    			Here also the stability of the vector field near the origin depends on the sign of $\lambda$.  Again as the expression of the flow on the center manifold is same as (\ref{eq29}). So we can conclude as above that for $\lambda<0$ the origin is a stable node,i.e., stable in nature and for $\lambda>0$ the origin is unstable due to its saddle nature.  The vector fields near the origin on $uw$-plane and $vw$-plane are shown as in FIG.\ref{24} and FIG.\ref{25} respectively.    Hence, the critical point $A_3$ is a stable node for $\lambda<0$ and a saddle node for $\lambda>0$.\bigbreak
\newpage
\subsubsection*{Case-(ii)$~$\underline{$\mu\neq0$ and $\lambda=0$}}
In this case the autonomous system $(\ref{eq9}-\ref{eq11})$ changes into
\begin{eqnarray}
x'&=&-3x+\frac{3}{2}x(1-x^2-y^2)-\frac{\mu}{2}z(1+x^2-y^2),\label{eq33} \\
y'&=&\frac{3}{2}y(1-x^2-y^2),\label{eq34}  \\
z'&=&-xz^2.\label{eq35} 
\end{eqnarray}
We have also three critical points corresponding to the above autonomous system, in which two are space of critical points.  The critical points for this autonomous system are $C_1(0, 0, 0)$, $C_2(0,1, z_c)$ and $C_3(0,-1,z_c)$ where $z_c$ is any real number.    Corresponding to the critical points $C_0$, $C_1$ and $C_2$ the eigenvalues of the Jacobian matrix, value of cosmological parameters and the nature of the critical points are same as $A_1$, $A_2$ and $A_3$ respectively.			

\begin{figure}[h]
	\centering
	\includegraphics[width=0.6\textwidth]{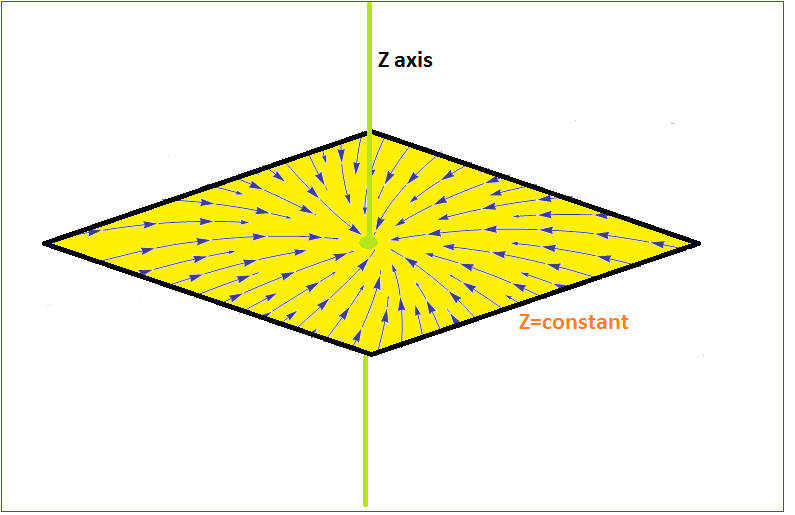}
	\caption{Vector field near about every point on $Z-$axis for the critical points $C_2$ and $C_3$.}
	\label{z_c}
\end{figure}

\begin{center}
	$1.~Critical~Point~C_1$ 
\end{center}
The Jacobian matrix $J(C_1)$ for the autonomous system $(\ref{eq33}-\ref{eq35})$ at this critical point is same as (\ref{eq12}). So, all the eigenvalues and the corresponding eigenvectors are also same as for $J(C_1)$.  If we put forward argument like the stability analysis of the critical point $A_1$ then the center manifold can be expressed as $(\ref{eq18}-\ref{eq19})$ and the flow on the center manifold is determined by $(\ref{eq20})$.  So the stability of the vector field near the origin is same as for the critical point $A_1$.
\begin{center}
	$2.~Critical~Point~C_2$ 
\end{center}
The Jacobian matrix at the critical point $C_2$ can be put as

\begin{equation}\renewcommand{\arraystretch}{1.5}	
J(C_2)=\begin{bmatrix}
-3 & ~~\mu z_c & 0\\	
~~0  &  -3 & 0\\
-z_c^2 & ~~0 & 0 
\end{bmatrix}.\label{eq36}	
\end{equation}
The eigenvalues of the above matrix are $-3$, $-3$, 0.  $\left[1, 0, \frac{z_c^2}{3}\right]^T$ and $[0, 1, 0]^T$ are the eigenvectors corresponding to the eigenvalue -3 and $[0, 0, 1]^T$ be the eigenvector corresponding to the eigenvalue 0. To apply CMT for a fixed $z_c$, first we transform the coordinates into a new system $x=X,~ y=Y+1,~ z=Z+z_c$, such that the critical point is at the origin and after that if we put forward argument as above to determine center manifold, then the center manifold can be written as
\begin{eqnarray}
X&=&0,\label{eq37}\\
Y&=&0\label{eq38}
\end{eqnarray} 
and the flow on the center manifold is determined by
\begin{eqnarray}
Z'&=&0.\label{eq39}
\end{eqnarray}
So, the center manifold is lying on the $Z$-axis and the flow on the center manifold can not be determined by (\ref{eq39}). Now,  if we project the vector field on the plane
which is parallel to $XY$-plane, i.e., the plane $Z=constant$(say), then the vector field is shown as in FIG.\ref{z_c}.  So every point on $Z$- axis is a stable star.

\begin{center}
	$2.~Critical~Point~C_3$ 
\end{center}
If we put forward argument as above to obtain the center manifold and the flow on the center manifold.  Then we will get the center manifold same as $(\ref{eq37}-\ref{eq38})$ and the flow on the center manifold is determined by (\ref{eq39}).  In this case also we will get the same vector field as FIG.\ref{z_c}.\bigbreak
From the above discussion, firstly we have seen that the space of critical points $C_2$ and $C_3$ are non-hyperbolic in nature but by using CMT we could not determine the vector field near those critical points and also the flow on the vector field.  So, in this case the last eqn.(\ref{eq35}) of the autonomous system $(\ref{eq33}-\ref{eq35})$ did not provide any special behaviour.  For this reason and as the expressions of $\Omega_\phi$, $\omega_\phi$ and $\omega_{total}$ depends only on $x$ and $y$ coordinates, we want to take only the first two equations of the autonomous system $(\ref{eq33}-\ref{eq35})$ and try to analyze the stability of the critical points which are lying on the plane, parallel to $xy-$plane, i.e., the plane $z=constant=c$ (say).  In $z=c$ plane the first two equations in $(\ref{eq33}-\ref{eq35})$ can be written as
\begin{eqnarray}
x'&=&-3x+\frac{3}{2}x(1-x^2-y^2)-\frac{\mu}{2}c(1+x^2-y^2),\label{eqn40} \\
y'&=&\frac{3}{2}y(1-x^2-y^2).\label{eqn41}  
\end{eqnarray}
In this case we have five critical points corresponding to the autonomous system $(\ref{eqn40}-\ref{eqn41})$.   The set of critical points, existence of critical points and the value of cosmological parameters are shown in Table \ref{T3} and the eigenvalues and the nature of critical points are shown in Table \ref{T4}.
\begin{table}[!]
	\caption{\label{T3}Table shows the set of critical points, existence of critical points and the value of cosmological parameters corresponding to the autonomous system $(\ref{eqn40}-\ref{eqn41})$. }
	\begin{tabular}{|c|c|c c |c c c c|}
		\hline
		\hline
		\begin{tabular}{@{}c@{}}$~~$\\$ CPs $\\$~$\end{tabular}  ~~  &  $ Existence $  ~~ & ~~$x$ ~~&~~ $y$& $\Omega_\phi$&~~$\omega_{\phi}$~~ &$\omega_{tot}$ & $~~~~q$ \\ \hline\hline
		\begin{tabular}{@{}c@{}}$~~$\\$ E_1 $\\$~$\end{tabular}  ~~  &  $For~all~\mu~and~c $&$0$&$~~~1$&$1$&$-1$&$-1$&$~~-1$  \\ \hline 
		\begin{tabular}{@{}c@{}}$~~$\\$ E_2 $\\$~$\end{tabular}   ~~  &  $For~all~\mu ~and~c $  ~~ & ~~$0$ &$~~-1$~~&$1$& $~-1$ ~~& $-1$ & $~~-1$\\ \hline 
		\begin{tabular}{@{}c@{}}$~~$\\$ E_3 $\\$~$\end{tabular}  ~~  & $For~all ~~\mu ~and~c $  ~~ & ~~$-\frac{\mu c}{3}$ ~~&~~$~~0$~&$-\frac{\mu^2 c^2}{9}$ & $~~-1$ ~~&~~ $-\frac{\mu^2 c^2}{9}$  &~~ $\frac{1}{2}\left(1-\frac{\mu^2 c^2}{3}\right)$\\ \hline 
		\begin{tabular}{@{}c@{}}$~~$\\$ E_4 $\\$~$\end{tabular}   ~~  & \begin{tabular}{@{}c@{}}$ For~c\neq 0~ and~$\\$for~all~\mu\in \left(-\infty,-\frac{3}{c}\right]\cup\left[\frac{3}{c},\infty\right)$ \end{tabular} ~~ & ~~$-\frac{3}{\mu c}$ ~~&~~ $\sqrt{1-\frac{9}{\mu^2 c^2}}$~~&~~$\left(1-\frac{18}{\mu^2 c^2}\right)$&$~~~~\frac{\mu^2 c^2}{18-\mu^2c^2}$~~&~~$-1$ ~~& ~~$~-1$\\ \hline 
		\begin{tabular}{@{}c@{}}$~~$\\$ E_5 $\\$~$\end{tabular}   ~~  & \begin{tabular}{@{}c@{}}$ For~c\neq 0~ and~$\\$for~all~\mu\in \left(-\infty,-\frac{3}{c}\right]\cup\left[\frac{3}{c},\infty\right)$ \end{tabular} ~~ & ~~$-\frac{3}{\mu c}$ ~~&~~ $-\sqrt{1-\frac{9}{\mu^2 c^2}}$~~&$\left(1-\frac{18}{\mu^2 c^2}\right)$ &~~$~~\frac{\mu^2 c^2}{18-\mu^2c^2}$~~&~~$-1$ ~~& ~~$~-1$\\ \hline 	
	\end{tabular}	
\end{table}

  \begin{table}[!]
	\caption{\label{T4}Table shows the eigenvalues $(\lambda_1, \lambda_2)$ of the Jacobian matrix corresponding to the critical points and the nature of all critical points $(E_1-E_5)$.}		 
	\begin{tabular}{|c|c c|c|}
		\hline
		\hline	
		\begin{tabular}{@{}c@{}}$~~$\\$ Critical~Points $\\$~$\end{tabular} &$ ~~\lambda_1 $ & $~~\lambda_2$ & $ Nature~~ of~~ Critical~~ points$ \\ \hline\hline
		\begin{tabular}{@{}c@{}}$~~$\\ $E_1$ \\$~$\end{tabular}  & $-3$ & $ -3  $&Hyperbolic\\ \hline
		\begin{tabular}{@{}c@{}}$~~$\\$ E_2 $\\$~$\end{tabular}  & $-3$ & $ -3  $& Hyperbolic\\ \hline
		\begin{tabular}{@{}c@{}}$~~$\\$ E_3 $\\$~$\end{tabular}  & $-\frac{3}{2}\left(1+\frac{\mu^2c^2}{9}\right)$ & $\frac{3}{2}\left(1-\frac{\mu^2c^2}{9}\right)$& \begin{tabular}{@{}c@{}}$~~$\\Non-hyperbolic for $\mu c=\pm3$\\and\\hyperbolic for $\mu c\neq\pm3$\\$~~$\end{tabular}\\ \hline
		\begin{tabular}{@{}c@{}}$~~$\\$ E_4 $\\$~$\end{tabular}  & \begin{tabular}{@{}c@{}}$~~$\\$\frac{-3+\sqrt{45-\frac{324}{\mu^2 c^2}}}{2}$ \\$~~$\end{tabular}&\begin{tabular}{@{}c@{}}$~~$\\  $\frac{-3-\sqrt{45-\frac{324}{\mu^2 c^2}}}{2}$\\$~~$\end{tabular}   &\begin{tabular}{@{}c@{}}$~~$\\Non-hyperbolic for $\mu c=\pm3$\\and\\hyperbolic for $\mu c\neq\pm3$\\$~~$\end{tabular}\\ \hline
		\begin{tabular}{@{}c@{}}$~~$\\$ E_5 $\\$~$\end{tabular}  & \begin{tabular}{@{}c@{}}$~~$\\$\frac{-3+\sqrt{45-\frac{324}{\mu^2 c^2}}}{2}$ \\$~~$\end{tabular}&\begin{tabular}{@{}c@{}}$~~$\\  $\frac{-3-\sqrt{45-\frac{324}{\mu^2 c^2}}}{2}$\\$~~$\end{tabular}   &\begin{tabular}{@{}c@{}}Non-hyperbolic for $\mu c=\pm3$\\and\\hyperbolic for $\mu c\neq\pm3$\end{tabular}\\ \hline
	\end{tabular}
\end{table}
\newpage
For avoiding similar arguments which we have mentioned for analyzing the stability of the above critical points, we only state the stability and the reason behind the stability of these critical points in a tabular form, which is shown as in Table \ref{T_stability}.  
\begin{table}[h]
	\caption{\label{T_stability}Table shows the stability and the reason behind the stability of the critical points $(E_1-E_5)$}
	\begin{tabular}{|c|c|c|}
	\hline
	\hline	
	\begin{tabular}{@{}c@{}}$~~$\\$ CPs $\\$~$\end{tabular} &$Stability$& $Reason~behind~the~stability$ \\ \hline\hline
	$E_1,~E_2$& Both are stable star & 	\begin{tabular}{@{}c@{}}$~~$\\As both eigenvalues $\lambda_1$ and $\lambda_2$ are negative and equal.  By Hartman-\\Grobman theorem we can conclude that the critical points $E_1$ and \\$E_2$ both are stable star.\\$~~$\end{tabular}\\ \hline
	$E_3$&\begin{tabular}{@{}c@{}}$~~$\\ Stable node for $\mu c=-3$,\\saddle node for $\mu c=3$ ,\\ stable node for $\mu c>3~or,~<-3$,\\saddle node for $-3<\mu c<3$ \\$~~$\end{tabular}&\begin{tabular}{@{}c@{}}$~~$\\For $\mu c=-3:$\\After shifting the this critical point into the origin by taking the\\ transformation $x= X-\frac{\mu c}{3}$, $y= Y$ and by using CMT, the CM \\is given by $X=Y^2+\mathcal{O}(Y^4) $ and the flow on the CM is determined \\by $ Y'=-\frac{3}{2}Y^3+\mathcal{O}(Y^5)$.  $Y'<0$ while $Y>0$ and for $Y<0$, $Y'>0$.\\ So, the critical point $E_3$ is a stable node (FIG.\ref{mu_c_3}(a)).\\$~~$\\ For $\mu c=3:$\\  The center manifold is given by $X=-Y^2+\mathcal{O}(Y^4) $ and the flow on\\ the center manifold is determined by $ Y'=\frac{3}{2}Y^3+\mathcal{O}(Y^5)$. $Y'<0$\\ while $Y<0$ and for $Y>0$, $Y'>0$.  So, the critical point $E_3$ is a\\ saddle node  (FIG.\ref{mu_c_3}(b)).\\$~~$\\For $\mu c>3~or,~\mu c<-3$:\\ Both of the eigenvalues $\lambda_1$ and $\lambda_2$ are negative and unequal.  So by\\ Hartman-Grobman theorem the critical point $E_3$ is a stable node.\\ $~~$\\ For $-3<\mu c<3:$\\$\lambda_1$ is negative and $\lambda_2$ is positive.  So by Hartman-Grobman theorem\\ the critical point $E_3$ is unstable node.\\$~~$\end{tabular} \\ \hline
	$E_4,~E_5$ &\begin{tabular}{@{}c@{}}$~~$\\Both are stable node for $\mu c=-3$,\\ saddle node for $\mu c=3$,\\  stable node for $\mu c>3~or,~<-3$\\$~~$\end{tabular}&	\begin{tabular}{@{}c@{}}$~~$\\For $\mu c=3$ and $\mu c=-3$:\\ The expression of the center manifold and the flow on the center\\ manifold is same as the expressions for $\mu c=-3$ and $\mu c=-3$\\ cases respectively for $E_3$.\\ $~~$\\ For $\mu c>3,~or~<-3$:\\ Both of the eigenvalues $\lambda_1$ and $\lambda_2$ are negative and unequal.\\  Hence, by Hartman-Grobman theorem we can conclude that the critical\\ points $E_4$ and $E_5$ both are unstable in nature.\\$~~$  \end{tabular}\\ \hline
\end{tabular}
\end{table}
Note that $\mu c\geq3$ and $\mu c\leq-3$ be the domain of existence of the critical point $E_4$ and $E_5$.  For this reason we did not determine the stability analysis of the critical points $E_4$ and $E_5$ for $\mu c\in (-3,3)$.
\begin{figure}[!]
	\centering
	\includegraphics[width=1\textwidth]{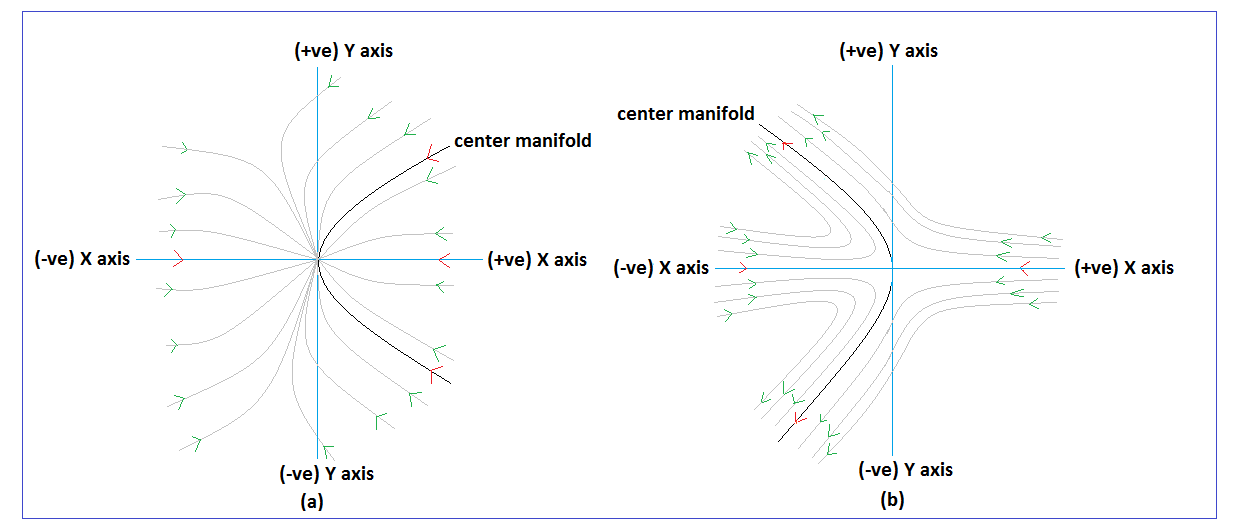}
	\caption{Vector field near near the origin for the critical point $E_3$.  L.H.S. for $\mu c=3$ and R.H.S. for $\mu c=-3$.}
	\label{mu_c_3}
\end{figure}

\newpage
\subsubsection*{Case-(iii)$~$\underline{$\mu=0$ and $\lambda\neq 0$}}	
\begin{figure}[h]
	\centering
	\includegraphics[width=0.7\textwidth]{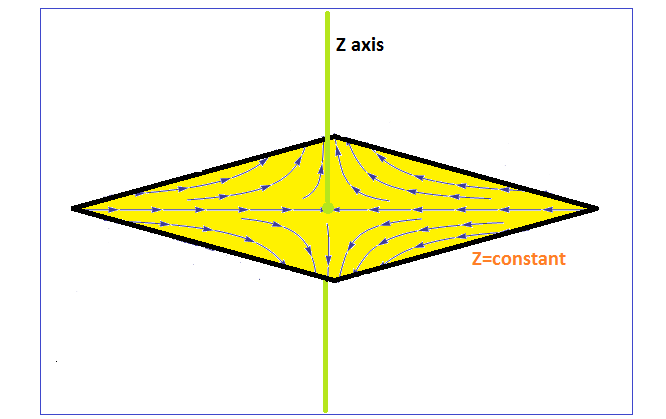}
	\caption{Vector field near about every point on $Z-$axis for the critical point $P_1$.}
	\label{saddle_z_c}
\end{figure}

In this case the autonomous system $(\ref{eq9}-\ref{eq11})$ changes into
\begin{eqnarray}
x'&=&-3x+\frac{3}{2}x(1-x^2-y^2)-\frac{\lambda y^2 z}{2},\label{eq40} \\
y'&=&\frac{3}{2}y(1-x^2-y^2)-\frac{\lambda xyz}{2},\label{eq41}  \\
z'&=&-xz^2. \label{eq42} 
\end{eqnarray}
Corresponding to the above autonomous system we have three space of critical points $P_1(0,0,z_c)$, $P_2(0,1,0)$ and $P_3(0,-1,0)$ where $z_c$ is any real number.  The value of cosmological parameters, eigenvalues of the Jacobian matrix at those critical points corresponding to the autonomous system $(\ref{eq40}-\ref{eq42})$ and the nature of critical points $P_1$, $P_2$ and $P_3$ are same as for the critical points $A_1$, $A_2$ and $A_3$ respectively, shown as in Table \ref{TI}. \newpage
\begin{center}
	$1.~Critical~Point~P_1$
\end{center}
The Jacobian matrix at the critical point $P_1$ can be put as
\begin{equation}
\renewcommand{\arraystretch}{1.5}	
J(P_1)=\begin{bmatrix}
-\frac{3}{2} & 0 & 0\\	
~~0  & \frac{3}{2}& 0\\
-z_c^2 & 0 & 0 
\end{bmatrix}.\label{eq45}	
\end{equation}
The eigenvalues of the above matrix are $-\frac{3}{2}$, $\frac{3}{2}$ and $0$ and  $\left[1, 0, \frac{2}{3}z_c^2\right]^T$, $[0, 1, 0]^T$ and $[0, 0, 1]^T$ are the corresponding eigenvectors respectively.
For a fixed $z_c$, first we shift the critical point $P_0$ to the origin by the coordinate transformation $x=X$, $y=Y$ and $z=Z+z_c$, if we put forward argument as above for non-hyperbolic critical points.  Then, the center manifold can be written as $(\ref{eq37}-\ref{eq38})$ and the flow on the center manifold is determined by (\ref{eq39}).  Similarly as above (the discussion of stability for the critical point $C_2$) we can conclude that the center manifold for the critical point $P_1$ is also lying in the $Z-$axis but flow on the center manifold can not be determined.   Now,  if we project the vector field on the plane
which is parallel to $XY$-plane, i.e., the plane $Z=constant$(say), then the vector field is shown as in FIG.\ref{saddle_z_c}.  So every point on $Z$- axis is a saddle node.\bigbreak
Again if we want to obtain the stability of the critical points in the plane which is parallel to $xy$-plane, i.e., $z=constant=c$(say), then we only take the first two equations (\ref{eq40}) and (\ref{eq41}) of the autonomous system $(\ref{eq40}-\ref{eq42})$ and also replace $z$ by $c$ in those two equations.  After that we can see that there exists three real and physically meaningful hyperbolic critical points $B_1(0,0)$, $B_2\left(-\frac{\lambda c}{6}, \sqrt{1+\frac{\lambda^2 c^2}{36}}\right)$ and $B_3\left(-\frac{\lambda c}{6}, -\sqrt{1+\frac{\lambda^2 c^2}{36}}\right)$.  So by obtaining the eigenvalues of the Jacobian matrix corresponding to the autonomous system at those critical points and using Hartman-Grobman theorem we only state the stability of all critical points and also write the value of cosmological parameters corresponding to these critical points in tabular form, which is shown as in Table \ref{TB}.\bigbreak
For the critical points $P_2$ and $P_3$ we have the same Jacobian matrix (\ref{eq21}) and if we will take the similar transformations (shifting and matrix) and then by using the similar arguments as $A_2$ and $A_3$ respectively, we conclude that the the stability of $P_2$ and $P_3$ is same as $A_2$ and $A_3$ respectively.
 \begin{table}[!]
	\caption{\label{TB}Table shows the eigenvalues $(\lambda_1, \lambda_2)$ of the Jacobian matrix, stability and value of cosmological parameters corresponding to the critical points and the nature of all critical points $(B_1-B_3)$.}		 
	\begin{tabular}{|c|c c|c|c c c c|}
		\hline
		\hline	
		\begin{tabular}{@{}c@{}}$~~$\\$ Critical~Points $\\$~$\end{tabular} &$ ~~\lambda_1 $ & $~~\lambda_2$ & $ Stability$&$~\Omega_\phi~$& ~~$\omega_{\phi}$~~ &$\omega_{tot}$ & ~~$q$  \\ \hline\hline
			\begin{tabular}{@{}c@{}}$~~$\\$B_1$\\$~$\end{tabular} &$-\frac{3}{2}$ & $\frac{3}{2}$&Stable star&$0$&Undetermined&$0$& $\frac{1}{2}$\\ \hline
			\begin{tabular}{@{}c@{}}$~~$\\$B_2,~B_3$\\$~$\end{tabular}&$-3\left(1+\frac{\lambda^2 c^2}{18}\right)$&$-3\left(1+\frac{\lambda^2 c^2}{36}\right)$&	\begin{tabular}{@{}c@{}}$~~$\\Stable star for $\lambda c=0$\\and\\stable node for $\lambda c\neq 0$\\$~~$\end{tabular}& $1$&$-\left(1+\frac{\lambda^2 c^2}{18}\right)$&$-\left(1+\frac{\lambda^2 c^2}{18}\right)$&$-\left(1+\frac{\lambda^2 c^2}{12}\right)$\\ \hline
		\end{tabular}
\end{table}
\newpage
\subsubsection*{Case-(iv)$~$\underline{$\mu=0$ and $\lambda=0$}}
In this case the autonomous system $(\ref{eq9}-\ref{eq11})$ changes into

\begin{eqnarray}
x'&=&-3x+\frac{3}{2}x(1-x^2-y^2),\label{eq49} \\
y'&=&\frac{3}{2}y(1-x^2-y^2),\label{eq50}  \\
z'&=&-xz^2\label{eq51}. 
\end{eqnarray}
Corresponding to the above autonomous system we have three space of critical points $S_1(0,0,z_c)$, $S_2(0,1,z_c)$ and $S_3(0,-1,z_c)$ where $z_c$ is any real number, which are exactly same as $C_1$, $C_2$ and $C_3$.  In this case also all critical points are non-hyperbolic in nature. By taking the possible shifting transformations (for $S_1~(x=X,y=Y,z=Z+z_c)$, for $S_2~(x=X,y=Y+1,z=Z+z_c)$ and for $S_3~(x=X,y=Y-1,z=Z+z_c)$ ) as above we can conclude that for all critical points the center manifold is given by $(\ref{eq37}-\ref{eq38})$ and the flow on the center manifold is determined by (\ref{eq39}), i.e., for all critical points the center manifold is lying on the $Z$-axis.  Again if we plot the vector field in $Z=constant$ plane, we can see that for the critical point $S_1$ every points on $Z$-axis is a saddle node (same as FIG.\ref{saddle_z_c}) and for $S_2$ and $S_3$ every points on $Z$-axis is a stable star (same as FIG.\ref{z_c}). 

\subsection{Model 2: Power-law potential and
	exponentially-dependent dark-matter particle mass \label{M2}}
	 In this consideration evolution equations in Section \ref{BES} can be converted to the autonomous system as follows 
\begin{eqnarray}
x'&=&-3x+\frac{3}{2}x(1-x^2-y^2)-\frac{\lambda y^2 z}{2}-\sqrt{\frac{3}{2}}\mu(1+x^2-y^2),\label{eq54} \\
y'&=&\frac{3}{2}y(1-x^2-y^2)-\frac{\lambda xyz}{2},\label{eq55}  \\
z'&=&-xz^2,\label{eq56} 
\end{eqnarray}
We have five critical points $L_1$, $L_2$, $L_3$, $L_4$ and $L_5$ corresponding to the above autonomous system.  The set of critical points, their existence and the value of cosmological parameters at those critical points are shown as in Table \ref{TPLE} and the eigenvalues of the Jacobian matrix corresponding to the autonomous system $(\ref{eq54}-\ref{eq56})$ at those critical points and the nature of the critical points are shown in Table \ref{TNE}.\par

Here we only concern about the stability of the critical points for $\mu\neq 0$ and $\lambda\neq 0$ because for another possible cases we will get the similar types result which we have obtained for Model $1$.
	\begin{table}[h]
	\caption{\label{TPLE}Table shows the set of critical points and their existence, value of cosmological parameters corresponding to that critical points. }
	\begin{tabular}{|c|c|c c c|c|c|c| c|}
		\hline
		\hline	
		\begin{tabular}{@{}c@{}}$~~$\\$~Critical ~Points$\\$~~$\end{tabular} &$Existence$&$x$&$y$&$z~~$& $~\Omega_\phi~$&$~\omega_\phi~$ &$~\omega_{tot}~$& $~q~$  \\ \hline\hline
	\begin{tabular}{@{}c@{}}$~~$\\	$L_1$\\$~~$\end{tabular}& For all $\mu$ and $\lambda$&0&1&0 & 1 & $-1$ & $-1$&$-1$\\ \hline
		\begin{tabular}{@{}c@{}}$~~$\\	$L_2$\\$~~$\end{tabular}& For all $\mu$ and $\lambda$&0&$-1$&0 & 1 & $-1$ & $-1$&$-1$\\ \hline
		$L_3$ & 	\begin{tabular}{@{}c@{}}$~~$\\For all \\$\mu\in\left(-\infty,-\sqrt{\frac{3}{2}}\right]\cup\left[\sqrt{\frac{3}{2}},\infty\right)$\\and all $\lambda$\\$~~$\end{tabular}&$-\frac{1}{\mu}\sqrt{\frac{3}{2}}$&$\sqrt{1-\frac{3}{2\mu^2}}$&0&$1-\frac{3}{\mu^2}$&$\frac{\mu^2}{3-\mu^2}$&$-1$&$-1$\\ \hline
		$L_4$ & 	\begin{tabular}{@{}c@{}}$~~$\\For all \\$\mu\in\left(-\infty,-\sqrt{\frac{3}{2}}\right]\cup\left[\sqrt{\frac{3}{2}},\infty\right)$\\and all $\lambda$\\$~~$\end{tabular}&$-\frac{1}{\mu}\sqrt{\frac{3}{2}}$&$-\sqrt{1-\frac{3}{2\mu^2}}$&0&$1-\frac{3}{\mu^2}$&$\frac{\mu^2}{3-\mu^2}$&$-1$&$-1$\\ \hline
			\begin{tabular}{@{}c@{}}$~~$\\	$L_5$\\$~~$\end{tabular}& For all $\mu$ and $\lambda$&$-\sqrt{\frac{2}{3}}\mu$&$0$&$0$&$-\frac{2}{3}\mu^2$ & $1$&$-\frac{2}{3}\mu^2$&$\frac{1}{2}\left(1-2\mu^2\right)$ \\ \hline
	\end{tabular}
\end{table}
\begin{table}[h]
	\caption{\label{TNE}The eigenvalues $(\lambda_1,\lambda_2,\lambda_3)$ of the Jacobian matrix corresponding to the autonomous system $(\ref{eq54}-\ref{eq56})$ at those critical points $(L_1-L_5)$ and the nature of the critical points}
	\begin{tabular}{|c|c c c|c|}
		\hline
		\hline	
		\begin{tabular}{@{}c@{}}$~~$\\$~Critical ~Points$\\$~~$\end{tabular} &$\lambda_1$&$\lambda_2$&$\lambda_3$&$Nature~ of~ critical~ Points$  \\ \hline\hline
		\begin{tabular}{@{}c@{}}$~~$\\$L_1$\\$~~$\end{tabular}&$-3$&$-3$&$0$&Non-hyperbolic\\ \hline
			\begin{tabular}{@{}c@{}}$~~$\\$L_2$\\$~~$\end{tabular}&$-3$&$-3$&$0$&Non-hyperbolic\\ \hline
		\begin{tabular}{@{}c@{}}$~~$\\$L_3$\\$~~$\end{tabular}&$-\frac{3}{2}\left(1+\frac{1}{\mu}\sqrt{-6+5\mu^2}\right)$&$-\frac{3}{2}\left(1-\frac{1}{\mu}\sqrt{-6+5\mu^2}\right)$&$0$&Non-hyperbolic\\ \hline
			\begin{tabular}{@{}c@{}}$~~$\\$L_4$\\$~~$\end{tabular}&$-\frac{3}{2}\left(1+\frac{1}{\mu}\sqrt{-6+5\mu^2}\right)$&$-\frac{3}{2}\left(1-\frac{1}{\mu}\sqrt{-6+5\mu^2}\right)$&$0$&Non-hyperbolic\\ \hline
			\begin{tabular}{@{}c@{}}$~~$\\$L_5$\\$~~$\end{tabular}&$-\frac{3}{2}$&$\frac{3}{2}$&$0$&Non-hyperbolic \\ \hline 
	\end{tabular}
\end{table}

\begin{center}
	$1.~Critical~Point~L_1$
\end{center}
The Jacobian matrix corresponding to the autonomous system $(\ref{eq54}-\ref{eq56})$ at the critical point $L_1$ can be put as
\begin{equation}
\renewcommand{\arraystretch}{1.5}
J(L_1)=\begin{bmatrix}
-3&\sqrt{6}\mu&-\frac{\lambda}{2}\\
~~0&-3&~~0\\
~~0 & ~~0& ~~0
\end{bmatrix}.
\end{equation}
The eigenvalues of $J(L_1)$ are $-3$, $-3$, $0$ and $[1,0,0]^T$, $\left[-\frac{\lambda}{6}, 0,1\right]^T$ are the eigenvectors corresponding to the eigenvalues $-3$ and $0$ respectively.  Since the algebraic multiplicity corresponding to the eigenvalue $-3$ is $2$ but the dimension of the eigenspace corresponding to that eigenvalue is $1$, i.e., algebraic multiplicity and geometric multiplicity corresponding to the eigenvalue $-3$ are not equal to each other.  So, the Jacobian matrix $J(L_1)$ is not diagonalizable.  To determine the center manifold for this critical point there only arises a problem for presence of the nonzero element in the top position of third column of the Jacobian matrix.  First we take the coordinate transformation $x=X,y=Y+1,z=Z$ which shift the critical point $L_1$ to the origin.  Now we introduce another coordinate system which will remove the term in the top position of the third column.  Since, there are only two linearly independent eigenvectors, so we have to obtain another linearly independent column vector that will help to construct the new coordinate system.  Since, $[0,1,0]^T$ be the column vector which is linearly independent to the eigenvectors of $J(L_1)$.  The new coordinate system $(u,v,w)$ can be written in terms of $(X,Y,Z)$ as (\ref{eq24})
	and in these new coordinate system the equations $(\ref{eq54}-\ref{eq56})$ are transformed into	
\begin{equation}\renewcommand{\arraystretch}{1.5}	
\begin{bmatrix}
u'\\
v'\\
w'
\end{bmatrix}\renewcommand{\arraystretch}{1.5}
=\begin{bmatrix}
-3&\sqrt{6}\mu&0\\
~~0&-3&~~0\\
~~0 & ~~0& ~~0
\end{bmatrix}
\begin{bmatrix}
u\\
v\\
w
\end{bmatrix}		
+\renewcommand{\arraystretch}{1.5}	
\begin{bmatrix}
non\\
linear\\
terms
\end{bmatrix}.	
\end{equation}	
By similar arguments which we have derived in the stability analysis of the critical point $A_2$,  the center manifold can be written as (\ref{eqn27}-\ref{eqn28})
and the flow on the center manifold is determined by (\ref{eq29}).
As the expression of center manifold and the flow are same as for the critical point $A_2$.  So the stability of the critical point $L_1$ is same as the stability of $A_2$.
\begin{center}
	$2.~Critical~Point~L_2$
\end{center}
After shifting the critical points to the origin (by taking the shifting transformations $(x=X,y=Y-1,z=Z)$ and the matrix transformation (\ref{eq24})) and by putting the forward arguments which we have mentioned for the analysis of $L_1$, the center manifold can be expressed as $(\ref{eqn30}-\ref{eqn31})$ and the flow on the center manifold is determined by (\ref{eqn32}).  So the stability of the critical point $L_2$ is same as the stability of $A_3$.

\begin{center}
	$3.~Critical~Point~L_3$
\end{center}
The Jacobian matrix corresponding to the autonomous system $(\ref{eq54}-\ref{eq56})$ at the critical point $L_3$ can be put as
\begin{equation}
	\renewcommand{\arraystretch}{3}
	J(L_3)=\begin{bmatrix}
		-\frac{9}{2\mu^2}&\sqrt{1-\frac{3}{2\mu^2}}\left(\frac{3}{\mu}\sqrt{\frac{3}{2}}+\sqrt{6}\mu\right)&-\frac{\lambda}{2}\left(1-\frac{3}{2\mu^2}\right)\\
		\frac{3}{\mu}\sqrt{\frac{3}{2}}\sqrt{1-\frac{3}{2\mu^2}}&-3\left(1-\frac{3}{2\mu^2}\right)&\frac{\lambda}{2\mu}\sqrt{\frac{3}{2}}\sqrt{1-\frac{3}{2\mu^2}}\\
		~~0 & ~~0& ~~0
	\end{bmatrix}.
\end{equation}
The eigenvalues corresponding to the Jacobian matrix $J(L_3)$ are shown in Table.\ref{TNE}.  From the existence of the critical point $L_3$ we can conclude that the eigenvalues of $J(L_3)$ always real.   Since the critical point $L_3$ exists for $\mu\leq -\sqrt{\frac{3}{2}}$ or $\mu\geq \sqrt{\frac{3}{2}}$, our aim is to define the stability in all possible regions of $\mu$ for at least one choice of $\mu$ in these region.  For this reason we will define the stability at four possible choices of $\mu$. We first determine the stability of this critical point at $\mu=\pm\sqrt{\frac{3}{2}}$.  Then for $\mu< -\sqrt{\frac{3}{2}}$, we shall determine the stability of $L_3$ at $\mu=-\sqrt{3}$ and for $\mu>\sqrt{\frac{3}{2}}$, we shall determine the stability of $L_3$ at $\mu=\sqrt{3}$.\par

\begin{figure}[h]
	\centering
	\includegraphics[width=1\textwidth]{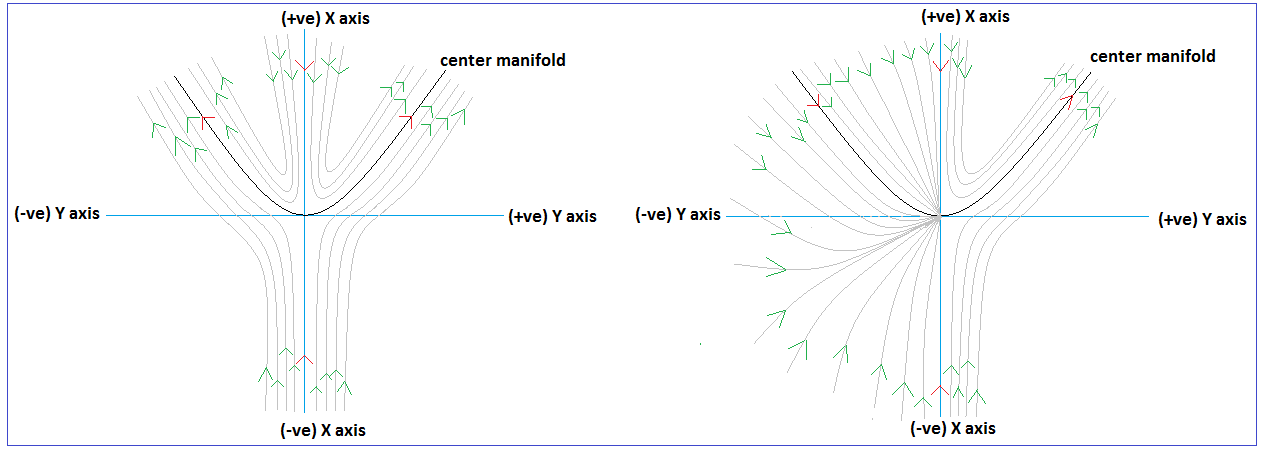}
	\caption{Vector field near the origin when $\mu=\sqrt{\frac{3}{2}}$, for the critical point $L_3$. L.H.S. phase plot is for $\lambda=1$ and R.H.S. phase plot is for $\lambda=2$.}
	\label{L_21}
\end{figure}

\begin{figure}[h]
	\centering
	\includegraphics[width=1\textwidth]{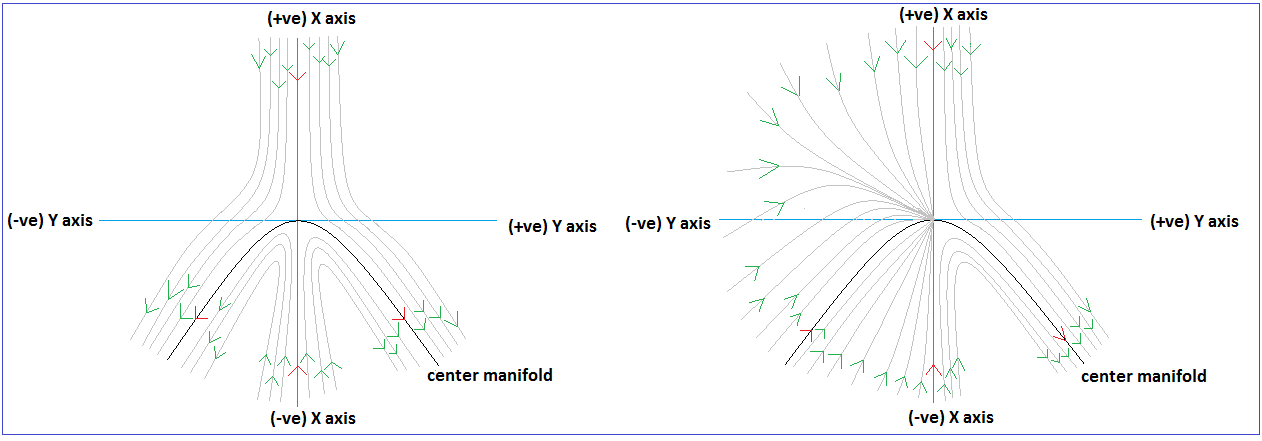}
	\caption{Vector field near the origin when $\mu=-\sqrt{\frac{3}{2}}$, for the critical point $L_3$. L.H.S. phase plot is for $\lambda=1$ and R.H.S. phase plot is for $\lambda=2$.}
	\label{L_2_1_1}
\end{figure}

For $\mu=\pm\sqrt{\frac{3}{2}}$, the Jacobian matrix $J(L_3)$ converts into
$$
\begin{bmatrix}
-3&0&0\\~~0&0&0\\~~0&0&0
\end{bmatrix}
$$
and as the critical point $L_3$ converts into $(\mp 1,0,0)$, first we take the transformation $x=X\mp 1, y= Y, z=Z$ so that $L_3$ moves into the origin.  As the critical point is non-hyperbolic in nature we use CMT for determining the stability of this critical point.  From center manifold theory there exist a continuously differentiable function
$h:$$\mathbb{R}^2$$\rightarrow$$\mathbb{R}$ such that $X=h(Y,Z)=aY^2+bYZ+cZ^2+higher~order~terms,$ where $a,~b,~c~ \epsilon~\mathbb{R}$. \\
Now differentiating both side with respect to $N$, we get
\begin{eqnarray}
\frac{dX}{dN}=[2aY+bZ ~~~~ bY+2cZ]\begin{bmatrix}
\frac{dY}{dN}\\
~\\
\frac{dZ}{dN}\\
\end{bmatrix}\label{equn52}
\end{eqnarray}
Comparing L.H.S. and R.H.S. of (\ref{equn52}) we get,
$a=1$, $b=0$ and $c=0$, i.e., the center manifold can be written as
\begin{eqnarray}
X&=&\pm Y^2+higher~order~terms\label{eq65}
\end{eqnarray}
and the flow on the center manifold is determined by 
\begin{eqnarray}
\frac{dY}{dN}&=&\pm\frac{\lambda}{2}YZ+higher~order~terms,\label{eq66}\\
\frac{dZ}{dN}&=&\pm Z^2+higher~order~terms\label{eq67}.
\end{eqnarray}
 We only concern about the non-zero coefficients of the lowest power terms in CMT as we analyze arbitrary small neighborhood of the origin and here the lowest power term of the expression of center manifold depends only on $Y$.  So, we draw the vector field near the origin only on $XY$-plane, i.e., the nature of the vector field implicitly depends on $Z$ not explicitly.  Now we try to write the flow equations $(\ref{eq66}-\ref{eq67})$ in terms of $Y$ only.  For this reason, we divide the corresponding sides of (\ref{eq66}) by the corresponding sides of (\ref{eq67}) and then we will get 
 \begin{align*}
&\frac{dY}{dZ}=\frac{\lambda}{2}\frac{Y}{Z}\\ \implies& Z=\left(\frac{Y}{C}\right)^{2/\lambda},~~\mbox{where $C$ is a positive arbitrary constant}
 \end{align*}
After substituting this any of $(\ref{eq66})$ or $(\ref{eq67})$, we get
\begin{align}
\frac{dY}{dN}=\frac{\lambda}{2C^{2/\lambda}}Y^{1+2/\lambda}
\end{align}
As the power of $Y$ can not be negative or fraction, so we have only two choices of $\lambda$, $\lambda=1$ or $\lambda=2$.  For $\lambda=1$ or, $\lambda=2$ both of the cases the origin is a saddle node, i.e., unstable in nature (FIG.\ref{L_21} is for $\mu=\sqrt{\frac{3}{2}}$ and FIG.\ref{L_2_1_1} is for $\mu=-\sqrt{\frac{3}{2}}$).  Hence, for $\mu=\pm \sqrt{\frac{3}{2}}$, in the old coordinate system the critical point $L_3$ is unstable due to its saddle nature.\bigbreak

\begin{figure}[h]
	\centering
	\includegraphics[width=1\textwidth]{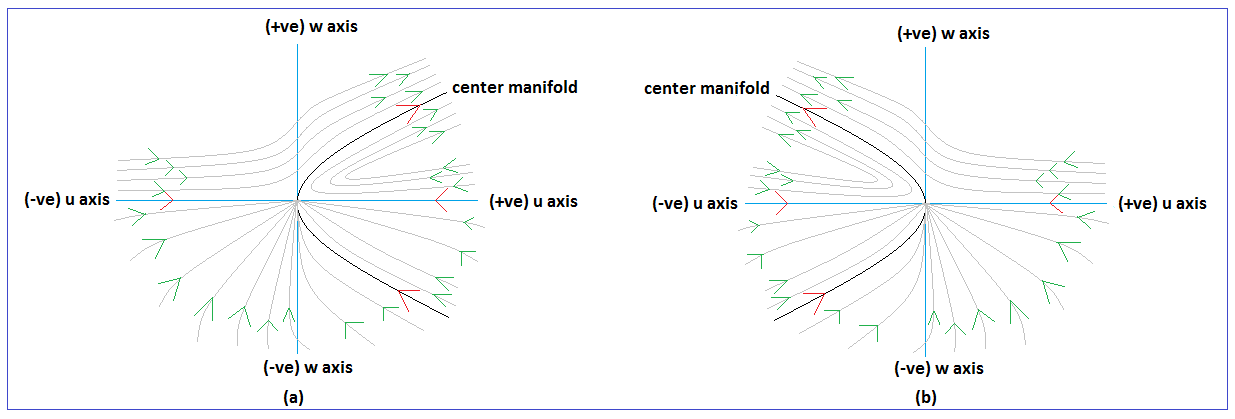}
	\caption{Vector field near the origin in $uw$-plane when $\mu=\sqrt{3}$, for the critical points $L_3$ and $L_4$.  For the critical point $L_3$, the phase plot (a) is for $\lambda<0$ or $\lambda>\frac{4}{\sqrt{3}-1}$ and the phase plot (b) is for $0<\lambda<\frac{4}{\sqrt{3}-1}$. For the critical point $L_4$, the phase plot (a) is for $0<\lambda<\frac{4}{\sqrt{3}-1}$ and the phase plot (b) is for $\lambda<0$ or $\lambda>\frac{4}{\sqrt{3}-1}$.}
	\label{L_22}
\end{figure}

\begin{figure}[h]
	\centering
	\includegraphics[width=1\textwidth]{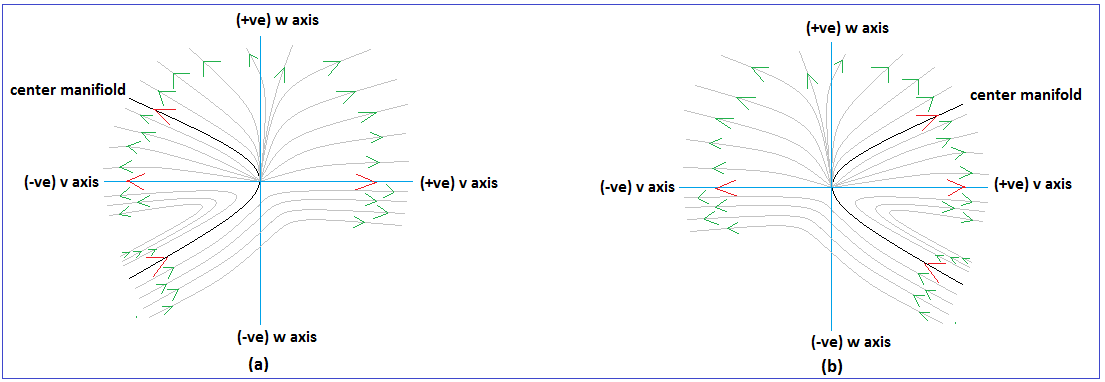}
	\caption{Vector field near the origin in $vw$-plane when $\mu=\sqrt{3}$, for the critical points $L_3$ and $L_4$. For the critical point $L_3$, the phase plot (a) is for $\lambda<-\frac{4}{\sqrt{3}+1}$ or $\lambda>0$ and the phase plot (b) is for $-\frac{4}{\sqrt{3}+1}<\lambda<0$.  For the critical point $L_4$, the phase plot (a) is for $-\frac{4}{\sqrt{3}+1}<\lambda<0$ and the phase plot (b) is for $\lambda<-\frac{4}{\sqrt{3}+1}$ or $\lambda>0$.}
	\label{L_2_2}
\end{figure}

For $\mu=\sqrt{3}$, the Jacobian matrix $J(L_3)$ converts into
$$	\renewcommand{\arraystretch}{1.5}
\begin{bmatrix}
-\frac{3}{2}&~~\frac{9}{2}&-\frac{\lambda}{4}\\~~\frac{3}{2}&-\frac{3}{2}&~~\frac{\lambda}{4}\\~~0&~~0&~~0
\end{bmatrix}.
$$
The eigenvalues of the above Jacobian matrix are $-\frac{3}{2}(1+\sqrt{3})$, $-\frac{3}{2}(1-\sqrt{3})$ and $0$ and the corresponding eigenvectors are $[-\sqrt{3},1,0]^T$, $[\sqrt{3},1,0]^T$ and $\left[-\frac{\lambda}{6},0,1\right]^T$ respectively.  As for $\mu=\sqrt{3}$, the critical point $L_3$ converts into $\left(-\frac{1}{\sqrt{2}},\frac{1}{\sqrt{2}},0\right)$; so first we take the transformations $x= X-\frac{1}{\sqrt{2}}$, $y= Y+\frac{1}{\sqrt{2}}$ and $z= Z$ which shift the critical point to the origin.  By using the eigenvectors of the above Jacobian matrix, we introduce a new coordinate system $(u,v,w)$ in terms of $(X,Y,Z)$ as
\begin{equation}\renewcommand{\arraystretch}{1.5}	
\begin{bmatrix}
u\\
v\\
w
\end{bmatrix}\renewcommand{\arraystretch}{1.5}
=\begin{bmatrix}
-\frac{1}{2\sqrt{3}} & \frac{1}{2} & -\frac{\lambda}{12\sqrt{3}} \\	
\frac{1}{2\sqrt{3}} & \frac{1}{2} & \frac{\lambda}{12\sqrt{3}}\\
0 & 0 & 1
\end{bmatrix}\renewcommand{\arraystretch}{1.5}
\begin{bmatrix}
X\\
Y\\
Z
\end{bmatrix}	
\end{equation}		
and in these new coordinates the equations $(\ref{eq54}-\ref{eq56})$ are transformed into	
\begin{equation}	\renewcommand{\arraystretch}{1.5}
\begin{bmatrix}
-u'+v'\\
u'+v'\\
w'
\end{bmatrix}
=\begin{bmatrix}
\frac{3}{2}(1+\sqrt{3})& -\frac{3}{2}(1-\sqrt{3}) & 0 \\	
 -\frac{3}{2}(1+\sqrt{3}) & -\frac{3}{2}(1-\sqrt{3}) & 0 \\
~~0 & ~~0 & 0
\end{bmatrix}
\begin{bmatrix}
u\\
v\\
w
\end{bmatrix}		
+	
\begin{bmatrix}
non\\
linear\\
terms
\end{bmatrix}.	
\end{equation}	
Now if we add $1$st and $2$nd equation of the above matrix equation and then divide both sides by $2$, then we get $v'$.  Again, if we subtract $1$st equation from $2$nd equation and divide both sides by $2$, we get $u'$.  Finally,  in matrix form in the new coordinate system the autonomous system can be written as
\begin{equation}	\renewcommand{\arraystretch}{1.5}
\begin{bmatrix}
u'\\
v'\\
w'
\end{bmatrix}
=\begin{bmatrix}
-\frac{3}{2}(1+\sqrt{3})& 0 & 0 \\	
0 & -\frac{3}{2}(1-\sqrt{3}) & 0 \\
0 & ~~0 & 0
\end{bmatrix}
\begin{bmatrix}
u\\
v\\
w
\end{bmatrix}		
+	
\begin{bmatrix}
non\\
linear\\
terms
\end{bmatrix}.	
\end{equation}
If we put similar arguments which we have mentioned for the analysis of $A_2$, then the center manifold can be expressed as
\begin{align}
u&=\frac{2}{3(1+\sqrt{3})}\left\{\frac{(\sqrt{3}-1)\lambda^2-4\lambda}{48\sqrt{6}}\right \}w^2+\mathcal{O}(w^3),\label{eqn72}\\
v&=-\frac{2}{3(\sqrt{3}-1)}\left\{\frac{(\sqrt{3}+1)\lambda^2+4\lambda}{48\sqrt{6}}\right \}w^2+\mathcal{O}(w^3)\label{eqn73}
\end{align}
and the flow on the center manifold is determined by
\begin{align}
w'&=\frac{1}{\sqrt{2}}w^2+\mathcal{O}(w^3)\label{eqn74}.
\end{align}
From the flow equation we can easily conclude that the origin is a saddle node and unstable in nature.  The vector field near the origin in $uw$-plane is shown as in FIG.\ref{L_22} and the vector field near the origin in $vw$-plane is shown as in FIG.\ref{L_2_2}.  Hence, in the old coordinate system $(x,y,z)$, for $\mu=\sqrt{3}$ the critical point $L_3$ is unstable due to its saddle nature.

Lastly, for $\mu=-\sqrt{3}$, we have the same eigenvalues $-\frac{3}{2}(1+\sqrt{3})$, $-\frac{3}{2}(1-\sqrt{3})$ and $0$ and the corresponding eigenvectors are $[\sqrt{3},1,0]^T$, $[-\sqrt{3},1,0]^T$ and $\left[-\frac{\lambda}{6},0,1\right]^T$ respectively of $J(L_3)$.  After putting corresponding arguments which we have mentioned for $\mu=\sqrt{3}$ case, then we will get the same expressions $(\ref{eqn72}-\ref{eqn73})$ for center manifold and (\ref{eqn74}) for flow on the center manifold.  So, for this case also we conclude that the critical point $L_3$ is a saddle node and unstable in nature.
\newpage
\begin{center}
	$4.~Critical~Point~L_4$
\end{center}
The Jacobian matrix corresponding to the autonomous system $(\ref{eq54}-\ref{eq56})$ at the critical point $L_4$ can be put as
\begin{equation}
\renewcommand{\arraystretch}{3}
J(L_4)=\begin{bmatrix}
-\frac{9}{2\mu^2}&-\sqrt{1-\frac{3}{2\mu^2}}\left(\frac{3}{\mu}\sqrt{\frac{3}{2}}+\sqrt{6}\mu\right)&-\frac{\lambda}{2}\left(1-\frac{3}{2\mu^2}\right)\\
-\frac{3}{\mu}\sqrt{\frac{3}{2}}\sqrt{1-\frac{3}{2\mu^2}}&-3\left(1-\frac{3}{2\mu^2}\right)&-\frac{\lambda}{2\mu}\sqrt{\frac{3}{2}}\sqrt{1-\frac{3}{2\mu^2}}\\
~~0 & ~~0& ~~0
\end{bmatrix}.
\end{equation}
For this critical point also we analyze the stability for the above four choices of $\mu$, i.e., $\mu=\pm\sqrt{\frac{3}{2}}$, $\mu=\sqrt{3}$ and $\mu=-\sqrt{3}$. \par
  For $\mu=\pm\sqrt{\frac{3}{2}}$, we will get the same expressions of center manifold (\ref{eq65}) and the flow on the center manifold $(\ref{eq66}-\ref{eq67})$.  So, for this case the critical point $L_4$ is unstable due to its saddle nature. \par
For $\mu=\sqrt{3}$, after putting corresponding arguments as $L_3$, the center manifold can be written as
\begin{align}
u&=\frac{2}{3(1+\sqrt{3})}\left\{\frac{(1-\sqrt{3})\lambda^2+4\lambda}{48\sqrt{6}}\right \}w^2+\mathcal{O}(w^3),\label{eqn76}\\
v&=\frac{2}{3(\sqrt{3}-1)}\left\{\frac{(\sqrt{3}+1)\lambda^2+4\lambda}{48\sqrt{6}}\right \}w^2+\mathcal{O}(w^3)\label{eqn77}
\end{align}
and the flow on the center manifold is determined by
\begin{align}
w'&=\frac{1}{\sqrt{2}}w^2+\mathcal{O}(w^3)\label{eqn78}.
\end{align}
From the flow equation we can conclude that the origin is a saddle node and hence in the old coordinate system $L_4$ is a saddle node, i.e., unstable in nature.  The vector field near the origin in $uw$-plane is shown as in FIG.\ref{L_22} and the vector field near the origin in $vw$-plane is shown as in FIG.\ref{L_2_2}.\par
For $\mu=-\sqrt{3}$ we also get the same expression of center manifold and flow equation as for $\mu=\sqrt{3}$ case.

\begin{figure}[h]
	\centering
	\includegraphics[width=1\textwidth]{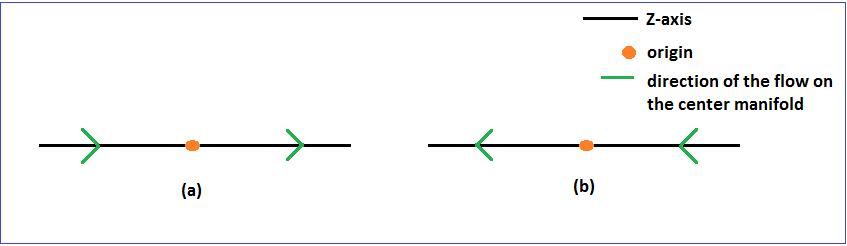}
	\caption{Flow on the center manifold near the origin for the critical point $L_5$.  (a) is for $\mu>0$ and (b) is for $\mu<0$.}
	\label{z_center_manifold}
\end{figure}

\begin{center}
	$5.~Critical~Point~L_5$
\end{center}
First we shift the critical point $L_5$ to the origin by the transformation $x= X-\sqrt{\frac{2}{3}}\mu$, $y=Y$ and $z= Z$.  For avoiding similar arguments which we have mentioned for the above critical points, we only state the main results center manifold and the flow equation for this critical point.  The center manifold for this critical point can be written as
\begin{align}
X&=0,\\Y&=0
\end{align}
and the flow on the center manifold can be obtained as
\begin{align}
\frac{dZ}{dN}=\sqrt{\frac{2}{3}}\mu Z^2 +\mathcal{O}(Z^3).
\end{align}
From the expressions of the center manifold we can conclude that the center manifold is lying on the $Z$-axis.  From the flow on the center manifold FIG.\ref{z_center_manifold}, we conclude that the origin is unstable for both of the cases $\mu>0$ or $\mu<0$.
\subsection{Model 3: Exponential potential and
	power-law-dependent dark-matter particle mass \label{M3}}
In this case evolution equations in Section \ref{BES} can be written to the autonomous system as follows 
\begin{eqnarray}
x'&=&-3x+\frac{3}{2}x(1-x^2-y^2)-\sqrt{\frac{3}{2}}\lambda y^2-\frac{\mu}{2}z(1+x^2-y^2),\label{eq82} \\
y'&=&\frac{3}{2}y(1-x^2-y^2)-\sqrt{\frac{3}{2}}\lambda xy,\label{eq83}  \\
z'&=&-xz^2.\label{eq84} 
\end{eqnarray}
We have three physical meaningful critical points $R_1$, $R_2$ and $R_3$ corresponding to the above autonomous system.  The set of critical points, their existence and the value of cosmological parameters at those critical points corresponding to the autonomous system $(\ref{eq82}-\ref{eq84})$ shown in Table \ref{TPRE} and the eigenvalues of the Jacobian matrix corresponding to the autonomous system $(\ref{eq82}-\ref{eq84})$ at those critical points and the nature of the critical points are shown in Table \ref{TNRE}.\par

Here we also concern about the stability of the critical points for $\mu\neq 0$ and $\lambda\neq 0$ because for another possible cases we will get the similar types result which we have obtained for Model $1$.
	\begin{table}[h]
	\caption{\label{TPRE}Table shows the set of critical points and their existence, value of cosmological parameters corresponding to that critical points. }
	\begin{tabular}{|c|c|c c c|c|c|c| c|}
		\hline
		\hline	
		\begin{tabular}{@{}c@{}}$~~$\\$~Critical ~Points$\\$~~$\end{tabular} &$Existence$&$x$&$y$&$z~~$& $~\Omega_X~$&$~\omega_X~$ &$~\omega_{tot}~$& $~q~$  \\ \hline\hline
		\begin{tabular}{@{}c@{}}$~~$\\$R_1$\\$~~$\end{tabular}& For all $\mu$ and $\lambda$&$0$&$0$&$0$&$0$&Undetermined&$0$&$\frac{1}{2}$\\ \hline
			\begin{tabular}{@{}c@{}}$~~$\\$R_2$\\$~~$\end{tabular}&For all $\mu$ and $\lambda$&$-\frac{\lambda}{\sqrt{6}}$&$\sqrt{1+\frac{\lambda^2}{6}}$&$0$&$1$&$-1-\frac{\lambda^2}{3}$&$-1-\frac{\lambda^2}{3}$&$-\frac{1}{2}\left(2+\lambda^2\right)$\\ \hline
				\begin{tabular}{@{}c@{}}$~~$\\$R_3$\\$~~$\end{tabular}&For all $\mu$ and $\lambda$&$-\frac{\lambda}{\sqrt{6}}$&$-\sqrt{1+\frac{\lambda^2}{6}}$&$0$&$1$&$-1-\frac{\lambda^2}{3}$&$-1-\frac{\lambda^2}{3}$&$-\frac{1}{2}\left(2+\lambda^2\right)$\\ \hline
    \end{tabular}
\end{table}
\begin{table}[h]
	\caption{\label{TNRE}The eigenvalues $(\lambda_1,\lambda_2,\lambda_3)$ of the Jacobian matrix corresponding to the autonomous system $(\ref{eq82}-\ref{eq84})$ at those critical points $(R_1-R_3)$ and the nature of the critical points.}
	\begin{tabular}{|c|c c c|c|}
		\hline
		\hline	
		\begin{tabular}{@{}c@{}}$~~$\\$~Critical ~Points$\\$~~$\end{tabular} &$\lambda_1$&$\lambda_2$&$\lambda_3$&$Nature~ of~ critical~ Points$  \\ \hline\hline
		\begin{tabular}{@{}c@{}}$~~$\\$R_1$\\$~~$\end{tabular}&$-\frac{3}{2}$&$\frac{3}{2}$&$0$&Non-hyperbolic\\ \hline
			\begin{tabular}{@{}c@{}}$~~$\\$R_2$\\$~~$\end{tabular}&$-(3+\lambda^2)$&$-\left(3+\frac{\lambda^2}{2}\right)$&$0$&Non-hyperbolic\\ \hline
					\begin{tabular}{@{}c@{}}$~~$\\$R_3$\\$~~$\end{tabular}&$-(3+\lambda^2)$&$-\left(3+\frac{\lambda^2}{2}\right)$&$0$&Non-hyperbolic\\ \hline
    \end{tabular}
\end{table}

For avoiding similar types of argument, we only state the stability of every critical points and the reason behind the stability in the tabular form, which is shown as in Table \ref{T_R_stability}.

\begin{table}[!]
	\caption{\label{T_R_stability}Table shows the stability and the reason behind the stability of the critical points $(R_1-R_3)$.}
	\begin{tabular}{|c|c|c|}
		\hline
		\hline	
				\begin{tabular}{@{}c@{}}$~~$\\$ CPs $\\$~$\end{tabular} &$Stability$& $Reason~behind~the~stability$ \\ \hline\hline
		\begin{tabular}{@{}c@{}}$~~$\\$R_1$\\$~$\end{tabular}&\begin{tabular}{@{}c@{}}$~~$\\For $\mu>0$, $R_1$ is a saddle node\\$~~$\\ and \\$~~$\\for $\mu<0$, $R_1$ is a stable node\\$~$\end{tabular}&\begin{tabular}{@{}c@{}}$~~$\\After introducing the coordinate transformation (\ref{eq15}),\\ we will get the same expression of center manifold\\ $(\ref{eq18}-\ref{eq19})$ and the flow on the center manifold is\\ determined by $(\ref{eq20})$(FIG.\ref{A_1}).\\$~$\end{tabular}\\ \hline
		\begin{tabular}{@{}c@{}}$~~$\\$R_2,R_3$\\$~$\end{tabular}&\begin{tabular}{@{}c@{}}$~~$\\For $\lambda>0$ or $\lambda<0$, \\$~~$\\$R_2$ and $R_3$ both are unstable\\$~$\end{tabular}&	\begin{tabular}{@{}c@{}}$~~$\\After shifting $R_2$ and $R_3$ to the origin by using coordinate\\ transformation $\left(x=X-\frac{\lambda}{\sqrt{6}},y=Y+\sqrt{1+\frac{\lambda^2}{6}},z=Z\right)$ and\\ $\left(x=X-\frac{\lambda}{\sqrt{6}},y=Y-\sqrt{1+\frac{\lambda^2}{6}},z=Z \right)$ respectively,\\ we can conclude that the center manifold is lying on $Z$-axis\\ and the flow on the center manifold is determined by\\
		$\frac{dZ}{dN}=\frac{\lambda}{\sqrt{6}}Z^2+\mathcal{O}(Z^3)$.\\$~~$\\ The origin is unstable for both of the cases $\lambda>0$\\ (same as FIG.\ref{z_center_manifold}\textbf{(a)}) and $\lambda<0$ (same as FIG.\ref{z_center_manifold}\textbf{(b)}).\\$~$\end{tabular}\\ \hline
\end{tabular}
\end{table}

\subsection{Model 4: Exponential potential and
	exponentially-dependent dark-matter particle mass \label{M4}}
In this consideration evolution equations in Section \ref{BES} can be written to the autonomous system as follows 
\begin{eqnarray}
x'&=&-3x+\frac{3}{2}x(1-x^2-y^2)-\sqrt{\frac{3}{2}}\lambda y^2-\sqrt{\frac{3}{2}}\mu(1+x^2-y^2),\label{eq85} \\
y'&=&\frac{3}{2}y(1-x^2-y^2)-\sqrt{\frac{3}{2}}\lambda xy.\label{eq86}
\end{eqnarray}
We ignore the equation corresponding to the auxiliary variable $z$ in the above autonomous system because the R.H.S. expression of $x'$ and $y'$ does not depend on $z$.\par
\begin{figure}[h]
	\centering
	\includegraphics[width=1\textwidth]{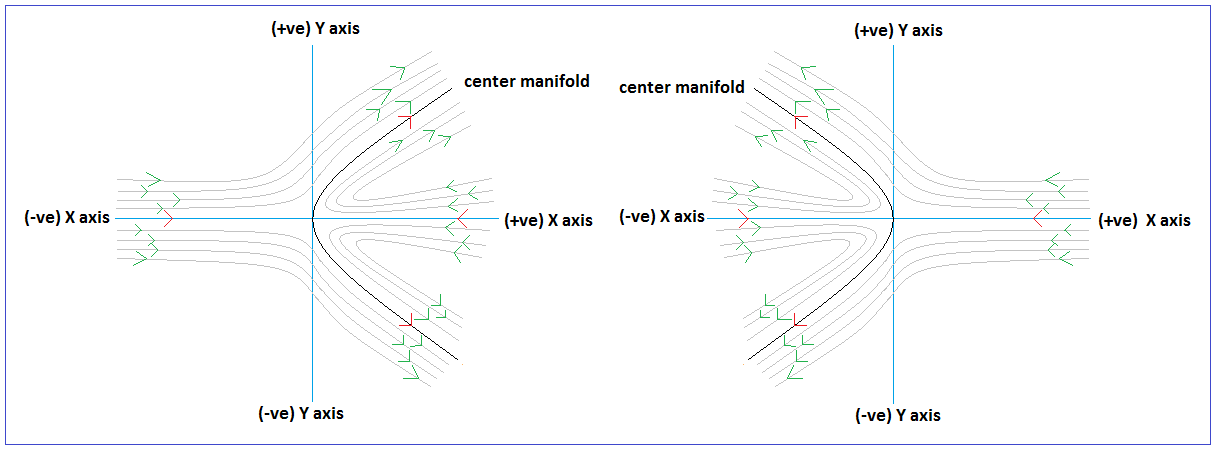}
	\caption{Vector field near the origin for the critical point $M_1$.  L.H.S. for $\mu >3$ and R.H.S. for $\mu<0$.}
	\label{M_1}
\end{figure}

 Corresponding to the above  autonomous system we have four critical points $M_1$, $M_2$, $M_3$ and $M_4$.   The set of critical points, their existence and the value of cosmological parameters at those critical points corresponding to the autonomous system $(\ref{eq85}-\ref{eq86})$ shown in Table \ref{TPME} and the eigenvalues of the Jacobian matrix corresponding to the autonomous system $(\ref{eq85}-\ref{eq86})$ at those critical points and the nature of the critical points are shown in Table \ref{TNME}.\bigbreak

\begin{table}[h]
	\caption{\label{TPME}Table shows the set of critical points and their existence, value of cosmological parameters corresponding to that critical points. }
	\begin{tabular}{|c|c|c c|c|c|c| c|}
		\hline
		\hline	
		\begin{tabular}{@{}c@{}}$~~$\\$~Critical ~Points$\\$~~$\end{tabular} &$Existence$&$x$&$y$& $~\Omega_X~$&$~\omega_X~$ &$~\omega_{tot}~$& $~q~$  \\ \hline\hline
		\begin{tabular}{@{}c@{}}$~~$\\	$M_1$\\$~~$\end{tabular}& For all $\mu$ and $\lambda$&$-\sqrt{\frac{2}{3}}\mu$&$0$&$-\frac{2}{3}\mu^2$ & $1$&$-\frac{2}{3}\mu^2$&$\frac{1}{2}\left(1-2\mu^2\right)$ \\ \hline
		\begin{tabular}{@{}c@{}}$~~$\\$M_2$\\$~~$\end{tabular}&For all $\mu$ and $\lambda$&$-\frac{\lambda}{\sqrt{6}}$&$\sqrt{1+\frac{\lambda^2}{6}}$&$1$&$-1-\frac{\lambda^2}{3}$&$-1-\frac{\lambda^2}{3}$&$-\frac{1}{2}\left(2+\lambda^2\right)$\\ \hline
	\begin{tabular}{@{}c@{}}$~~$\\$M_3$\\$~~$\end{tabular}&For all $\mu$ and $\lambda$&$-\frac{\lambda}{\sqrt{6}}$&$-\sqrt{1+\frac{\lambda^2}{6}}$&$1$&$-1-\frac{\lambda^2}{3}$&$-1-\frac{\lambda^2}{3}$&$-\frac{1}{2}\left(2+\lambda^2\right)$\\ \hline
    \begin{tabular}{@{}c@{}}$~~$\\$M_4$\\$~~$\end{tabular}&\begin{tabular}{@{}c@{}}$~~$\\For $\mu\neq\lambda$\\and\\ $\min\{\mu^2-\frac{3}{2},\lambda^2+3\}\geq\lambda\mu$\\$~~$\end{tabular}&$\frac{\sqrt{\frac{3}{2}}}{\lambda-\mu}$&$\frac{\sqrt{-\frac{3}{2}-\mu(\lambda-\mu)}}{|\lambda-\mu|}$&$\frac{\mu^2-\lambda\mu-3}{(\lambda-\mu)^2}$&$\frac{\mu(\lambda-\mu)}{\mu^2-\lambda\mu-3}$&$\frac{\mu}{\lambda-\mu}$&$\frac{1}{2}\left(\frac{\lambda+2\mu}{\lambda-\mu}\right)$\\ \hline 
\end{tabular}
\end{table}

\begin{table}[h]
	\caption{\label{TNME}The eigenvalues $(\lambda_1,\lambda_2)$ of the Jacobian matrix corresponding to the autonomous system $(\ref{eq85}-\ref{eq86})$ at those critical points $(M_1-M_4)$ and the nature of the critical points.}
	\begin{tabular}{|c|c c|c|}
		\hline
		\hline	
		\begin{tabular}{@{}c@{}}$~~$\\$~Critical ~Points$\\$~~$\end{tabular} &$\lambda_1$&$\lambda_2$&$Nature~ of~ critical~ Points$  \\ \hline\hline
		\begin{tabular}{@{}c@{}}$~~$\\$M_1$\\$~~$\end{tabular}&$-\left(\frac{3}{2}+\mu^2\right)$$~~$&$~~$$-\left(\mu^2-\frac{3}{2}\right)+\lambda\mu$& \begin{tabular}{@{}c@{}}$~~$\\Hyperbolic if $\left(\mu^2-\frac{3}{2}\right)\neq\lambda\mu$,\\$~~$ \\ non-hyperbolic if $\left(\mu^2-\frac{3}{2}\right)=\lambda\mu$\\$~~$\end{tabular}\\ \hline
		\begin{tabular}{@{}c@{}}$~~$\\$M_2$\\$~~$\end{tabular}&$-(3+\lambda^2)+\lambda\mu$&$-\left(3+\frac{\lambda^2}{2}\right)$&\begin{tabular}{@{}c@{}}$~~$\\Hyperbolic if $(\lambda^2+3)\neq\lambda\mu$,\\$~~$ \\ non-hyperbolic if $\left(\lambda^2+3\right)=\lambda\mu$\\$~~$\end{tabular}\\ \hline
			\begin{tabular}{@{}c@{}}$~~$\\$M_3$\\$~~$\end{tabular}&$-(3+\lambda^2)+\lambda\mu$&$-\left(3+\frac{\lambda^2}{2}\right)$&\begin{tabular}{@{}c@{}}$~~$\\Hyperbolic if $(\lambda^2+3)\neq\lambda\mu$,\\$~~$ \\ non-hyperbolic if $\left(\lambda^2+3\right)=\lambda\mu$\\$~~$\end{tabular}\\ \hline
			\begin{tabular}{@{}c@{}}$~~$\\$M_4$\\$~~$\end{tabular}&$\frac{a+d+\sqrt{(a-d)^2+4bc}}{2}$&$\frac{a+d-\sqrt{(a-d)^2+4bc}}{2}$&\begin{tabular}{@{}c@{}}$~~$\\Hyperbolic when $\mu^2-\frac{3}{2}>\lambda\mu$\\ and $\lambda^2+3>\lambda\mu$,\\$~~$\\non-hyperbolic when $\mu^2-\frac{3}{2}=\lambda\mu$\\ or $\lambda^2+3=\lambda\mu$\\$~~$\end{tabular}\\ \hline
    \end{tabular}
\end{table}
Note that for the critical point $M_4$ we have written the eigenvalues in terms of $a$, $b$, $c$ and $d$, where $a=-\frac{3}{2(\lambda-\mu)^2}(\lambda^2+3-\lambda\mu)$, $b=\mp\sqrt{\frac{3}{2}}\left(\frac{3}{(\lambda-\mu)^2}+2\right)\sqrt{-\frac{3}{2}-\mu(\lambda-\mu)}$, $c=\mp\sqrt{\frac{3}{2}}\left\{\frac{\lambda^2+3-\lambda\mu}{(\lambda-\mu)^2}\right\}
\sqrt{-\frac{3}{2}-\mu(\lambda-\mu)}$, $d=-\frac{3}{(\lambda-\mu)^2}\left\{\left(\mu^2-\frac{3}{2}\right)-\lambda\mu\right\}$.\par

Again, here we only state the stability of every critical points $(M_1-M_4)$ and the reason behind the stability in the tabular form, which is shown as in Table \ref{T_M_stability}.

\begin{table}[!]
	\caption{\label{T_M_stability}Table shows the stability and the reason behind the stability of the critical points $(M_1-M_4)$}
	\begin{tabular}{|c|c|c|}
		\hline
		\hline	
		\begin{tabular}{@{}c@{}}$~~$\\$ CPs $\\$~$\end{tabular} &$Stability$& $Reason~behind~the~stability$ \\ \hline\hline
	    \begin{tabular}{@{}c@{}}$~~$\\$ M_1 $\\$~$\end{tabular}& \begin{tabular}{@{}c@{}}$~~$\\Stable node for $\left(\mu^2-\frac{3}{2}\right)>\lambda\mu$\\ $~~$\\and\\$~~$\\ saddle node for $\left(\mu^2-\frac{3}{2}\right)\leq\lambda\mu$\\$~$\end{tabular}& \begin{tabular}{@{}c@{}}$~~$\\For $\left(\mu^2-\frac{3}{2}\right)>\lambda\mu$, as both eigenvalues \\of the Jacobian matrix at $M_1$ are negative, so by\\ Hartman-Grobman theorem we can conclude that\\ the critical point $M_1$ is a stable node.\\$~~$\\  For $\left(\mu^2-\frac{3}{2}\right)<\lambda\mu$, as one eigenvalue is positive\\ and another is negative, so by Hartman-Grobman theorem\\ we can conclude that the critical point $M_1$ is a saddle node.\\$~~$\\ For $\left(\mu^2-\frac{3}{2}\right)=\lambda\mu$, after shifting the critical point\\ $M_1$ to the origin by the coordinate transformation\\ $\left(x=X-\sqrt{\frac{2}{3}}\mu,y=Y\right)$, the center manifold can be written as \\$X=\frac{1}{\mu}\sqrt{\frac{3}{2}}Y^2+\mathcal{O}(Y^3)$\\ and the flow on the center manifold can be determined as\\ $\frac{dY}{dN}=\frac{9}{4\mu^2}Y^3+\mathcal{O}(Y^4)$.\\  Hence, for both of the cases $\mu>0$ and $\mu<0$ the origin\\ is a saddle node and unstable in nature (FIG.\ref{M_1}).\\$~~$\end{tabular}\\ \hline
	     \begin{tabular}{@{}c@{}}$~~$\\$ M_2,M_3 $\\$~$\end{tabular}& \begin{tabular}{@{}c@{}}$~~$\\Stable node for $\left(\lambda^2+3\right)>\lambda\mu$\\$~~$\\ and\\$~~$\\ saddle node for $\left(\lambda^2+3\right)\leq\lambda\mu$\\$~$\end{tabular}& \begin{tabular}{@{}c@{}}$~~$\\For $\left(\lambda^2+3\right)>\lambda\mu$, as both eigenvalues \\of the Jacobian matrix at $M_2$ are negative, so by\\ Hartman-Grobman theorem we can conclude that\\ the critical point $M_2$ is a stable node.\\$~~$\\  For $\left(\lambda^2+3\right)<\lambda\mu$, as one eigenvalue is positive\\ and another is negative, so by Hartman-Grobman theorem\\ we can conclude that the critical point $M_2$ is a saddle node.\\$~~$\\ For $\left(\lambda^2+3\right)=\lambda\mu$, after shifting the critical point\\ $M_1$ to the origin by the coordinate transformation\\ $\left(x=X-\frac{\lambda}{\sqrt{6}},y=Y\pm\sqrt{1+\frac{\lambda^2}{6}}\right)$, the center manifold can be\\ written as $~~Y=\mp\frac{1}{2\sqrt{1+\frac{\lambda^2}{6}}}X^2+\mathcal{O}(X^3)$\\ and the flow on the center manifold can be determined as\\ $\frac{dX}{dN}=\frac{\lambda}{2}\sqrt{\frac{3}{2}}\left\{1-\frac{6}{\lambda^2}\pm\frac{12}{\lambda^2}\left(1+\frac{\lambda^2}{6}\right)^{\frac{3}{2}}\right\}X^2+\mathcal{O}(X^4)$.\\  Hence, for all possible values $\lambda$ due to the even power $X$\\ in the R.H.S. of the flow equation, the origin is\\ a saddle node and unstable in nature.\\$~~$\end{tabular}\\ \hline
	     \begin{tabular}{@{}c@{}}$~~$\\$ M_4 $\\$~$\end{tabular}& \begin{tabular}{@{}c@{}}$~~$\\Saddle node for both of the cases, i.e.,\\ $\mu^2-\frac{3}{2}=\lambda\mu$ or $\lambda^2+3=\lambda\mu$\\$~$\end{tabular}& \begin{tabular}{@{}c@{}}$~~$\\ For $\mu^2-\frac{3}{2}=\lambda\mu$, as $M_4$ converts into\\ $M_1$, so we get the same stability like $M_1$.\\$~~$\\  For $\lambda^2+3=\lambda\mu$ as $M_4$ converts into $M_2$ and $M_3$, \\so we get the same stability like $M_2$ and $M_3$.\\$~$\end{tabular}\\\hline
   \end{tabular}
\end{table}
Also note that for hyperbolic case of $M_4$, the components of the Jacobian matrix $a,b,c$ and $d$ are very complicated and from the determination of eigenvalue, it is very difficult to provide any conclusion about the stability and for this reason we skip the stability analysis for this case.

\subsection{Model 5: Product of exponential and power-law potential and
	product of exponentially-dependent and power-law-dependent dark-matter particle mass \label{M5}}
In this consideration evolution equations in Section \ref{BES} can be written to the autonomous system as follows 
\begin{eqnarray}
x'&=&-3x+\frac{3}{2}x(1-x^2-y^2)-\sqrt{\frac{3}{2}}\lambda y^2-\frac{\lambda}{2}y^2z-\sqrt{\frac{3}{2}}\mu(1+x^2-y^2)-\frac{\mu}{2}z(1+x^2-y^2),\label{eqn80} \\
y'&=&\frac{3}{2}y(1-x^2-y^2)-\sqrt{\frac{3}{2}}\lambda xy-\frac{\lambda}{2}xyz,\label{eqn81}\\
z'&=&-xz^2\label{eqn82}
\end{eqnarray}
For determining the critical points corresponding to the above autonomous system, we first equate the R.H.S. of (\ref{eqn82}) with $0$.  Then we have either $x=0$ or $z=0$.  For $z=0$ then the above autonomous system converts in to the autonomous system of Model 4.  So, then we will get the similar types of result as Model 4.  When $x=0$, we have three physically meaningful critical points corresponding to the above autonomous system for $\mu\neq 0$ and $\lambda\neq 0$.  For another choices of $\mu$ and $\lambda$ like Model 1, we will get similar types of results.  The critical points are $N_1(0,0,-\sqrt{6})$, $N_2(0,1,-\sqrt{6})$ and $N_3(0,-1,-\sqrt{6})$ and all are hyperbolic in nature. As the $x$ and $y$ coordinates of these critical points are same as $A_1$, $A_2$ and $A_3$ and the value of cosmological parameters are not depending on $z$ coordinate, so we get the same result for the value of cosmological parameters as $A_1$, $A_2$ and $A_3$ respectively, which are presented in Table \ref{TI}.
\begin{center}
	$1.~Critical~Point~N_1$
\end{center}
The Jacobian matrix corresponding to the autonomous system (\ref{eqn80}-\ref{eqn82}) at the critical point $N_1$ has three eigenvalues $\frac{3}{2}$, $-\frac{1}{4}\left(3+\sqrt{9+48\mu}\right)$ and $-\frac{1}{4}\left(3-\sqrt{9+48\mu}\right)$ and the corresponding eigenvectors are $[0,1,0]^T$, $\left[\frac{1}{24}\left(3+\sqrt{9+48\mu}\right),0,1\right]^T$ and $\left[\frac{1}{24}\left(3-\sqrt{9+48\mu}\right),0,1\right]^T$ respectively.  As the critical point is hyperbolic in nature, so we use Hartman-Grobman theorem for analyzing the stability of this critical point.  From the determination of eigenvalues, we conclude that the stability of the critical point $N_1$ depends on $\mu$.  For $\mu<-\frac{9}{48}$, the last two eigenvalues are complex conjugate with negative real parts.  For $\mu\geq-\frac{9}{48}$, all eigenvalues are real.\par
\begin{figure}[h]
	\centering
	\includegraphics[width=0.4\textwidth]{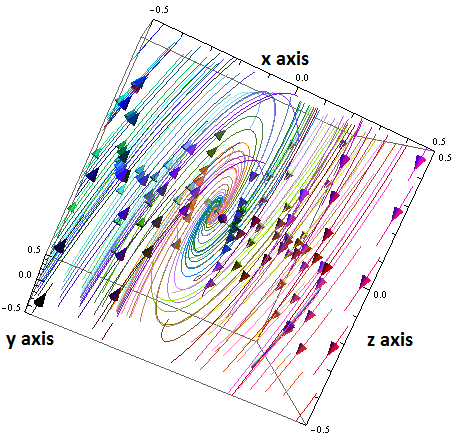}
	\caption{Phase portrait near the origin for the critical point $N_1$ in $xyz$ coordinate system.  This phase portrait is drawn for $\mu=-1$.}
	\label{focus_1}
\end{figure}

For $\mu<-\frac{9}{48}$, due to presence of negative real parts of last two eigenvalues, $yz$-plane is the stable subspace and as the first eigenvalue is positive, $x$-axis is the unstable subspace.  Hence, the critical point $N_1$ is saddle-focus, i.e., unstable in nature.  The phase portrait in $xyz$ coordinate system is shown as in FIG.\ref{focus_1}.\par
For $\mu\geq-\frac{9}{48}$, always we have at least one positive eigenvalue and at least one negative eigenvalue and hence we can conclude that the critical point $N_1$ is unstable due to its saddle nature.
\begin{center}
	$2.~Critical~Point~N_2~\&~ N_3$
\end{center}
The Jacobian matrix corresponding to the autonomous system $(\ref{eqn80}-\ref{eqn82})$ at the critical point $N_2$ and $N_3$ has three eigenvalues $-3$, $-\frac{1}{2}\left(3+\sqrt{9+12\lambda}\right)$ and $-\frac{1}{2}\left(3-\sqrt{9+12\lambda}\right)$ and the corresponding eigenvectors are $[0,1,0]^T$, $\left[\frac{1}{12}\left(3+\sqrt{9+12\lambda}\right),0,1\right]^T$ and $\left[\frac{1}{12}\left(3-\sqrt{9+12\lambda}\right),0,1\right]^T$ respectively.  From the determination of the eigenvalue, we conclude that the last two eigenvalues are complex conjugate while $\lambda<-\frac{3}{4}$ and the eigenvalues are real while $\lambda\geq-\frac{3}{4}$.\par
\begin{figure}[h]
	\centering
	\includegraphics[width=0.4\textwidth]{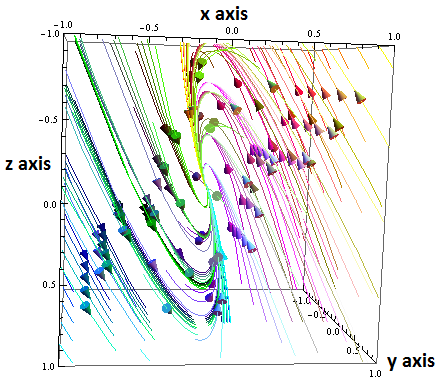}
	\caption{Phase portrait near the origin for the critical point $N_2$ and $N_3$ in $xyz$ coordinate system.  This phase portrait is drawn for $\lambda=-1$.}
	\label{focus_2}
\end{figure}

 
For $\lambda<-\frac{3}{4}$, we can see that the last two eigenvalues are complex with negative real parts and first eigenvalue is always negative.  Hence, by Hartman-Grobman theorem we conclude that the critical points $N_2$ and $N_3$ both are stable focus-node in this case.  The phase portrait in $xyz-$coordinate system is shown as in FIG.\ref{focus_2}.\par
For $-\frac{3}{4}\leq\lambda<0$, we can see that all eigenvalues are negative.  So, by Hartman-Grobman theorem we conclude that the critical points $N_2$ and $N_3$ both are stable node in this case.\par
For $\lambda>0$, we have two negative and one positive eigenvalues.  Hence, by Hartman-Grobman theorem we conclude that the critical points $N_2$ and $N_3$ both are saddle node and unstable in nature.\bigskip

\section{Bifurcation Analysis by Poincar\'{e} index and Global Cosmological evolution \label{BAPGCE}}

The flat potential plays a crucial role to obtain the bouncing solution.  After the bounce, the flat potential naturally allows the universe to penetrate the slow-roll inflation regime, as a result of that making the bouncing universe compatible with observations.\par

\begin{figure}[h]
	\centering
	\includegraphics[width=.4\textwidth]{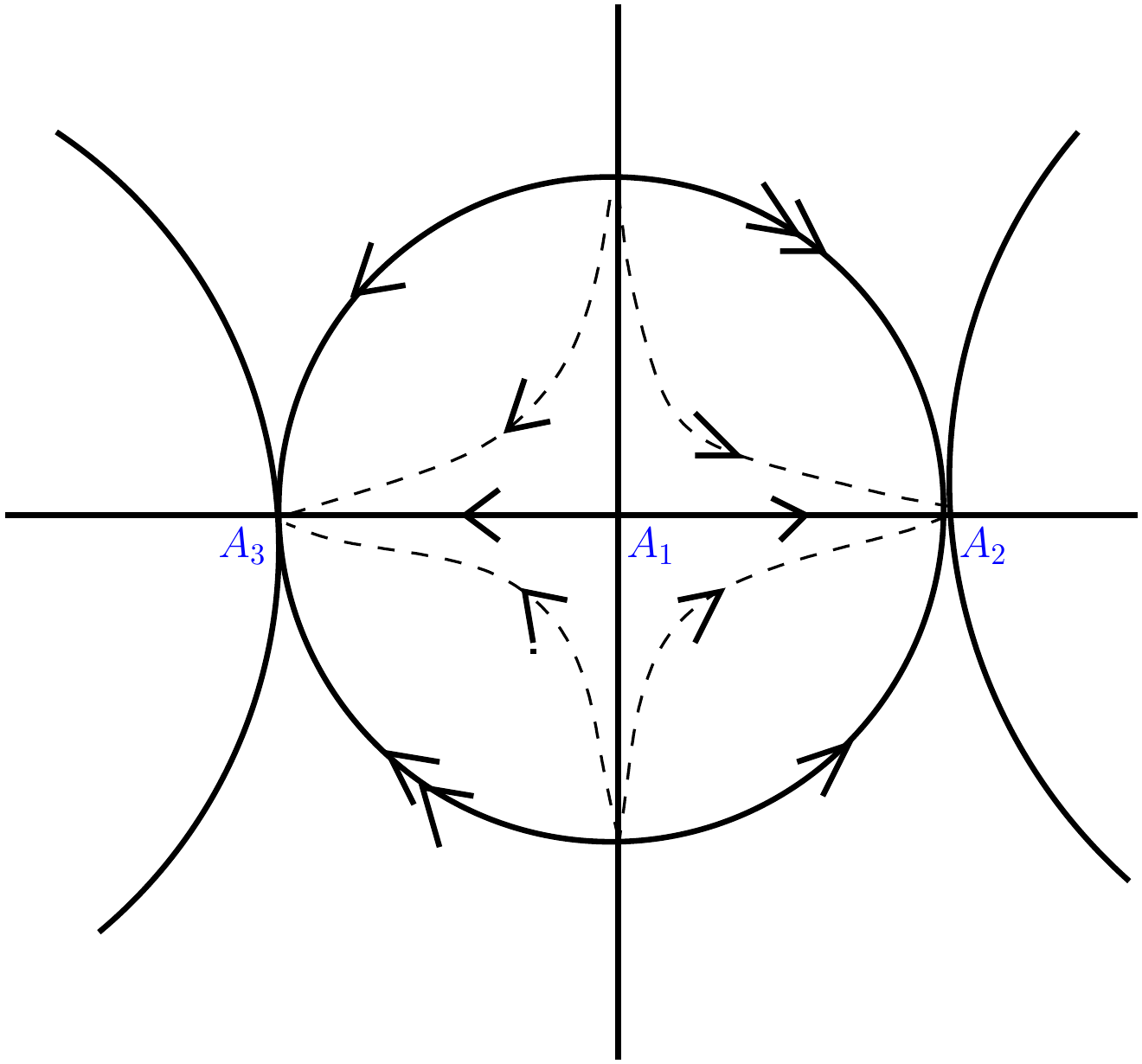}
	\caption{Vector field on the projective plane by antipodal points identified of the disk.}
	\label{hyperbola}
\end{figure}

\begin{figure}[htbp!]
	\begin{subfigure}{0.34\textwidth}
		\includegraphics[width=.9\linewidth]{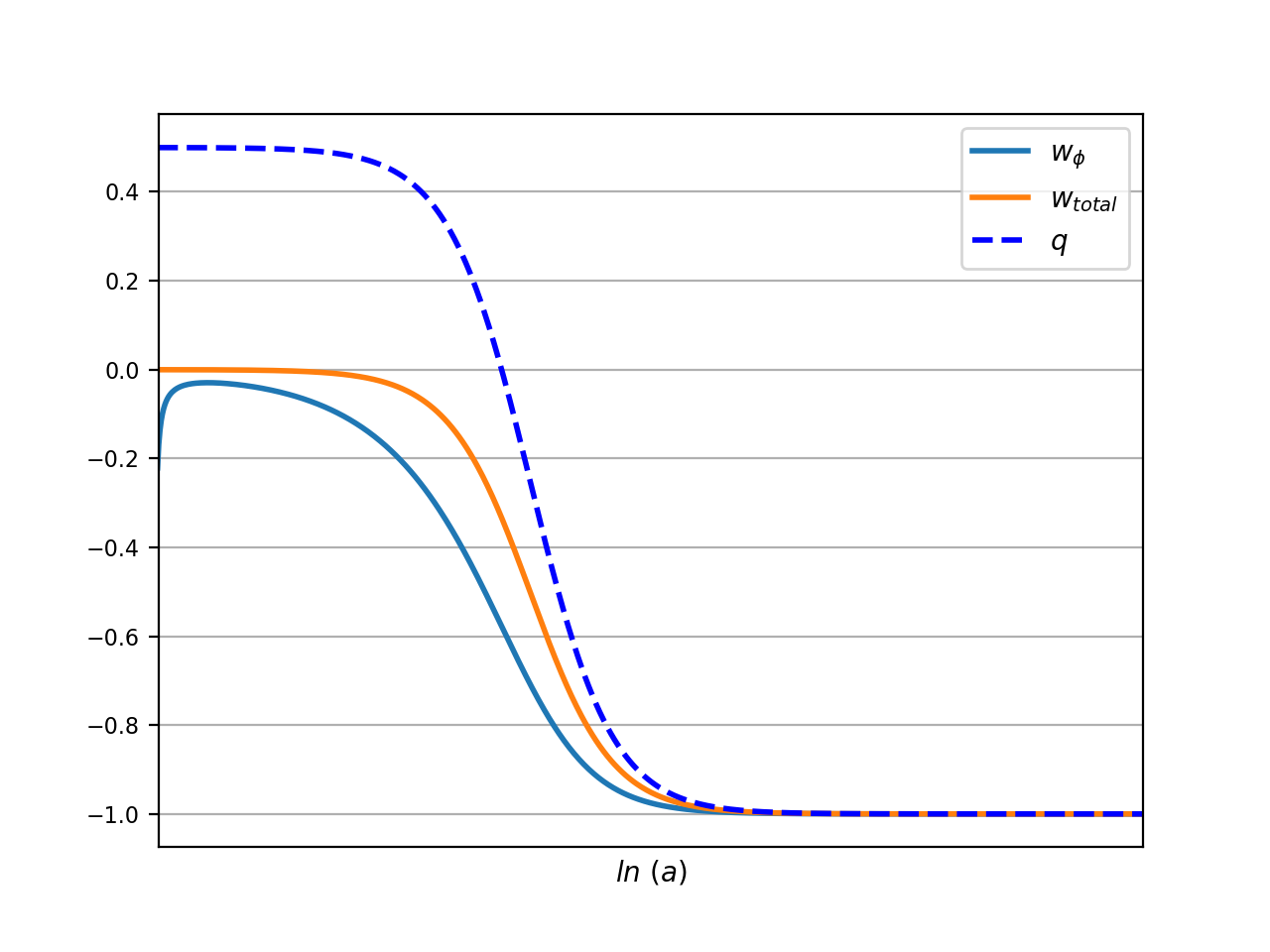}
		\caption{}
		\label{fig:A1}
	\end{subfigure}%
	\begin{subfigure}{0.34\textwidth}
		\includegraphics[width=.9\linewidth]{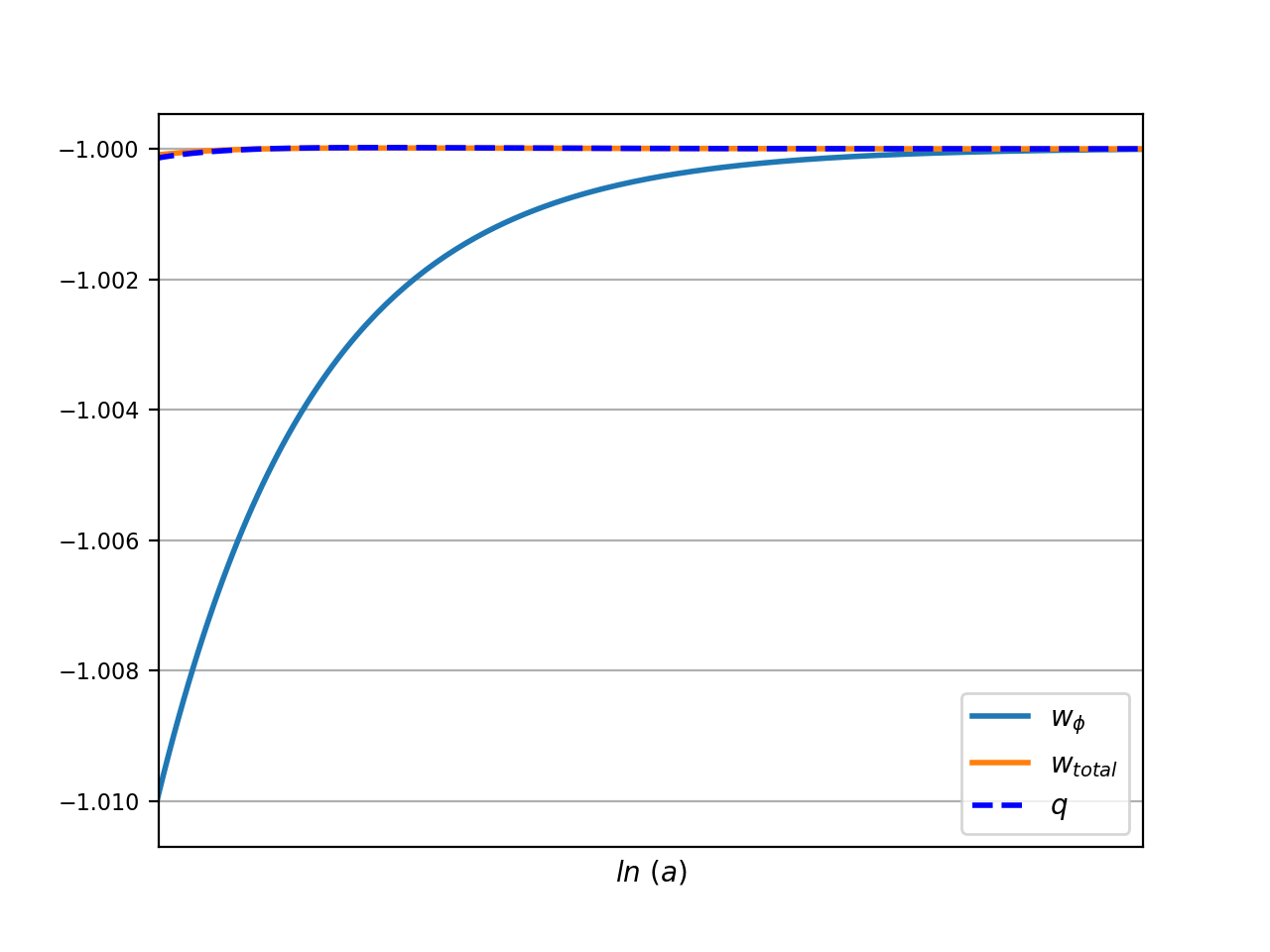}
		\caption{}
		\label{fig:A2}
	\end{subfigure}%
	\begin{subfigure}{.34\textwidth}
		\includegraphics[width=.9\linewidth]{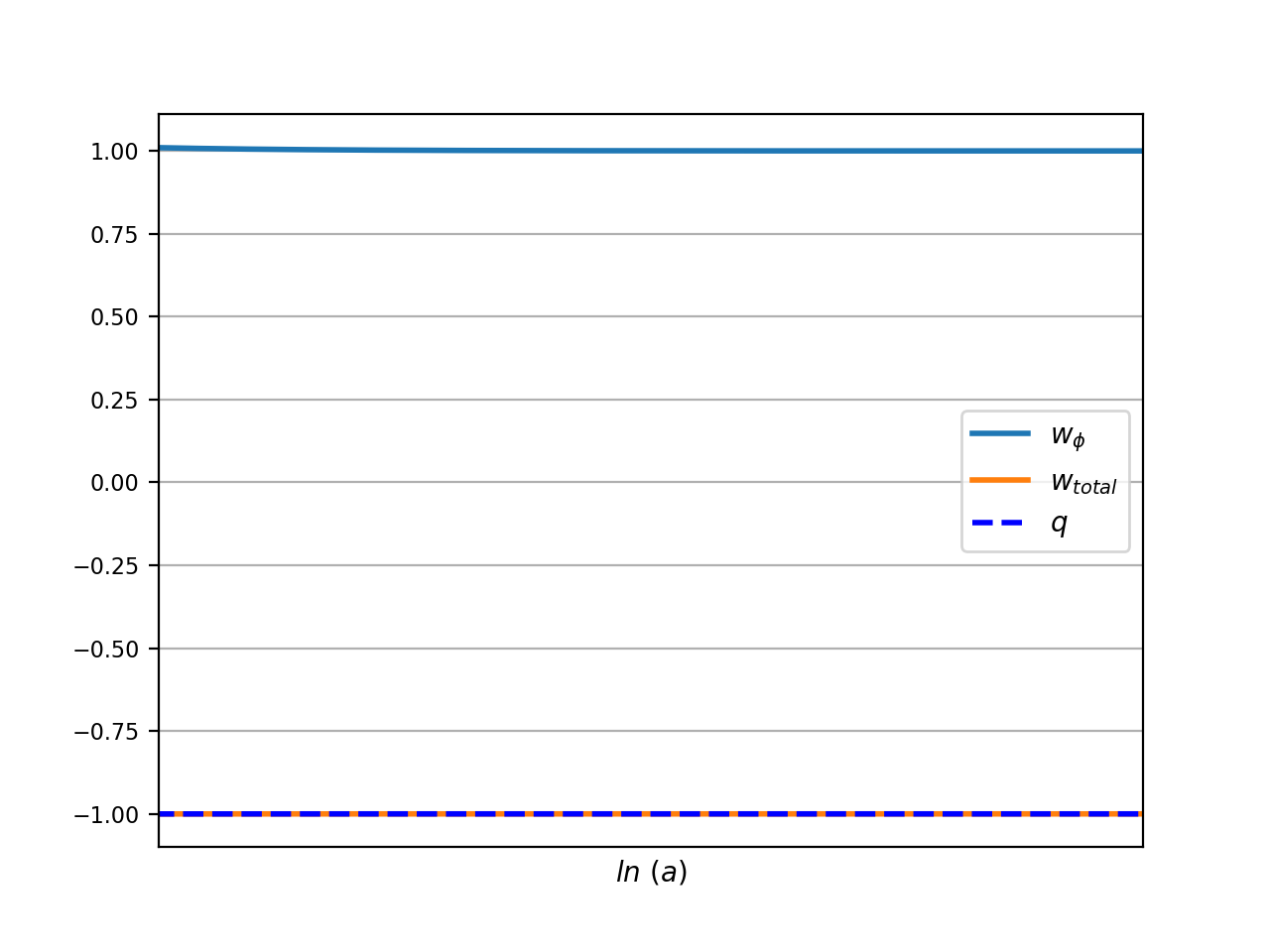}
		\caption{}
		\label{fig:A3}
	\end{subfigure}
	
	\caption{\label{Model1}\textit{Model 1}: Qualitative evolution of the physical variables $\omega_{total}$, $\omega_{\phi}$ and $q$ for perturbation of the parameters ($\lambda$ \& $\mu$) near the bifurcation values for three sets of initial conditions. (a) The initial condition near the point $A_1$. (b) The initial condition near the point $A_2$. (c) The initial condition near the point $A_3$.  We observe that the limit of the physical parameter $\omega_{total}\rightarrow -1$.  In early or present time the scalar field may be in phantom phase but the field is attracted to the de-Sitter phase.}
	
\end{figure}

In Model 1 (\ref{M1}),  for the inflationary scenario, we consider $\lambda$ and $\mu$ very small positive number so that $V(\phi) \approx V_0$ and $M_{DM} \approx M_0$.  The Eqn. (\ref{eq11}) mainly regulate the flow along $Z$-axis.  Due to Eqn. (\ref{eq11}) the overall 3-dimensional phase space splits up into two compartments and the $ZY$-plane becomes the separatrix.  In the right compartment, for $x>0$, we have $z' <0$ and $z'>0$ in the left compartment.  on the $ZY$ plane $z' \approx 0$.  For $\lambda \neq 0$ and $\mu \neq 0$, all critical points are located on the Y-axis.  As all cosmological parameters can be expressed in terms of $x$ and $y$, so we rigorously inspect the vector field on $XY$-plane.  Due to Eqn. (\ref{eq4}), the viable phase-space region (say $S$) satisfies $y^2-x^2 \leqslant 1$  which is inside of a hyperbola centered at the origin (FIG.\ref{hyperbola}).  On the $XY$-plane $z' \approx 0$.  So on the $XY$-plane, by Hartman-Grobman theorem we can conclude there are four hyperbolic sectors around $A_1$ ($\alpha$-limit set) and one parabolic sector around each of $A_2$ and $A_3$ ($\omega$-limit sets).  So, by Bendixson theorem, it is to be noted that, the index of $A_1|_{XY}$ is $1$ and the index of $A_2|_{XY}$ and $A_3|_{XY}$ is $-1$.   If the initial position of the universe is in left compartment and near to the $\alpha$-limit, then the universe remains in the left compartment and moves towards $\omega$-limit set asymptotically  at late time.  Similar phenomenon happens in right compartment also.   The universe experiences a fluid dominated non-generic evolution near $A_1$ for $\mu>0$ and a generic evolution for $\mu<0$.  For sufficiently flat potential, near $A_2$ and $A_3$, a scalar field dominated non-generic and generic evolution occur for $\lambda>0$ and $\lambda<0$ respectively (see FIG. \ref{Model1}).

The Poinca\'{e} index theorem \cite{0-387-95116-4} helps us to determine Euler Poincar\'{e} characteristic which is $\chi(S)=n-f-s$, where $n$, $f$, $s$ are the number of nodes, foci and saddle on $S$.  Henceforward we consider index as  Poinca\'{e} index. So for the vector field  of case-(i)$|_{XY-plane}$, $\chi(S)=1$.  This vector field can define a vector field on the projective plane, i.e., in 3-dimensional phase-space, if we consider a closed disk the $XY$-plane of radius one and centered at the origin, then we have the same vector field on the projective plane by antipodal point identified.\par
For $z=constant (\neq 0)$ plane the above characterization of vector field changes as a vertical flow along $Z$-axis regulate the character of the vector field.  Using Bendixson theorem \cite{0-387-95116-4} we can find the index of nonhyperbolic critical point by restricting the vector field on a suitable two-dimensional subspace.\par
If we restrict ourselves on $XZ$-plane, $A_1$ is saddle in nature for $\mu > 0$.  On the $XZ$ plane the index of $A_1$ is -1 for $\mu>0$ as four hyperbolic sectors are separated by two separatices around $A_1$.  For $\mu<0$, there is only one parabolic sector and the index is zero (FIG.\ref{A_1}).  On the $YZ$ plane $A_1$ swap its index with $XZ$ plane depending on the sign of $\mu$.\par 
On the uw-plane $A_2$ and $A_3$ have index 1 for $\lambda>0$ and -1 for $\lambda \leqslant 0$.  On the uw-plane $A_2$ and $A_3$ have index -1 for $\lambda>0$ and 1 for $\lambda < 0$.  At $\lambda=0$, the index of $A_2$ is 0 but the index of $A_3$ is 1.  On uv-plane the index $A_2$ or $A_3$ is 1 and does not depend on $\lambda$.  On the (uw)-plane around $A_2$ the number of hyperbolic sector is four and there is no elliptic sector. So the index of $A_2$ and $A_3$ $(origin)|_{uw~plane}/ _{vw~plane}$  is -1 for $\lambda>0$ and for $\lambda<0$ the index is 1 as there is no hyperbolic or elliptic orbit.\par
A set of non-isolated equilibrium points is said to be normally hyperbolic if the only eigenvalues with zero real parts are those whose corresponding eigenvectors are tangent to the set.  For the case (ii) to case (iv), we get normally hyperbolic critical points as the eigenvector $[0~ 0~ 1]^T$ (in new $(u,v,w)$ coordinate system) corresponding to only zero eigenvalue, is tangent to the line of critical points. The stability of a set which is normally hyperbolic can be completely classified by considering the signs of the eigenvalues in the remaining directions.  So the character of the flow of the phase space for each $z=constant$ plane is identical to the $XY$-plane in the previous case.  Thus the system (\ref{eq9}-\ref{eq11}) is structurally unstable \cite{0-387-95116-4} at $\lambda=0$ or $\mu=0$ or both.  On the other hand, the potential changes its character from runaway to non-runaway as $\lambda$ crosses zero from positive to negative.  Thus $\lambda=0$ and $\mu=0$ are the bifurcation values\cite{1950261}.\bigbreak

Model 2 (\ref{M2}) contains five critical points $L_1-L_5$.  For $\lambda>0$, the flow is unstable and for $\lambda<0$ the flow on the center manifold is stable.  Around $L_2$, the character of the vector field same as $L_1$.  For $\mu=\pm \sqrt{\frac{3}{2}}$, the flow on the center manifold at $L_3$ or $L_4$ depends on the sign of $\lambda$ (FIG.\ref{L_21} \& FIG.\ref{L_2_1_1}).  On the other hand,  $\mu>\sqrt{\frac{3}{2}}$ or $\mu< \sqrt{\frac{3}{2}}$ the flow on the center manifold does not depend on $\lambda$.  For $\mu >0$, the flow on the center manifold at $L_5$ moves increasing direction of $z$.  On the other hand, for $\mu <0$, the flow on the center manifold is in decreasing direction of $z$.    The index of $L_1$ is same as $A_2$. 
For $\mu=\pm \sqrt{\frac{3}{2}}$ and $\lambda=1$, the index of $L_2|_{XY plane}$ is -1 as there are only four hyperbolic sectors.  But for $\lambda=2$, there are two hyperbolic and one parabolic sectors, so the index is zero. 
The index of $L_3$ is same as $L_2$.  The index of $L_4$ on $ZX$ or $XY$ plane is zero as there are two hyperbolic and one parabolic sector for each $\mu>0$ and $\mu<0$.  So it is to be noted that, for $\lambda=0, \pm \sqrt{\frac{3}{2}} $ and $\mu=0$  the system is structurally unstable.  
\begin{figure}[htbp!]
	\begin{subfigure}{0.34\textwidth}
		\includegraphics[width=.9\linewidth]{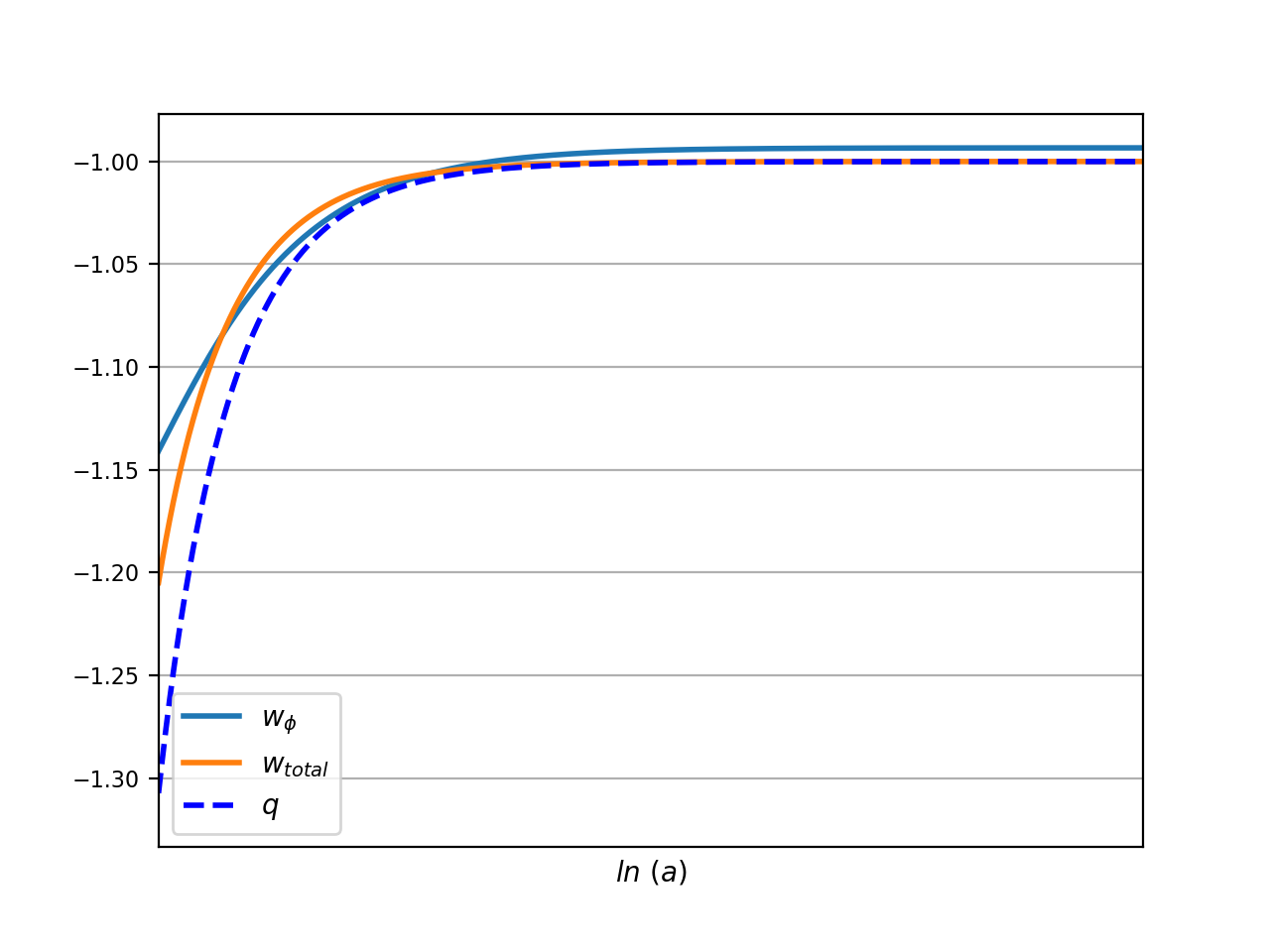}
		\caption{}
		\label{fig:L1}
	\end{subfigure}%
	\begin{subfigure}{0.34\textwidth}
		\includegraphics[width=.9\linewidth]{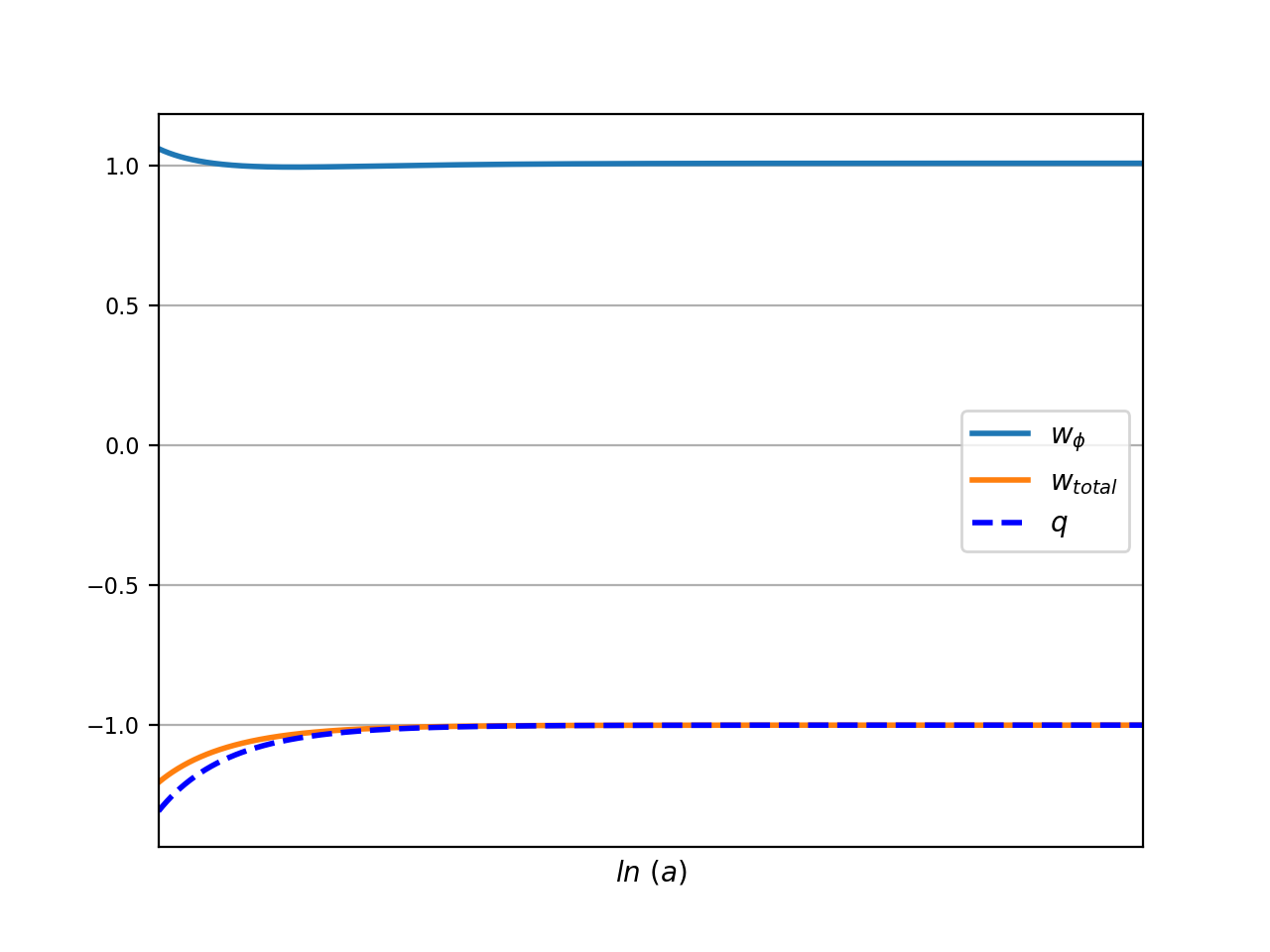}
		\caption{}
		\label{fig:L2}
	\end{subfigure}%
	\begin{subfigure}{.34\textwidth}
		\includegraphics[width=.9\linewidth]{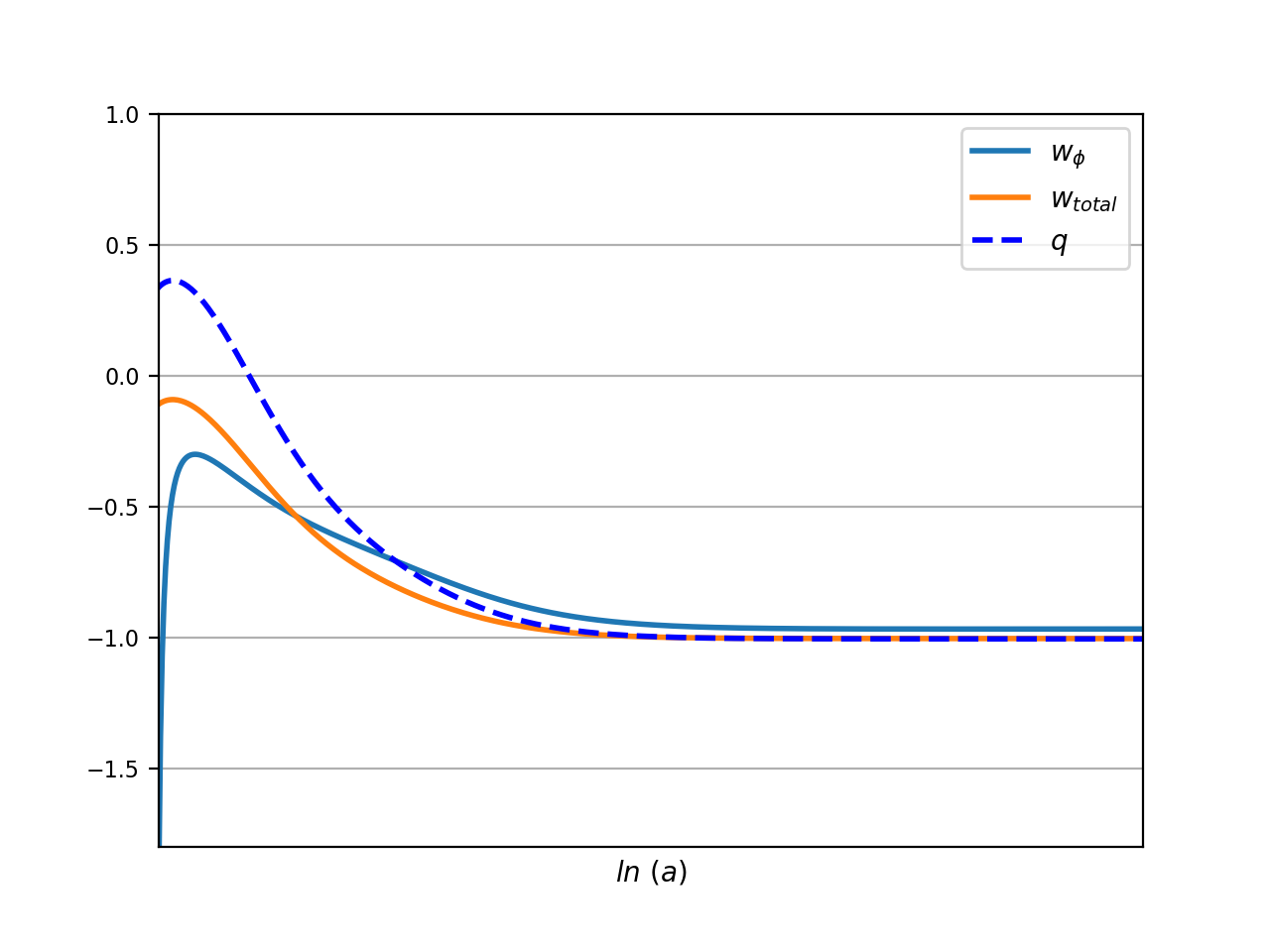}
		\caption{}
		\label{fig:L3n}
	\end{subfigure}
	
	\begin{subfigure}{0.34\textwidth}
		\includegraphics[width=.9\linewidth]{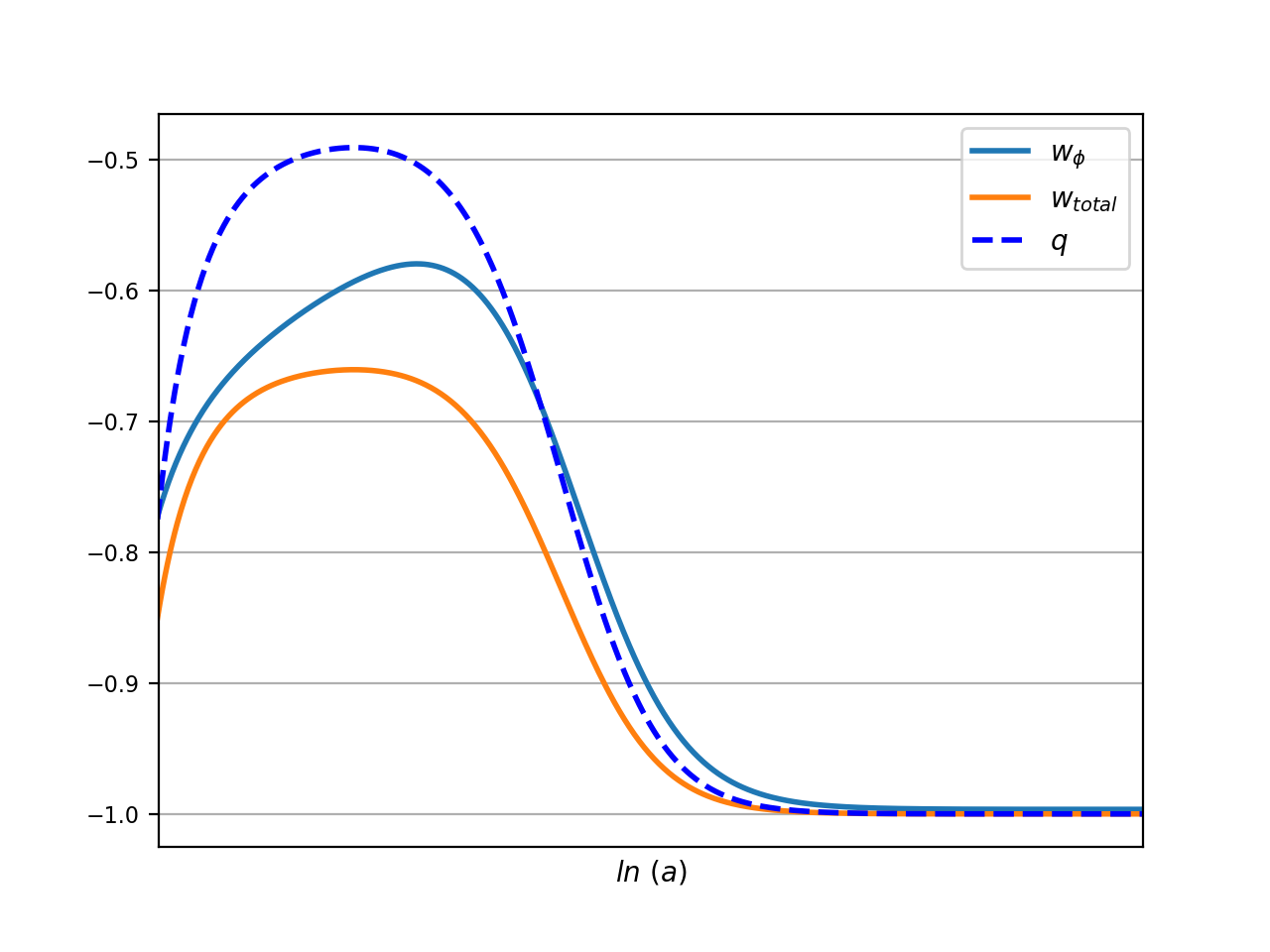}
		\caption{}
		\label{fig:L3n}
	\end{subfigure}%
	\begin{subfigure}{0.34\textwidth}
		\includegraphics[width=.9\linewidth]{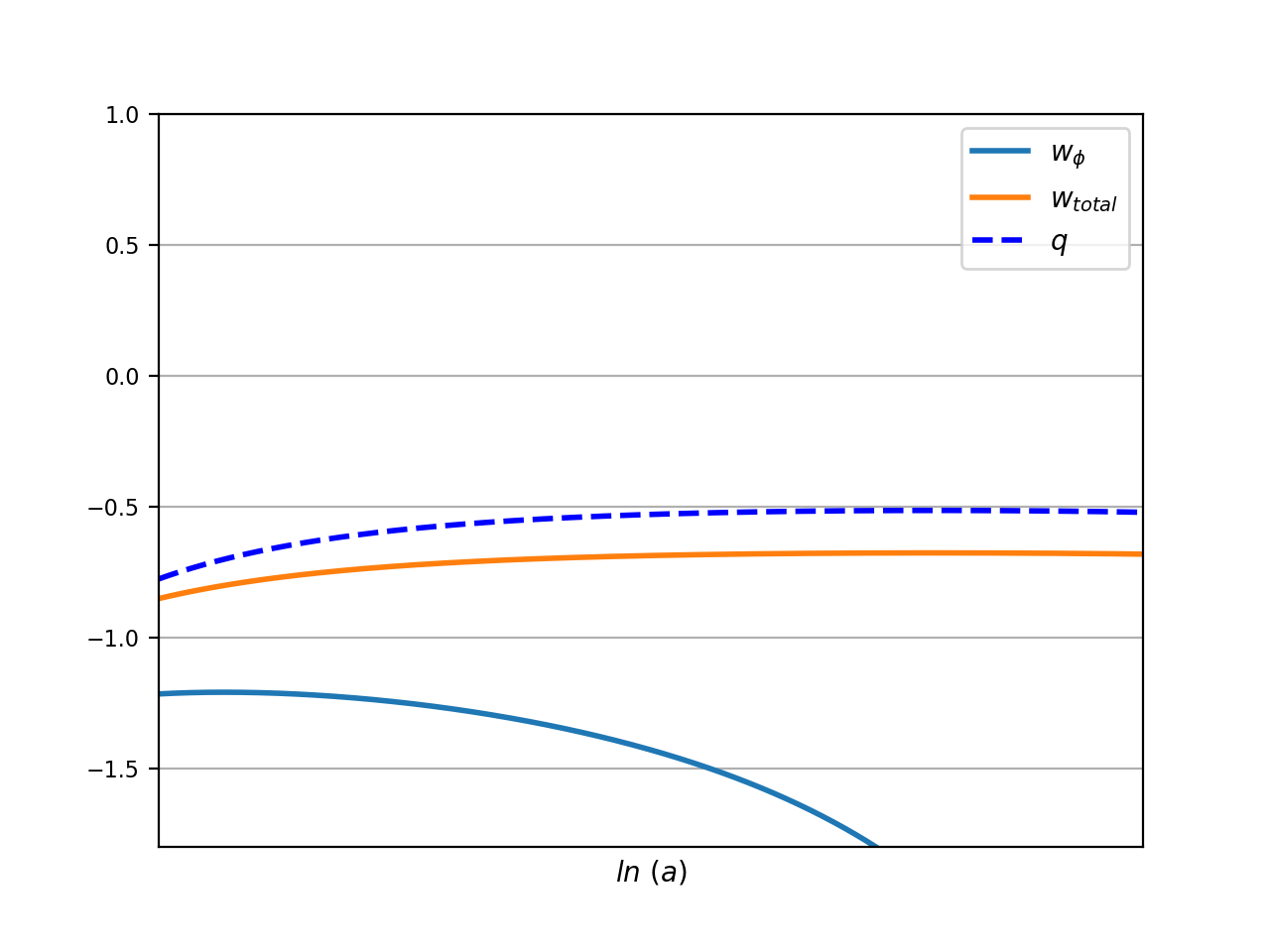}
		\caption{}
		\label{fig:L4p}
	\end{subfigure}%
	\begin{subfigure}{.34\textwidth}
		\includegraphics[width=.9\linewidth]{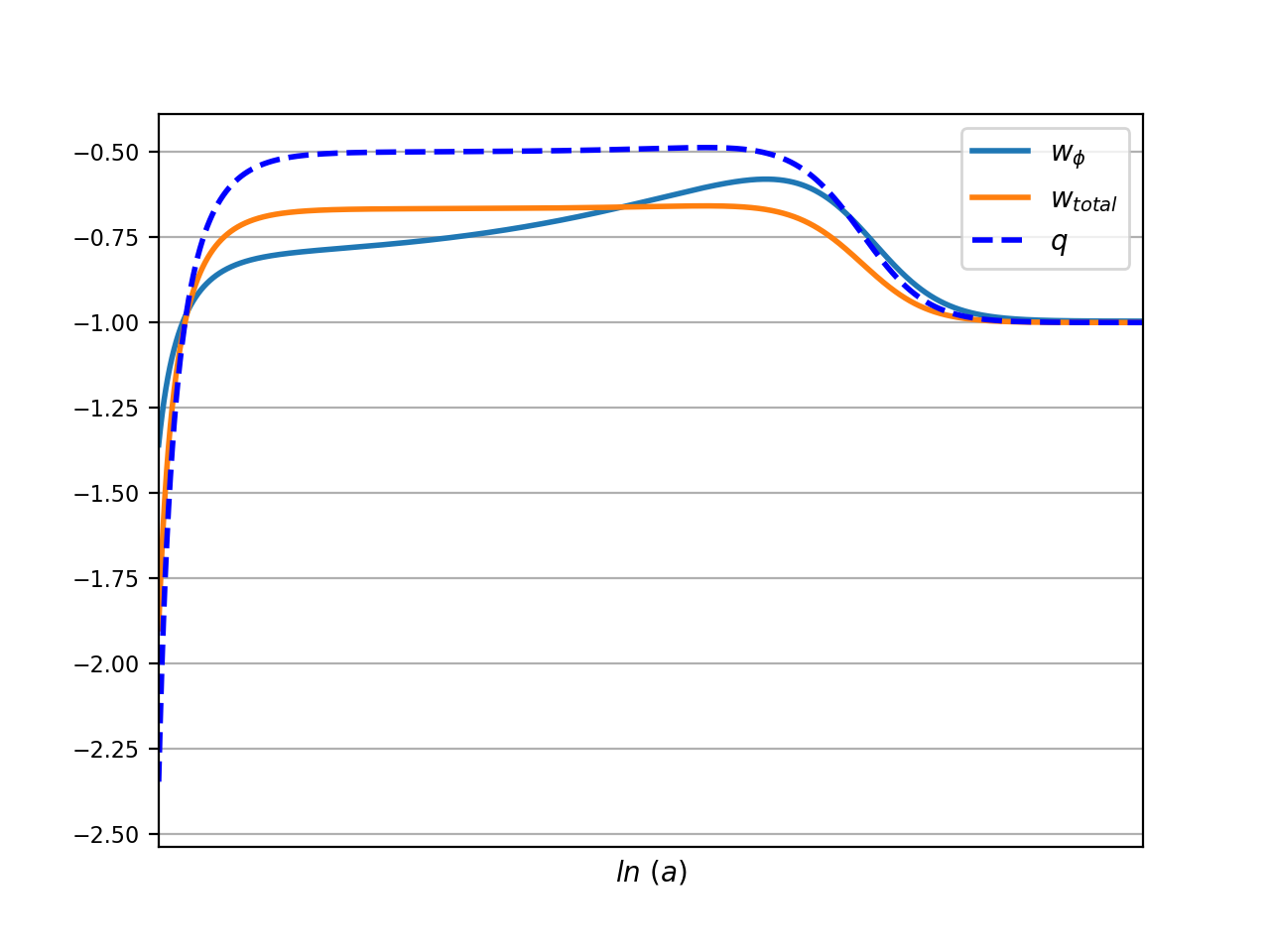}
		\caption{}
		\label{fig:L5}
	\end{subfigure}

	\caption{\label{Model2}\textit{Model 2}: Some interesting qualitative evolution of the physical variables $\omega_{total}$, $\omega_{\phi}$ and $q$ for perturbation of the parameters ($\lambda$ \& $\mu$) near the bifurcation values for six sets of initial conditions. (a) The initial position near the point $L_1$. (b) The initial position near the point $L_2$. (c) The initial position near the point $L_3$ and $\mu<-\sqrt{\frac{3}{2}}$. (d) The initial position near the point $L_3$ and $\mu>\sqrt{\frac{3}{2}}$. (e) The initial position near the point $L_4$ and $\mu>\sqrt{\frac{3}{2}}$. (f) The initial position near the point $L_5$ and $\mu>0$.  We observe that the limit of the physical parameter $\omega_{total}\rightarrow -1$.  In early or present time the scalar field may be in phantom phase but the field is attracted to the de-Sitter phase except for (b) and (e).  In (e) the scalar field crosses phantom boundary line and enters into the phantom phase in late timeand would cause Big-Rip.}
\end{figure}

The universe experiences a scalar field dominated non-generic evolution near $L_1$ and $L_2$ for $\lambda>0$ and a scalar field dominated generic evolution for $\lambda<0$ or on the z-nullcline.  Near $L_3$ and $L_4$, a scalar field dominated non-generic evolution of the universe occur at $\mu \approx \pm \sqrt{\frac{3}{2}}$.  At $\mu \approx 0$ a scaling non-generic evolution occur near $L_5$ (see FIG.\ref{Model2}).
\bigbreak

\begin{figure}[htbp!]
	\begin{subfigure}{0.34\textwidth}
		\includegraphics[width=.9\linewidth]{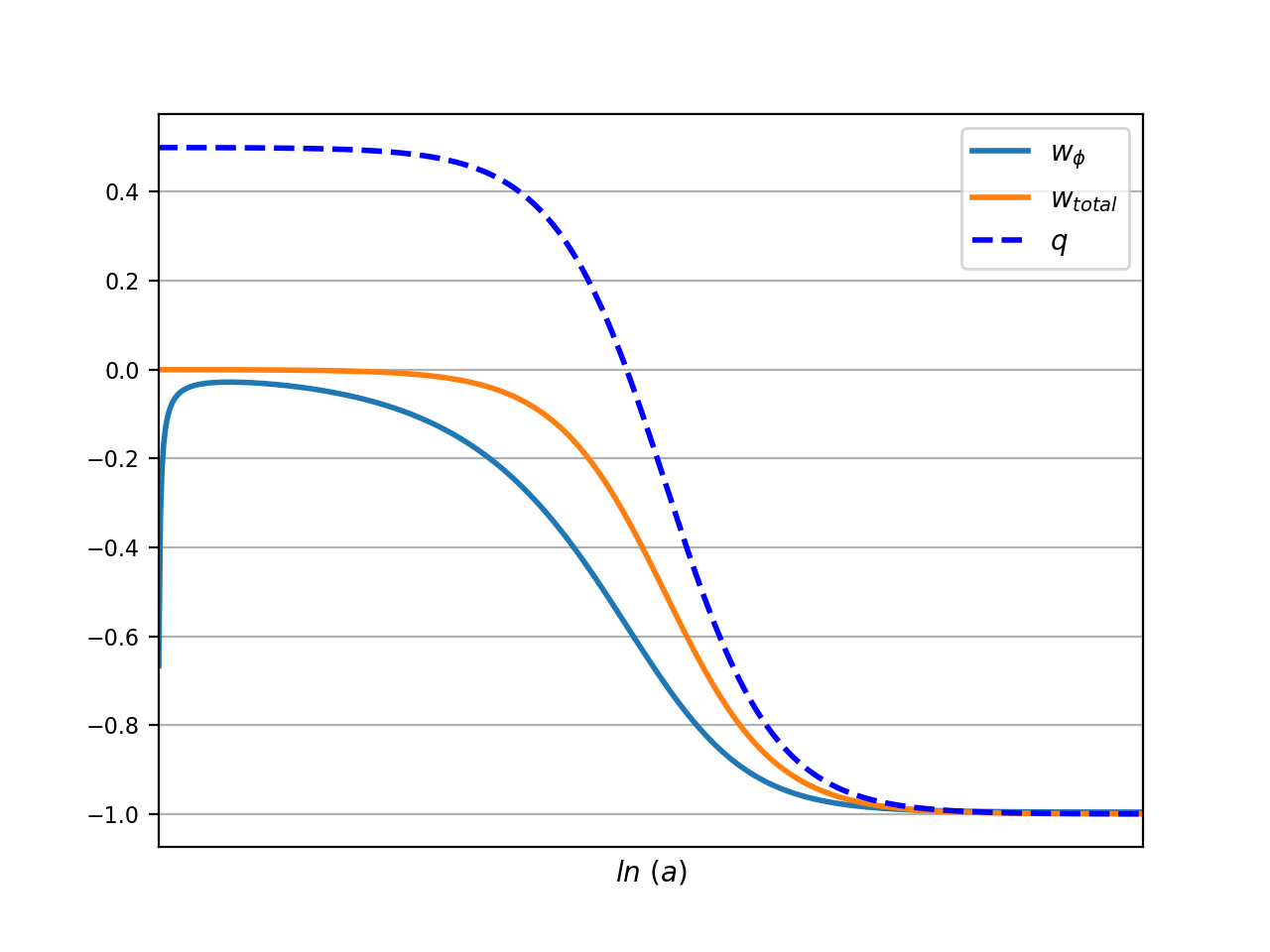}
		\caption{}
		\label{fig:R1}
	\end{subfigure}%
	\begin{subfigure}{0.34\textwidth}
		\includegraphics[width=.9\linewidth]{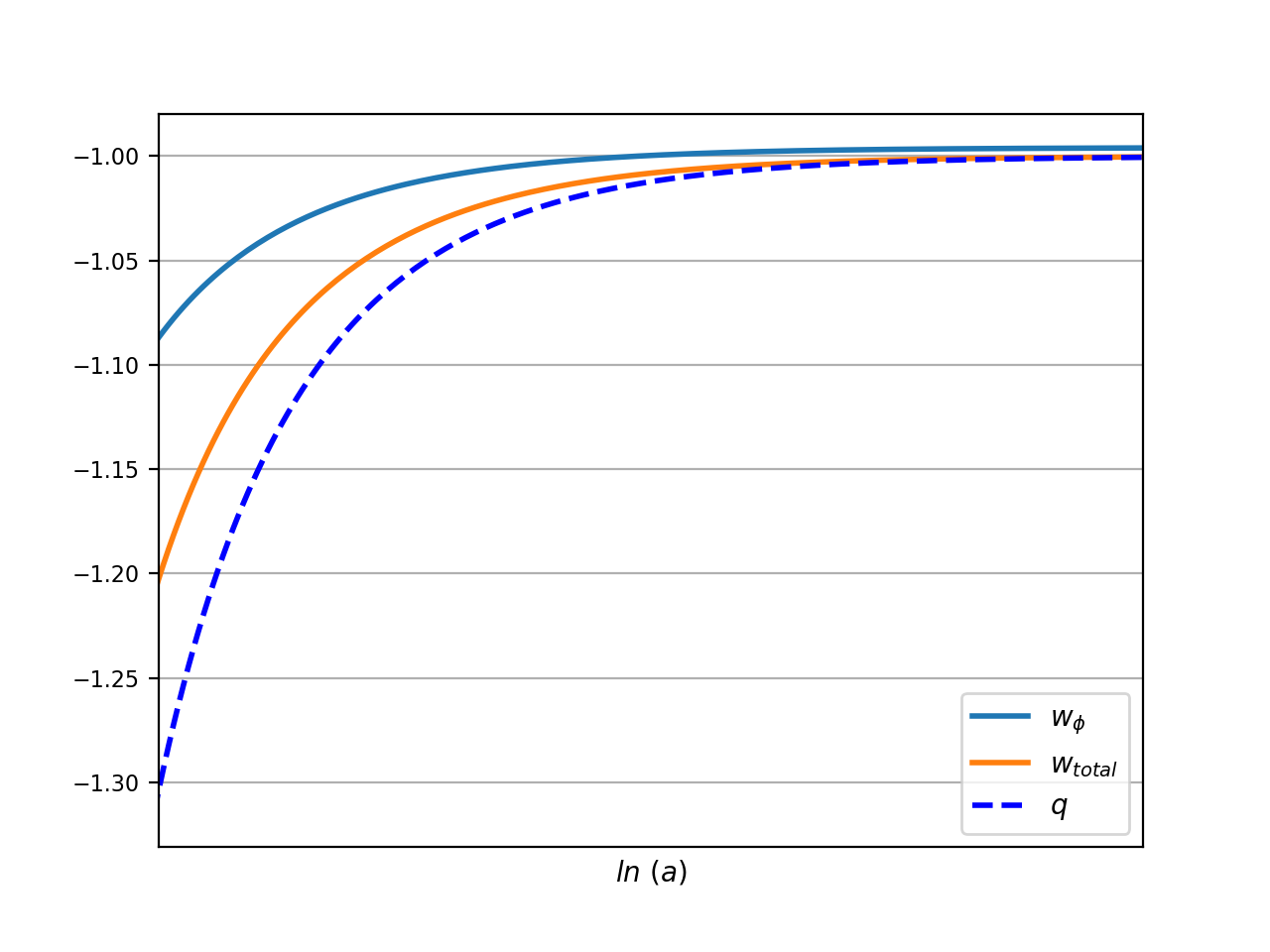}
		\caption{}
		\label{fig:R2}
	\end{subfigure}%
	\begin{subfigure}{.34\textwidth}
		\includegraphics[width=.9\linewidth]{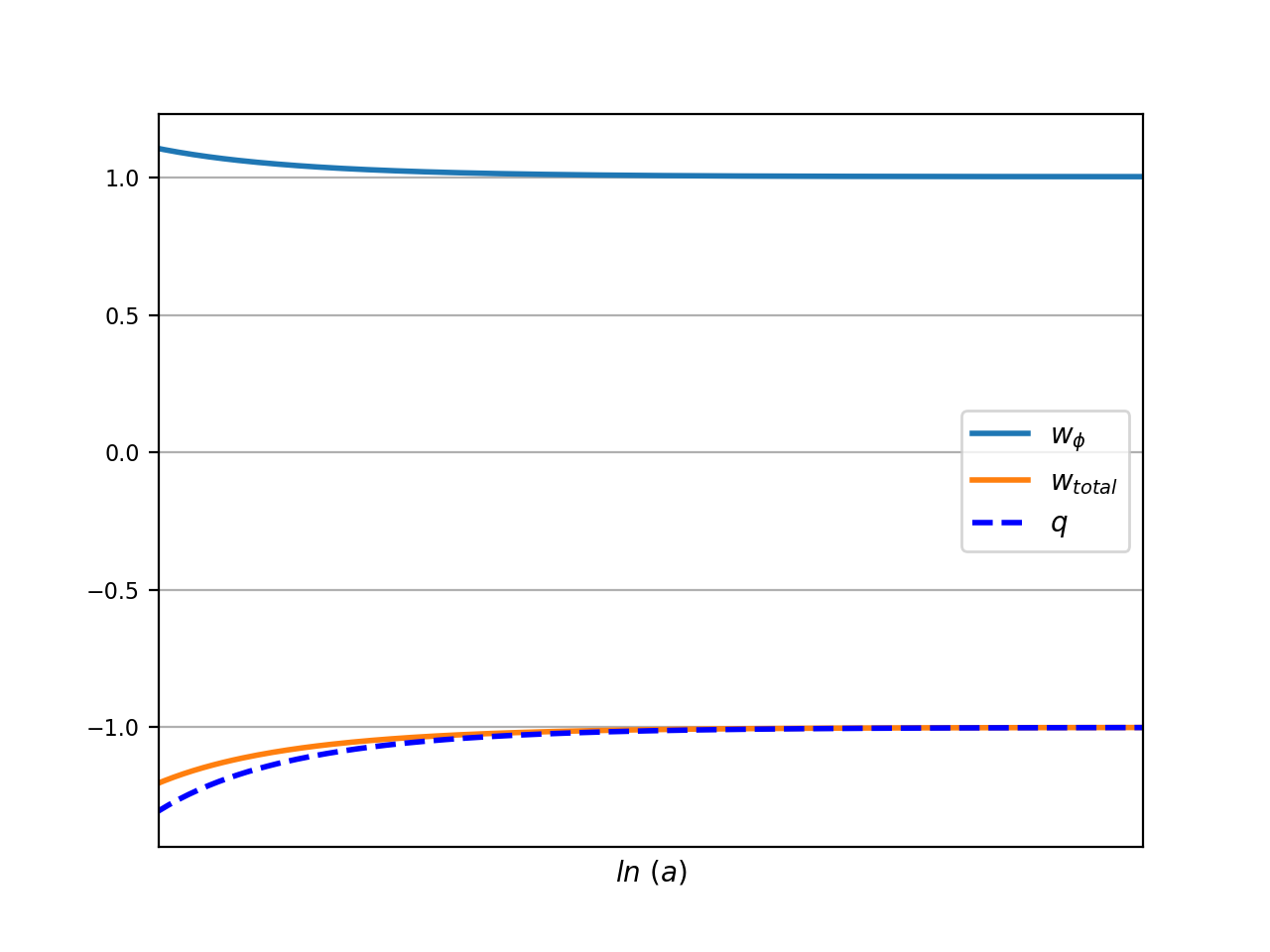}
		\caption{}
		\label{fig:R3}
	\end{subfigure}

	\caption{Qualitative evolution of the physical variables $\omega_{total}$, $\omega_{\phi}$ and $q$ for perturbation of the parameters ($\lambda$ \& $\mu$) near the bifurcation values each of \textit{Model 3}, \textit{Model 4} and \textit{Model 5} for three sets of initial conditions.  The initial positions in (a), (b) and (c) are near \underline{$R_1$, $R_2$ and $R_3$} (\underline{$M_1$, $M_2$ and $M_3$}/\underline{$N_1$, $N_2$ and $N_3$}) respectively. \label{fig:Model3} }
\end{figure}

Model 3 (\ref{M3}) contains three critical points $R_1-R_3$. $R_1$ is saddle for all values of $\mu$. On the $xy$ plane the index of $R_1$ is same as $A_1$. On the projection of the $xy$-plane $R_2$ and $R_3$ are stable nodes for all values of $\lambda$. On the center manifold at $R_2$ or $R_3$, the flow is increasing direction along $z$-axis and the flow is decreasing direction along $z$-axis for $\lambda<0$.   On the $XZ$ or $YZ$ plane, the index of $R_2$ or $R_3$  is zero as around each of them there are two hyperbolic and one parabolic sectors.Thus we note that, for $\mu=0$ and $\lambda=0$, the stability of the system bifurcate.\\  
We observe that no scaling solutions or a tracking solutions exist in this specific model like in the quintessence
theory. However, the critical points which describe the de Sitter solution do not exist in the case of quintessence for
the exponential potential; the universe experiences a fluid dominated non-generic evolution near critical point $R_1$ and a scalar field dominated non-generic evolution near critical point $R_2$ and $R_3$.  For sufficiently flat potential, early or present phantom/non-phantom universe is attracted to $\Lambda$CDM cosmological model (see FIG. \ref{fig:Model3}).\bigbreak

Model 4 (\ref{M4}) contains four critical points  $M_1-M_4$.  $M_1-M_3$ are stable node for $\left(\lambda^2+3\right)>\lambda\mu$ (index 1) and saddle node (index zero) for $\left(\lambda^2+3\right)\leq\lambda\mu$, i.e., the stability of the system bifurcate at $\left(\lambda^2+3\right)=\lambda\mu$.   Thus we find a generic evolution for $\left(\lambda^2+3\right)\neq \lambda\mu$ and no-generic otherwise.  The kinetic dominated solution ($M_1$) and scalar field dominated solutions ($M_2$ and $M_3$) are stable for $\left(\lambda^2+3\right)>\lambda\mu$.   For the energy density, near $M_2$ and $M_3$, we observe that at late times the scalar field dominates $\Omega_X=\Omega_\phi \rightarrow 1$ and $\Omega_m \rightarrow 0$, while the parameter for the equation of state $\omega_{tot}$ have the limits $\omega_{tot} \rightarrow -1$ for sufficiently flat potential.\bigbreak

Model 5 (\ref{M5}) contains three critical points $N_1$, $N_2$, $N_3$.   For $\mu< -\frac{3}{16}$,  the Shilnikov's saddle index \cite{Shilnikov} of $N_1$ is $\nu_{N_1}=\frac{\rho_{N_1}}{\gamma_{N_1}}=0.5$ and saddle value is $\sigma_{N_1}=-\rho_{N_1}+\gamma_{N_1}=0.75$.  As So Shilnikov condition \cite{Shilnikov} is satisfied as $\nu_{N_1}<1$ and $\sigma_{N_1}>0$.  The second Shilnikov's saddle value $\sigma^{(2)}_{N_1}=-2\rho_{N_1}+\gamma_{N_1}=0$.  So, by L. Shilnikov's theorem (Shilnikov, 1965) \cite{Shilnikov}  there are countably many saddle periodic orbits in a neighborhood of the homoclinic loop of the saddle-focus $N_1$.  As $\nu_{N_1}$ is invariant for any choice of $\mu$, so Shilnikov's bifurcation does not appear.  For $-\frac{3}{16}<\mu < 0$, the vector field near $N_1$ is saddle in character.  On the other hand, $N_1$ is saddle for $\mu>0$.  So, $\mu=0$ is a bifurcation value for the bifurcation point $N_1$.  Similarly, $\lambda=0$ is a bifurcation point for the bifurcation points $N_2$ and $N_3$.  We observe scalar field dominated solutions near $N_2$ and $N_3$ which exists at bifurcation value, i.e., for sufficiently flat universe and attracted to $\Lambda$CDM cosmological model.  \\


\section{Brief discussion and concluding remarks \label{conclusion}}
The present work deals with a detailed dynamical system analysis of the interacting DM and DE cosmological model in the background of FLRW geometry.  The DE is chose as a phantom scalar field with self-interacting potential while varying mass (a function of the scalar field) DM is chosen as dust.  The potential of the scalar field and the varying mass of DM are chosen as exponential or power-law form (or a product of them) and five possible combination of them are studied.\bigbreak
\textbf{Model 1: $V(\phi)=V_0\phi^{-\lambda}, M_{_{DM}}(\phi)=M_0\phi^{-\mu}$}\par
For case (i), i.e., $\mu\neq 0, \lambda\neq 0$; there are three non-hyperbolic critical points $A_1$, $A_2$, $A_3$ of which $A_1$ corresponds to DM dominated decelerating phase (dust era) while $A_2$ and $A_3$ purely DE dominated and they represent the $\Lambda$CDM model (i.e., de-Sitter phase) of the universe.\par
For case (ii), i.e., $\mu\neq 0, \lambda=0$; there is one critical point and two space of critical points.  The cosmological consequence of these critical points are similar to case (i).\par
For case (iii), i.e., $\mu= 0, \lambda\neq 0$; there is one  space of critical points and two distinct critical points.  But as before the cosmological analysis is identical to case (i).\par
For the fourth case, i.e., $\mu=0, \lambda=0$; there are three space of critical points $(S_1,S_2,S_3)$ which are all non-hyperbolic in nature and are identical to the critical points in case (ii).  Further, considering the vector fields in $Z=constant$ plane, it is found that for the critical point $S_1$, every point on $Z-$ axis is a saddle node while for critical points $S_2$ and $S_3$ every point on $Z-$axis is a stable star.\bigbreak
\textbf{Model 2: $V(\phi)=V_0\phi^{-\lambda}, M_{_{DM}}(\phi)=M_1e^{-\kappa\mu\phi}$}\par
The autonomous system for this model has five non-hyperbolic critical points $L_i$, $i=1,\ldots,5$.  For $L_1$ and $L_2$, the cosmological model is completely DE dominated and the model describes cosmic evolution at the phantom barrier.  The critical points $L_3$ and $L_4$ are DE dominated cosmological solution ($\mu^2>3$) representing the $\Lambda$CDM model.  The critical point $L_5$ corresponds to ghost (phantom) scalar field and it describes the cosmic evolution in phantom domain ($2\mu^2>3$).\bigbreak
\textbf{Model 3: $V(\phi)=V_1e^{-\kappa\lambda\phi}, M_{_{DM}}(\phi)=M_0\phi^{-\mu}$}\par
There are three non-hyperbolic critical points in this case.  The first one (i.e., $R_1$) is purely DM dominated cosmic evolution describing the dust era while the other two critical points (i.e., $R_2$, $R_3$) are fully dominated by DE and both describe the cosmic evolution in the phantom era.\bigbreak
\textbf{Model 4: $V(\phi)=V_1 e ^{-\kappa\lambda\phi}, M_{_{DM}}(\phi)=M_1e^{-\kappa\mu\phi}$}\par
The autonomous system so formed in this case has four critical points $M_i$, $i=1,\ldots,4$ which may be hyperbolic/non-hyperbolic depending on the parameters involved.  The critical point $M_1$ represents DE as ghost scalar field and it describes the cosmic evolution in the phantom domain.  For the critical points $M_2$ and $M_3$, the cosmic evolution is fully DE dominated and is also in the phantom era.  The cosmic era corresponding to the critical point $M_4$ describes scaling solution where both DM and DE contribute to the cosmic evolution.\bigbreak
\textbf{Model 5: $V(\phi)=V_2\phi^{-\lambda} e ^{-\kappa\lambda\phi}, M_{_{DM}}(\phi)=M_2\phi^{-\mu}e^{-\kappa\mu\phi}$}\par
This model is very similar to either model $4$ or model $1$, depending on the choices of the dimensionless variables $x$ and $z$.  For $z=0$, the model reduces to model $4$ while for $x=0$ the model is very similar to model $1$ and hence the cosmological analysis is very similar to that.\par
Finally, using Poincar\'{e} index theorem, Euler Poincar\'{e} characteristic is determined for bifurcation analysis of the above cases from the point of view of the cosmic evolution described by the equilibrium points.  Lastly, inflationary era of cosmic evolution is studied by using bifurcation analysis.

\begin{acknowledgements}
	The author Soumya Chakraborty is grateful to CSIR, Govt. of India for giving Junior Research Fellowship (CSIR Award No: 09/096(1009)/2020-EMR-I) for the Ph.D work.
	The author S. Mishra is grateful to CSIR, Govt. of India for giving Senior Research Fellowship (CSIR Award No: 09/096 (0890)/2017-EMR-I) for the Ph.D work.  The author Subenoy Chakraborty is thankful to Science and Engineering Research Board (SERB) for awarding MATRICS Research Grant support (File No: MTR/2017/000407).\\ 
\end{acknowledgements}

		 \bibliographystyle{unsrt}
		 \bibliography{references}\bigbreak 	

\begin{thebibliography}{10}

\bibitem{Riess:1998cb}
Adam~G. Riess et~al.
\newblock {Observational evidence from supernovae for an accelerating universe
  and a cosmological constant}.
\newblock {\em Astron. J.}, 116:1009--1038, 1998.

\bibitem{Perlmutter:1998np}
S.~Perlmutter et~al.
\newblock {Measurements of $\Omega$ and $\Lambda$ from 42 high redshift
  supernovae}.
\newblock {\em Astrophys. J.}, 517:565--586, 1999.

\bibitem{Spergel:2003cb}
D.N. Spergel et~al.
\newblock {First year Wilkinson Microwave Anisotropy Probe (WMAP) observations:
  Determination of cosmological parameters}.
\newblock {\em Astrophys. J. Suppl.}, 148:175--194, 2003.

\bibitem{Allen:2004cd}
S.W. Allen, R.W. Schmidt, H.~Ebeling, A.C. Fabian, and L.~van Speybroeck.
\newblock {Constraints on dark energy from Chandra observations of the largest
  relaxed galaxy clusters}.
\newblock {\em Mon. Not. Roy. Astron. Soc.}, 353:457, 2004.

\bibitem{Riess:2004nr}
Adam~G. Riess et~al.
\newblock {Type Ia supernova discoveries at z \ensuremath{>} 1 from the Hubble
  Space Telescope: Evidence for past deceleration and constraints on dark
  energy evolution}.
\newblock {\em Astrophys. J.}, 607:665--687, 2004.

\bibitem{RevModPhys.61.1}
Steven Weinberg.
\newblock The cosmological constant problem.
\newblock {\em Rev. Mod. Phys.}, 61:1--23, Jan 1989.

\bibitem{Caldwell:2003vq}
Robert~R. Caldwell, Marc Kamionkowski, and Nevin~N. Weinberg.
\newblock {Phantom energy and cosmic doomsday}.
\newblock {\em Phys. Rev. Lett.}, 91:071301, 2003.

\bibitem{Vikman:2004dc}
Alexander Vikman.
\newblock {Can dark energy evolve to the phantom?}
\newblock {\em Phys. Rev. D}, 71:023515, 2005.

\bibitem{Nojiri:2005sr}
Shin'ichi Nojiri and Sergei~D. Odintsov.
\newblock {Inhomogeneous equation of state of the universe: Phantom era, future
  singularity and crossing the phantom barrier}.
\newblock {\em Phys. Rev. D}, 72:023003, 2005.

\bibitem{Saridakis:2009pj}
Emmanuel~N. Saridakis.
\newblock {Phantom evolution in power-law potentials}.
\newblock {\em Nucl. Phys. B}, 819:116--126, 2009.

\bibitem{Setare:2008mb}
M.R. Setare and E.N. Saridakis.
\newblock {Braneworld models with a non-minimally coupled phantom bulk field: A
  Simple way to obtain the -1-crossing at late times}.
\newblock {\em JCAP}, 03:002, 2009.

\bibitem{Feng:2004ad}
Bo~Feng, Xiu-Lian Wang, and Xin-Min Zhang.
\newblock {Dark energy constraints from the cosmic age and supernova}.
\newblock {\em Phys. Lett. B}, 607:35--41, 2005.

\bibitem{Guo:2004fq}
Zong-Kuan Guo, Yun-Song Piao, Xin-Min Zhang, and Yuan-Zhong Zhang.
\newblock {Cosmological evolution of a quintom model of dark energy}.
\newblock {\em Phys. Lett. B}, 608:177--182, 2005.

\bibitem{Feng:2004ff}
Bo~Feng, Mingzhe Li, Yun-Song Piao, and Xinmin Zhang.
\newblock {Oscillating quintom and the recurrent universe}.
\newblock {\em Phys. Lett. B}, 634:101--105, 2006.

\bibitem{Amendola:2006qi}
Luca Amendola, Miguel Quartin, Shinji Tsujikawa, and Ioav Waga.
\newblock {Challenges for scaling cosmologies}.
\newblock {\em Phys. Rev. D}, 74:023525, 2006.

\bibitem{Chen:2008ft}
Xi-ming Chen, Yun-gui Gong, and Emmanuel~N. Saridakis.
\newblock {Phase-space analysis of interacting phantom cosmology}.
\newblock {\em JCAP}, 04:001, 2009.

\bibitem{Nunes:2004wn}
Nelson~J. Nunes and D.F. Mota.
\newblock {Structure formation in inhomogeneous dark energy models}.
\newblock {\em Mon. Not. Roy. Astron. Soc.}, 368:751--758, 2006.

\bibitem{Clifton:2007tn}
T.~Clifton and John~D. Barrow.
\newblock {The Ups and downs of cyclic universes}.
\newblock {\em Phys. Rev. D}, 75:043515, 2007.

\bibitem{Xu:2012jf}
Chen Xu, Emmanuel~N. Saridakis, and Genly Leon.
\newblock {Phase-Space analysis of Teleparallel Dark Energy}.
\newblock {\em JCAP}, 07:005, 2012.

\bibitem{Zhang:2005jj}
Hong-Sheng Zhang and Zong-Hong Zhu.
\newblock {Interacting chaplygin gas}.
\newblock {\em Phys. Rev. D}, 73:043518, 2006.

\bibitem{Fadragas:2014mra}
Carlos~R. Fadragas and Genly Leon.
\newblock {Some remarks about non-minimally coupled scalar field models}.
\newblock {\em Class. Quant. Grav.}, 31(19):195011, 2014.

\bibitem{Gonzalez:2007ht}
Tame Gonzalez and Israel Quiros.
\newblock {Exact models with non-minimal interaction between dark matter and
  (either phantom or quintessence) dark energy}.
\newblock {\em Class. Quant. Grav.}, 25:175019, 2008.

\bibitem{Anderson:1997un}
Greg~W. Anderson and Sean~M. Carroll.
\newblock {Dark matter with time dependent mass}.
\newblock In {\em {1st International Conference on Particle Physics and the
  Early Universe}}, pages 227--229, 9 1997.

\bibitem{PhysRevLett.64.123}
T.~Damour, G.~W. Gibbons, and C.~Gundlach.
\newblock Dark matter, time-varying g, and a dilaton field.
\newblock {\em Phys. Rev. Lett.}, 64:123--126, Jan 1990.

\bibitem{Farrar:2003uw}
Glennys~R. Farrar and P.James~E. Peebles.
\newblock {Interacting dark matter and dark energy}.
\newblock {\em Astrophys. J.}, 604:1--11, 2004.

\bibitem{Hoffman:2003ru}
Mark~B. Hoffman.
\newblock Cosmological constraints on a dark matter -- dark energy interaction.
\newblock {\em arXiv: Astrophysics}, 7 2003.

\bibitem{Zhang:2005rg}
Xin Zhang.
\newblock {Coupled quintessence in a power-law case and the cosmic coincidence
  problem}.
\newblock {\em Mod. Phys. Lett. A}, 20:2575, 2005.

\bibitem{Berger:2006db}
Micheal~S. Berger and Hamed Shojaei.
\newblock {Interacting dark energy and the cosmic coincidence problem}.
\newblock {\em Phys. Rev. D}, 73:083528, 2006.

\bibitem{PhysRevD.66.043528}
Luca Amendola and Domenico Tocchini-Valentini.
\newblock Baryon bias and structure formation in an accelerating universe.
\newblock {\em Phys. Rev. D}, 66:043528, Aug 2002.

\bibitem{PhysRevD.67.103523}
Massimo Pietroni.
\newblock Brane worlds and the cosmic coincidence problem.
\newblock {\em Phys. Rev. D}, 67:103523, May 2003.

\bibitem{PhysRevD.75.083506}
Luca Amendola, Gabriela~Camargo Campos, and Rogerio Rosenfeld.
\newblock Consequences of dark matter-dark energy interaction on cosmological
  parameters derived from type ia supernova data.
\newblock {\em Phys. Rev. D}, 75:083506, Apr 2007.

\bibitem{Amendola:1999er}
Luca Amendola.
\newblock {Coupled quintessence}.
\newblock {\em Phys. Rev. D}, 62:043511, 2000.

\bibitem{Comelli:2003cv}
D.~Comelli, M.~Pietroni, and A.~Riotto.
\newblock {Dark energy and dark matter}.
\newblock {\em Phys. Lett. B}, 571:115--120, 2003.

\bibitem{PhysRevD.69.063517}
Urbano Fran\ifmmode~\mbox{\c{c}}\else \c{c}\fi{}a and Rogerio Rosenfeld.
\newblock Age constraints and fine tuning in variable-mass particle models.
\newblock {\em Phys. Rev. D}, 69:063517, Mar 2004.

\bibitem{10.1140/epjc/s10052-019-6839-8}
Sudip Mishra and Subenoy Chakraborty.
\newblock {Stability and bifurcation analysis of interacting f(T) cosmology}.
\newblock {\em Eur. Phys. J.}, C79(4):328, 2019.

\bibitem{1950261}
Sudip Mishra and Subenoy Chakraborty.
\newblock {A non-canonical scalar field cosmological model: Stability and
  bifurcation analysis}.
\newblock {\em Mod. Phys. Lett.}, A34(32):1950261, 2019.

\bibitem{1812.01975}
Sudip Mishra and Subenoy Chakraborty.
\newblock {Dynamical system analysis of Einstein–Skyrme model in a
  Kantowski–Sachs spacetime}.
\newblock {\em Annals Phys.}, 406:207--219, 2019.

\bibitem{Leon:2009dt}
Genly Leon and Emmanuel~N. Saridakis.
\newblock {Phantom dark energy with varying-mass dark matter particles:
  acceleration and cosmic coincidence problem}.
\newblock {\em Phys. Lett. B}, 693:1--10, 2010.

\bibitem{Chakraborty:2020vkp}
Soumya Chakraborty, Sudip Mishra, and Subenoy Chakraborty.
\newblock {Dynamical system analysis of three-form field dark energy model with
  baryonic matter}.
\newblock {\em Eur. Phys. J.}, C80(9):852, 2020.

\bibitem{1111.6247}
Christian~G. Boehmer, Nyein Chan, and Ruth Lazkoz.
\newblock {Dynamics of dark energy models and centre manifolds}.
\newblock {\em Phys. Lett. B}, 714:11--17, 2012.

\bibitem{0-387-95116-4}
Lawrence Perko.
\newblock {\em {Differential equations and Dynamical systems}}.
\newblock {Springer-Verlag, New York. Inc.}, {Third} edition, 1991.

\bibitem{Shilnikov}
LP~Shilnikov and Andrey Shilnikov.
\newblock Shilnikov bifurcation.
\newblock {\em Schopedia}, 2007.
\newblock \url{https://www.researchgate.net/publication/220580167}.

\end{thebibliography}
\end{document}